\documentclass[twocolumn]{autart}
\usepackage{amsfonts,amsmath,amssymb}
\usepackage{lineno,hyperref}
\usepackage{subfigure}
\usepackage{algorithm}
\usepackage{algorithmic}
\usepackage{savesym}
\savesymbol{AND}
\usepackage{color}
\usepackage{graphicx}
\graphicspath{{./figures_eps/}}
\usepackage{cases}
\usepackage{comment}
\newtheorem{mydef}{Definition}
\newtheorem{mythm}{Theorem}
\newtheorem{myprop}{Proposition}
\newtheorem{mylem}{Lemma}
\newtheorem{myas}{Assumption}
\newcommand{\rsec}[1]{Section\,\ref{#1}}
\newtheorem{myrem}{Remark}

\newcommand{\rfig}[1]{Fig.\,\ref{#1}} 
 
\newcommand{\req}[1]{\eqref{#1}} 
 
\newcommand{\rtab}[1]{Table\,\ref{#1}}
\newcommand{\rlem}[1]{Lemma\,\ref{#1}}
\newcommand{\rdef}[1]{Definition\,\ref{#1}}
\newcommand{\rrem}[1]{Remark\,\ref{#1}}
\newcommand{\ralg}[1]{Algorithm\,\ref{#1}}
\newcommand{\ras}[1]{Assumption\,\ref{#1}}
\newcommand{\rline}[1]{line\,\ref{#1}}
\newcommand{\rprop}[1]{Proposition\,\ref{#1}}

\renewcommand{\labelenumi}{(\roman{enumi})}
\renewcommand\thefootnote{\arabic{footnote})}
\newcommand{\qedwhite}{\hfill \ensuremath{\Box}}
\begin{document}

\begin{frontmatter}

\title{Learning-based Symbolic Abstractions for \\ Nonlinear Control Systems\thanksref{footnoteinfo}}
\thanks[footnoteinfo]{This work was supported by JST ERATO Grant Number JPMJER1603, Japan, JST CREST Grant Number JPMJCR2012, Japan, and by JSPS KAKENHI Grant Number 21K14184.}
%Event-Triggered Design for \\ Nonlinear Model Predictive Controllers: \\A New Stability and Reduced Evaluations \tnoteref{mytitlenote}}
\author[First]{Kazumune Hashimoto}\ead{hashimoto@eei.eng.osaka-u.ac.jp},
\author[Second]{Adnane Saoud}\ead{Adnane.Saoud@l2s.centralesupelec.fr},
\author[Third]{Masako Kishida}\ead{kishida@nii.ac.jp},
\author[Forth]{Toshimitsu Ushio}\ead{ushio@sys.es.osaka-u.ac.jp},
\author[Fifth]{Dimos V. Dimarogonas}\ead{dimos@kth.se}
\address[First]{Graduate School of Engineering, Osaka University, Suita, Japan}\address[Second]{ Laboratoire des Signaux et Syst$\grave{e}$mes, Universit${\rm \acute{e}}$ Paris-Saclay, CNRS, CentraleSup${\rm \acute{e}}$lec} \address[Third]{National Institute of Informatics (NII), Tokyo, Japan.}
\address[Forth]{Graduate School of Engineering and Science, Osaka University, Toyonaka, Japan}\address[Fifth]{School of Electrical Engineering, KTH Royal Institute of Technology, Stockholm, Sweden.}

\begin{abstract}
Symbolic models or abstractions are known to be powerful tools for the control design of cyber-physical systems (CPSs) with logic specifications. In this paper, we investigate a novel \textit{learning-based} approach to the construction of symbolic models for nonlinear control systems. In particular, the symbolic model is constructed based on \textit{learning} the un-modeled part of the dynamics from training data based on state-space exploration, and the concept of an alternating simulation relation that represents behavioral relationships with respect to the original control system. Moreover, we aim at achieving \textit{safe exploration}, meaning that the trajectory of the system is guaranteed to be in a safe region for all times while collecting the training data. In addition, we provide some techniques to reduce the computational load\textcolor{black}{, in terms of memory and computation time, }of constructing the symbolic models and the safety controller synthesis, so as to make our approach practical. Finally, a numerical simulation illustrates the effectiveness of the proposed approach. 
\end{abstract}


\end{frontmatter}

%% main text
\section{Introduction}
In cyber-physical systems (CPS), computational devices are tightly integrated with physical processes. Embedded computers monitor the behavior of the physical processes through sensors, and usually control them through actuators using feedback loops. Nowadays, CPSs are ubiquitous in modern control engineering, including automobiles, aircraft, building control systems, chemical plants, transportation systems, and so on. Many CPSs are safety critical or mission critical: it must ensure that the system operates correctly meeting the satisfaction of safety or some desired specifications. Formal methods are known to provide essential tools for the design of CPSs, as they give theoretical or rigorous mathematical \textit{proofs} that the system works correctly meeting the desired specification \cite{seshia2015formal}. While the formal methods have been originally developed in software engineering that aims at finding bugs or security vulnerabilities in the software, the methodologies have been recently recognized to be useful in other applications, including the control design of CPSs. In particular, one of the most successful methods that interface the formal methods and the control design of CPSs is the so-called \textit{symbolic control}, see, e.g., \cite{pola2019control}. The main objective of the symbolic control is to design controllers for CPSs with logic specifications (as detailed below). In such approaches, \textit{symbolic models} or \textit{abstractions} are constructed based on the original control systems. Roughly speaking, while the original control system is represented in a \textit{continuous} state (and input) space, the symbolic model is represented in a \textit{discrete} state (and input) space, while preserving the behavior of the original control system. As such, controllers can be designed based on several algorithmic techniques from supervisory control of discrete event systems, such as a safety/reachability game \cite{tabuadabook2009}. 

The symbolic approach is known to be a powerful tool for the control design of CPSs in the following three ways. First, it allows us to synthesize controllers for general nonlinear dynamical systems with state and input constraints. Second, by constructing the symbolic model, we can take into account the constraints that are imposed on the cyber part with regard to the digital platform, such as a quantization effect. Third, it allows us to synthesize controllers under various control specifications, including safety, reachability, or more complex ones such as those expressed by linear temporal logic (LTL) formulas or automata on infinite strings. As previously mentioned, symbolic models or abstractions are constructed such that they are represented in a discrete state space while preserving the behavior of the original control system. More formally, behavioural relationships such as the concept of approximate (bi-)simulation relation, see, e.g.,  \cite{tabuadabook2009,girard2012a,girard2009a,girard2010a,pola2009symbolic,zamani2012a,meyer2018,hashimoto2019d,rungger2016c,mizoguchi,outputsymbolic}, are used to relate the behaviours of the original control system and its symbolic model. 
%More formally, the symbolic models are constructed based on behavioral relationships between the two systems, such as the concept of (bi)simulation relations, see, e.g., \cite{tabuadabook2009,girard2012a,girard2009a,girard2010a,pola2009symbolic,zamani2012a,meyer2018,hashimoto2019d,rungger2016c,mizoguchi}. 
For example, \cite{girard2009a} employs an approximate bisimulation relation to construct the symbolic model for nonlinear, incrementally asymptotically stable systems. \cite{zamani2012a} employs an approximate alternating simulation relation, so that the symbolic models can be constructed for general (incrementally forward complete) nonlinear systems without any assumption on stability. Moreover, \cite{rungger2016c,mizoguchi} characterize the notion of robustness for input-output dynamically stable systems based on the concept of a contractive approximate simulation relation. 

In this paper, we focus on investigating the construction of symbolic models for nonlinear control systems. 
In particular, we consider the case where the dynamics of the plant includes state-dependent, \textit{un-modeled} dynamics. In contrast to the aforecited abstraction schemes, we propose a \textit{learning-based} solution to this problem, in which the symbolic model is constructed based on learning the un-modeled dynamics from training data. Moreover, we aim at achieving \textit{safe-exploration}, meaning that the trajectory of the system stays inside a safe set for all times while collecting the training data. Achieving safe exploration is particularly useful for safety critical CPSs, see, e.g., \cite{seshia2016towards}. 
More technically, as a starting point of our approach, we employ the Gaussian process (GP) regression \cite{rasmussen} in order to estimate the un-modeled dynamics from training data. 
As we will see later, it is shown that, under some smoothness assumption on the un-modeled dynamics, an error bound on the un-modeled dynamics can be derived based on the result from \cite{srinivas2}. \textcolor{black}{Note that, in contrast to previous approaches of learning-based controller synthesis with the GP regression (e.g., \cite{felix1,felix2}) that make use of an error (or regret) bound that involves an information gain, here we will make use of a {deterministic} error bound that does not involve the information gain, which has been also derived in \cite{srinivas2} (for details, see \rlem{boundlemma} and \rrem{comppreviousrem} in this paper).} Based on this error bound and the concept of an approximate alternating simulation relation \cite{zamani2012a}, we then provide an approach to construct the symbolic model. To achieve the safe exploration, we also provide a safety controller synthesis via a safety game \cite{tabuadabook2009}. 
Finally, we provide an overall algorithm that collects the training data from scratch and constructs the symbolic model. Along with this algorithm, we provide several techniques to reduce the computational load of constructing the symbolic model and the safety controller synthesis. 
In particular, we provide a {lazy} abstraction scheme, in which the transitions of the symbolic model are updated only around the region where the training data is collected. 

\textit{(Related works)}: 
\textcolor{black}{The approach presented in this paper is related to previous literature in terms of {symbolic control} (or temporal logics) and controller synthesis for dynamical systems learned by training data. In what follows, we discuss how our approach differs from previous works and highlight our main contributions.} 

As previously mentioned, there have been a wide variety of symbolic control techniques for dynamical systems, e.g.,  \cite{girard2012a,girard2009a,girard2010a,pola2009symbolic,zamani2012a,meyer2018b,tabuada2008c,outputsymbolic}; however, most of the previous approaches typically assume that the dynamics of the plant is completely \textit{known} or they consider uniform disturbance that is not learned from training data. 
\textcolor{black}{The learning-based approach is advantageous over the uniform disturbance-based approach in the following sense. 
%in contrast to the uniform disturbance approach is that we can find a \textit{larger} region that guarantees safety (i.e., the controlled invariant set), especially as the state exploration progresses. 
In the uniform disturbance-based approach, every transition of the symbolic model is defined by taking the \textit{worst case effect} of the un-modeled function, since the un-modeled function will not be learned from data. On the other hand, in the learning-based approach, the un-modeled function will be learned and thus its uncertainty will decrease as the state exploration progresses. 
Hence, the symbolic model will have fewer redundant transitions than the uniform disturbance-based approach, and this leads to obtaining a larger region that guarantees safety (i.e., controlled invariant set).} 
%, which means that we can enlarge a region that guarantees safety,especially as the iteration progresses. 
\textcolor{black}{To the best of our knowledge, there are only few works of symbolic or temporal logic control for a dynamical system that is partially unknown and is learned by training data (e.g., by the GP regression) \cite{jackson,gangchen}. In \cite{jackson}, the authors provided a way to obtain a finite abstraction using interval Markov decision processes (IMDPs) with the unknown dynamics learned by the GP regression. The abstraction has been then utilized for safety verification.}
{\color{black} Our approach is different from this previous work in the following sense: first, while the proposed approach in~\cite{jackson} makes it possible to provide probabilistic guarantees, in this paper we are able to provide deterministic guarantees. Second, while in~\cite{jackson} the symbolic model is constructed to deal with only safety specifications, in this paper, we are constructing the symbolic abstraction in the more general sense of alternating simulation relations, i.e., we can refine a controller for the symbolic model into a controller for the original system for any specification, and not just safety. Finally, for the particular class of safety specifications, the authors in \cite{jackson} provided a way of constructing a symbolic model with \textit{given initial} training data, and they did not provide an approach to update the symbolic model when new training data are collected online, and which is the main issue in learning-based control, since the objective is to exploit the new training data collected online. In contrast, our approach provides a new computationally efficient approach to collect new training data while reducing the computation load to update the symbolic model and to synthesize safety controllers.}
%As we will see later in Section~5.1, this will be achieved by deriving a deterministic error bound that becomes tighter as the number of the training data increases.
\textcolor{black}{
In \cite{gangchen}, the authors provided a way to detect faults using signal temporal logic (STL) for partially unknown dynamical systems and these are learned by the GP regression. However, the problem setup considered in \cite{gangchen} is different from the one considered in this paper. Specifically, while \cite{gangchen} considered a \textit{monitoring} scheme in which they monitor behaviors of the system without control inputs and check if a given STL formula is satisfied, this paper considers a \textit{synthesis} scheme in which we find a controller to satisfy certain specifications (expressed by, e.g., temporal logic formulas). The proposed approach is also different; while our approach aims at constructing symbolic models from training data while guaranteeing a given specification, the approach in \cite{gangchen} provided a monitoring scheme by employing a robustness degree of STL formulas.} 

\textcolor{black}{Apart from the use of symbolic control, various learning-based controller synthesis techniques with the GP regression have been proposed. Most of the previous works aim at synthesizing controllers to achieve stability/tracking \cite{umlauft2019feedback,modelfree4,modelfree5,hashimoto2021a}, or to guarantee safety \cite{felix1,felix2,safelearning,safelearning1,safelearning2,safelearning10,safelearning11,safelearning12,felix3}. 
Since we here construct a safety controller to achieve a safe exploration, our approach is particularly related to the second category, i.e., \cite{felix1,felix2,safelearning,safelearning1,safelearning2,safelearning10,safelearning11,safelearning12,felix3}. For example, the authors in \cite{felix1} (resp. \cite{felix2}) proposed an approach to learn a region of attraction (ROA) using safety controllers for continuous-time systems: $\dot{x}(t) = f(x(t), u(t)) + g(x(t), u(t))$ (resp. discrete-time systems: $x(k+1)=h(x(k),u(k))+g(x(k), u(k))$), where the function $g(\cdot)$ is unknown and it is learned by the GP regression. 
The assumptions on the unknown function $g(\cdot)$ that are made in \cite{felix1} and \cite{felix2} are the {same} as the ones we are using in this paper, namely the fact that the unknown function lies in the reproducing kernel Hilbert space (RKHS).
However, the authors in \cite{felix1,felix2} assumed the existence of a \textit{known} Lyapunov function for the nominal system $\dot{x}(t) = f(x(t), u(t))$ ($x(k+1)=h(x(k),u(k))$), for which the computation may be difficult for general nonlinear systems. The approach presented in this paper allows us to deal with more general complex specifications (including safety) without requiring the existence of a Lyapunov function. 
The approaches presented in \cite{safelearning10,safelearning11,safelearning12} used a control barrier function and \cite{safelearning2} used a Hamilton-Jacobi-Issac (HJI) equation to synthesize safety controllers with the GP regression. The proposed approach presented in this paper is significantly different from \cite{safelearning10,safelearning11,safelearning12,safelearning2} in the following sense. First, note that while the goal of the previous work is to derive a safety controller, our \textit{main} goal is to construct a symbolic model. Constructing the symbolic model is beneficial since it allows not only to compute a safety controller, but also controllers from more general complex specifications, such as those expressed by temporal logic specifications and automata on infinite strings. {Moreover, while in \cite{safelearning10,safelearning11} (resp. \cite{safelearning12}), the use of barrier functions makes it only possible to deal with the class of polynomial dynamical systems (resp. input affine systems), the proposed approach in this paper makes it possible to deal with general nonlinear systems, and this is achieved by employing the symbolic models.} Besides, while solving the HJI equation generally requires a heavy computational load, \cite{safelearning2} did not provide a way of speeding up the computation of solving the HJI equation when a new set of training data is obtained. On the other hand, we here propose a way of reducing the computation load to update the symbolic model as well as synthesize safety controllers even if a new set of training data is obtained online.}

\noindent 
\textit{Notation.} 
Let $\mathbb{N}$, $\mathbb{N}_{\geq a}$, $\mathbb{N}_{>a}$, $\mathbb{N}_{a:b}$ be the sets of integers, integers larger than or equal to $a$, integers larger than $a$, and integers from $a$ to $b$ respectively. 
Let $\mathbb{R}$, $\mathbb{R}_{\geq a}$, $\mathbb{R}_{>a}$ be the sets of reals, reals larger than or equal to $a$ and reals larger than $a$, respectively. 
Given $a, b \in \mathbb{R}$ with $a \leq b$, let $ [a, b]$ be the interval set from $a$ to $b$. 
Given $a, b \in \mathbb{R}_{\geq 0}$, we let $[a \pm b] = [a - b, a + b]$. 
Denote by $\|x\|_{\infty}$ the infinity norm of a vector $x$. 
Given $x\in\mathbb{R}^{n}, \varepsilon \in \mathbb{R}_{\geq 0}$, let $\mathcal{B}_{\varepsilon} (x) \subset \mathbb{R}^n$ be the ball set given by $\mathcal{B}_{\varepsilon} (x) = \{ x \in\mathbb{R}^n\ |\ \| x \|_\infty \leq \varepsilon\}$. 
Given $\mathcal{X} \subseteq \mathbb{R}^n$ and $\eta >0$, denote by $[\mathcal{X}]_\eta \subset \mathbb{R}^n$ the lattice in $\mathcal{X}$ with the quantization parameter $\eta$, i.e.,
%\begin{equation*}
$[\mathcal{X}]_\eta =  \{ x\in\mathcal{X} \ |\ x_i = a_i \eta ,\ a_i\in \mathbb{N},\ i = 1, 2, \ldots, n \}$, 
%\end{equation*}
where $x_i \in \mathbb{R}$ is the $i$-th element of $x$. Given $x \in \mathbb{R}^n$, $\mathcal{X}\subseteq \mathbb{R}^n$, denote by $\mathsf{Nearest}_{\mathcal{X}} (x)$ the closest points in $\mathcal{X}$ to $x$, i.e., $\mathsf{Nearest}_{\mathcal{X}} (x) = {\arg\min}_{x' \in \mathcal{X}} \| x - x' \|_\infty$. Given $\mathcal{X} \subset \mathbb{R}^n$, we let $\mathsf{Interior}_\varepsilon (\mathcal{X}) = \left \{ x \in \mathcal{X}\ |\ \mathcal{B}_{\varepsilon} (x) \subseteq \mathcal{X} \right \}$.

\section{Preliminaries}
In this section we recall some basic concepts of the Gaussian Process (GP) regression \cite{rasmussen}, transition systems and approximate alternating simulation relations \cite{zamani2012a}. 

\subsection{\textcolor{black}{Gaussian process regression}}\label{gpsec}
\textcolor{black}{Consider a nonlinear function $h : \mathbb{R}^{n_x} \rightarrow \mathbb{R}$ perturbed by additive noise as $y = h ({ x}) + v$, where ${ x} \in \mathbb{R}^{n_x}$ is the input, $y \in \mathbb{R}$ is the output, and $v \sim \mathcal{N} (0, \sigma^2)$ is the Gaussian distributed white noise. 
Given $m : \mathbb{R}^{n_x} \rightarrow \mathbb{R}$ and some kernel function $\mathsf{k}:\mathbb{R}^{n_x} \times \mathbb{R}^{n_x} \rightarrow \mathbb{R}_{\geq 0}$, suppose that, for any finite number of inputs $X = [x_1, \ldots, x_T]$ ($x_t \in \mathbb{R}^{n_x}$, $t\in\{1, \ldots, T\}$), the joint probability distribution of the corresponding outputs ${y} = [y_1, y_2, \ldots, y_T]^\mathsf{T}$ follows the multivariate Gaussian distribution: $y \sim \mathcal{N}(M, K)$, where $M = [m(x_1), \ldots, m(x_T)]$ and $K_{tt'} = \mathsf{k}(x_t, x_{t'})$, $t, t' \in \{1, \ldots, T\}$ ($K_{tt'}$ denotes the $(t, t')$-element of $K$). 
Then, we say that the function $h$ follows a \textit{Gaussian process} (GP) \cite{rasmussen}, and it is denoted by $h(x) \sim \mathcal{GP}(m(x), \mathsf{k}(x, x'))$.} %The functions $m(\cdot)$ and $\mathsf{k}(\cdot)$ are called the \textit{mean} and the \textit{covariance} function, respectively.} 

\textcolor{black}{In the GP \textit{regression} problem, we start by assuming a GP {prior}: $h(x) \sim \mathcal{GP}(m(x), \mathsf{k}(x, x'))$. Let $\mathcal{D} = \{{ x}_t, y_t\}^T _{t=1}$ denote a training data set. Then, using Bayes rule, the \textit{posterior} distribution of the output for an arbitrary input ${ x} \in \mathbb{R}^{n_x}$ follows the Gaussian distribution, i.e., ${\rm Pr}(y | { x}, \mathcal{D}) = \mathcal{N} (\mu ({ x}; \mathcal{D}), \sigma^2 ({ x}; \mathcal{D}))$. Here, the mean $\mu ({ x}; \mathcal{D})$ and the variance $\sigma^2 ({ x}; \mathcal{D}))$ are given by
\begin{align}
{\mu} ({ x}; \mathcal{D} ) &= m(x) + \mathsf{{k}}^{*\mathsf{T}} _T ({ x}) ({K} + \sigma^2 {I})^{-1} ({Y}-M), \\ {\sigma}^2 ({ x}; \mathcal{D} )& = \mathsf{k}({ x} , { x} ) - \mathsf{k}^{*\mathsf{T}} _T ({ x}) ({K} + \sigma^2 {I})^{-1} \mathsf{k}^* _T ({ x}),
\end{align} 
where $I$ is the identity matrix of appropriate dimension, and $\mathsf{k}^*_{T} ({ x}) = \left [\mathsf{k} ({ x} ,{ x}_1), \ldots, \mathsf{k} ({ x} , { x}_T)\right]^\mathsf{T}$.} 

\subsection{Transition system, alternating simulation relation} 
We provide the notion of a transition system, which will be useful to describe a control system formalized later in this paper. 
\begin{mydef}\label{transystem}
\normalfont
A \textit{transition system} is a quadruple $S = (\mathcal{X}, x_0, \mathcal{U}, G)$, where: 
\begin{itemize}
\item $\mathcal{X}$ is a set of states; 
\item $x_0 \in \mathcal{X}$ is an initial state;  
\item $\mathcal{U}$ is a set of inputs; 
\item $G : \mathcal{X} \times \mathcal{U} \rightarrow 2^{\mathcal{X}}$ is a transition map. \qedwhite 
\end{itemize} 
\end{mydef}
Roughly speaking, we denote by $x' \in G (x, u)$ if and only if the system evolves from $x$ to $x'$ by applying the control input $u$. The state $x'$ is called a $u$-\textit{successor} of $x$. Moreover, we denote by $\mathcal{U} (x)$ the set of all inputs $u \in \mathcal{U}$, for which $G(x, u) \neq \varnothing$. 

Next, we shall recall the notion of an \textit{approximate alternating simulation relation}\cite{pola2009symbolic,zamani2012a}, which is a well-known concept to represent behavioral relationships on the similarity between two transition systems. 
%which is in particular useful to synthesize the controller. 
\begin{mydef}[$\varepsilon$-ASR]\label{asr2}
\normalfont 
Let $S_{a} = (\mathcal{X}_a, x_{a0}, \mathcal{U}_a, {G}_{a})$ and ${S}_b = (\mathcal{X}_b, x_{b0}, \mathcal{U}_b, {G}_{b})$ be two transition systems. 
Given $\varepsilon \in \mathbb{R}_{\geq 0}$, a relation $R (\varepsilon) \subseteq \mathcal{X}_a \times \mathcal{X}_b$ is called an \textit{$\varepsilon$-approximate Alternating Simulation Relation} (or $\varepsilon$-ASR for short) from $S_{a}$ to ${S}_b$, if the following conditions are satisfied: 

\begin{enumerate}
\renewcommand{\labelenumi}{(C.\arabic{enumi})}
\setlength{\leftskip}{0.1cm}
\item %For every ${x}_{a0} \in \mathcal{X}_{a0}$, there exists ${x}_{b0} \in {X}_{b0}$ such that 
$({x}_{a0}, {x}_{b0}) \in R (\varepsilon)$; 
\item For every $(x_a, x_b) \in R (\varepsilon)$, we have $\| x_a - x_b \|_\infty \leq \varepsilon$; 
\item For every $({x}_a, x_b) \in R (\varepsilon)$ and for every ${u}_a \in \mathcal{U}_a (x_a)$, there exist $u_b \in \mathcal{U}_b (x_b)$, such that the following holds: for every $x'_b \in G_b(x_b, u_b)$, there exists $x' _a \in G_a (x_a, u_a)$, such that $(x'_a, x'_b) \in R (\varepsilon)$. \qedwhite 
\end{enumerate}
\end{mydef}
%In general, the transition system $S_b$ is regarded as a \textit{concrete} system having more states and transitions than those of $S_a$. 
%More specifically, 
The transition system $S_a$ serves as the \textit{abstract} expression of $S_b$, in the sense that every transition of $S_b$ can be \textit{approximately simulated} by those of $S_a$ according to (C.1)--(C.3) in \rdef{asr2}. 
The concept of an $\varepsilon$-ASR is particularly useful to synthesize a controller for the transition system $S_b$, based on the controller for $S_a$. 
That is, once we obtain $S_a$ that guarantees the existence of an $\varepsilon$-ASR from $S_a$ to $S_b$, we can synthesize a controller for $S_b$ by \textit{refining} a controller for $S_a$ that can be synthesized by algorithmic techniques from discrete event systems, see, e.g., \cite{tabuadabook2009}.

\section{Problem formulation}\label{problemstatement}
In this section, we describe a control system that we seek to consider, provide the notion of a controlled invariant set, and describe the goal of this paper. 
\subsection{System description}
Let us consider the following nonlinear systems: 
\begin{align}\label{dynamics}
&{x}_{t+1} = f({ x}_t, { u}_t) + d({ x}_t) + v_t,\\ 
&x_0 = \bar{x}, \ { u}_t \in \mathcal{U},\ v_t \in \mathcal{V}, 
\end{align}
for all $t\in\mathbb{N}_{\geq 0}$, where $x_t \in \mathbb{R}^{n_x}$ is the state, $u_t\in \mathbb{R}^{n_u}$ is the control input, $v_t \in \mathbb{R}^{n_x}$ is the additive noise, and $\bar{x} \in \mathbb{R}^{n_x}$ is the initial state. Moreover, $\mathcal{U} \subset \mathbb{R}^{n_u}$ and $\mathcal{V} \subset \mathbb{R}^{n_x}$ are the set of control inputs and the additive noise, respectively. It is assumed that $\mathcal{U}$ is compact and $\mathcal{V}$ is given by $\mathcal{V} = \{v \in \mathbb{R}^{n_x}\ |\ \|v\|_\infty \leq \sigma_v \}$ for a given $\sigma_v >0$. 
%As shown in \req{dynamics}, it is assumed that the control input, and the additive noise are constrained in the sets $\mathcal{X} \subset \mathbb{R}^{n_x}$, $\mathcal{U} \subset \mathbb{R}^{n_u}$, $\mathcal{V} \subset \mathbb{R}^{n_x}$, respectively. Regarding these sets, we assume that: (i) both $\mathcal{X}$ and $\mathcal{U}$ are compact; (ii) $\mathcal{X}$ can be either convex or \textit{non-convex}; (iii) $\mathcal{U}$ is convex, and (iv) $\mathcal{V}$ is given by $\mathcal{V} = \{v \in \mathbb{R}^{n_x}\ |\ \|v\|_\infty \leq \sigma_v \}$ where $\sigma_v >0$ is a given constant. 
Moreover, $f: \mathbb{R}^{n_x} \times \mathbb{R}^{n_u} \rightarrow \mathbb{R}^{n_x}$ is the \textit{known} function that captures the modeled (or nominal) dynamics, and $d: \mathbb{R}^{n_x} \rightarrow \mathbb{R}^{n_x}$ is the state-dependent, \textit{unknown} deterministic function that captures the un-modeled dynamics. 
Regarding the function $f$, we assume the following Lipschitz continuity: 
\begin{myas}\label{lipschitz}
\normalfont
The function $f$ is Lipschitz continuous in $x\in\mathbb{R}^{n_x}$, i.e., given $L_f \in \mathbb{R}_{\geq 0}$, $\| f({ x}_1, { u}) - f({ x}_2, { u}) \|_{\infty} \leq L_f \| { x}_1 - { x}_2\|_{\infty}$, $\forall { x}_1, { x}_2 \in \mathbb{R}^{n_x}, \forall {u}\in \mathcal{U}$. \qedwhite %\textcolor{black}{It is assumed that $L_f$ is known.} \qedwhite
\end{myas}

Regarding the unknown function $d_i$, $i \in \mathbb{N}_{1:n_x}$, in this paper we provide a certain \textit{smoothness} assumption (see, e.g., \cite{felix1,felix2}): 
\begin{myas}\label{rkhsd}
\normalfont 
\textcolor{black}{For each $i\in\mathbb{N}_{1:n_x}$, let $\mathsf{k}_i : \mathbb{R}^{n_x} \times \mathbb{R}^{n_x} \rightarrow \mathbb{R}_{\geq 0}$ be a given, continuously differentiable kernel function and $\mathcal{H}_{\mathsf{k}_i}$ be the reproducing kernel Hilbert space (RKHS) corresponding to $\mathsf{k}_i$ with the induced norm denoted by $\|\cdot\|_{\mathsf{k}_i}$. 
Then, for each $i\in\mathbb{N}_{1:n_x}$, it is assumed that $d_i \in \mathcal{H}_{\mathsf{k}_i}$. \textcolor{black}{Moreover, an upper bound of the RKHS norm $\|d_i\|_{\mathsf{k}_i} \leq B_i$ is available.}} \qedwhite 
%\textcolor{black}{For each $i\in\mathbb{N}_{1:n_x}$, the function $d_i$ lies in the reproducing kernel Hilbert space (RKHS) corresponding to a given, continuously differentiable kernel function $\mathsf{k}_i : \mathbb{R}^{n_x} \times \mathbb{R}^{n_x} \rightarrow \mathbb{R}_{\geq 0}$.} \qedwhite 
%let $\mathsf{k}_i : \mathbb{R}^{n_x} \times \mathbb{R}^{n_x} \rightarrow \mathbb{R}_{\geq 0}$ be a given kernel and $\mathcal{H}_{\mathsf{k}_i}$ be the reproducing kernel Hilbert space (RKHS) corresponding to $\mathsf{k}_i$ with the induced norm denoted by $\|\cdot\|_{\mathsf{k}_i}$. Then, for each $i\in\mathbb{N}_{1:n_x}$, it is assumed that $d_i \in \mathcal{H}_{\mathsf{k}_i}$ with a bounded norm: $\|d_i\|_{\mathsf{k}_i} < \infty$.} \qedwhite 
%\textcolor{black}{For each $i\in\mathbb{N}_{1:n_x}$, let $\mathsf{k}_i : \mathbb{R}^{n_x} \times \mathbb{R}^{n_x} \rightarrow \mathbb{R}_{\geq 0}$ be a given kernel and $\mathcal{H}_{\mathsf{k}_i}$ be the reproducing kernel Hilbert space (RKHS) corresponding to $\mathsf{k}_i$ with the induced norm denoted by $\|\cdot\|_{\mathsf{k}_i}$. Then, for each $i\in\mathbb{N}_{1:n_x}$, it is assumed that $d_i \in \mathcal{H}_{\mathsf{k}_i}$ with a bounded norm: $\|d_i\|_{\mathsf{k}_i} < \infty$.} \qedwhite 
\end{myas}
\ras{rkhsd} implies that each $d_i : \mathbb{R}^{n_x} \rightarrow \mathbb{R}$ is characterized of the form $d_i ({x}) = \sum^\infty _{n=1} \alpha_n \mathsf{k}_i ({ x}, { x}_n)$, where ${x}_n \in \mathbb{R}^{n_x}$, $n \in \mathbb{N}_{>0}$ are the representer points and $\alpha_n \in \mathbb{R}$, $n \in \mathbb{N}_{>0}$ are the parameters that it is necessary to decay sufficiently fast as $n$ increases. 
\textcolor{black}{%For each $i\in\mathbb{N}_{1:n_x}$, let $\|\cdot \|_{\mathsf{k}_i}$ denote an induced norm of the RKHS corresponding to $\mathsf{k}_i$. 
The induced norm is given by $\|d_i\|^2 _{\mathsf{k}_i} = \sum_{n=1}^\infty \sum_{n'=1}^\infty \alpha_n \alpha_{n'} \mathsf{k}_i ({ x}_n, { x}_{n'})$. In general, obtaining a large enough, yet not too conservative bound for $\|d_i\|_{\mathsf{k}_i}$ is hard; nevertheless, 
%Note that, while we do not have direct access to $\|d_i\|_{\mathsf{k}_i}$ if $d_i$ is unknown, 
 there exist several ways to compute an upper bound of $\|d_i\|_{\mathsf{k}_i}$ (see, e.g., \cite{emilio2021}). The overview of how to compute an upper bound of $\|d_i\|_{\mathsf{k}_i}$ is given in Appendix~A, and we refer the interested reader to \cite{emilio2021} for a more detailed discussion.} 
%\begin{myrem}[On the computation of $B_i$]
%While we do not have a direct access to $B_i$ if $d_i$ is unknown, $B_i$ can be estimated from training data (see, e.g., \cite{hoge}). 
%\end{myrem}

\ras{rkhsd} allows us to show the following result: 
\begin{mylem}\label{lipschiz}
\normalfont
\textcolor{black}{Suppose that \ras{rkhsd} holds. Then, it follows that $|d_i (x_1) - d_i (x_2)| \leq L_i  \sqrt{\| x_1- x_2 \|_\infty}$, 
for all $x_1, x_2 \in \mathcal{X}$, where $L_i  = B_i \sqrt{2 \|\partial \mathsf{k}_i /\partial x \|_\infty}$.} \qedwhite 
\end{mylem}
The proof is given in Appendix~B. 

\begin{myrem}
\normalfont 
\textcolor{black}{\textit{(On selecting $\mathsf{k_i}$)} \ras{rkhsd} implies that the kernel function $\mathsf{k}_i$ should be chosen apriori. 
Potential candidates of this kernel function are:
%This kernel function $\mathsf{k}_i$ may be chosen such that any continuous function can be estimated arbitrary well by a function that lies in the RKHS.
%the unknown function can be estimated arbitrary well by a function that lies in the RKHS.
%The following kernels are known to fulfill such a property \cite{universalkernel}: 
%To guarantee such property, it is known that some kernel functions are useful to be employed: 
$\mathsf{k}_i(x, y) = e^{-a \|x-y\| ^2}, \mathsf{k}_i(x, y) = (b + \|x-y\|^2)^{-a}$, where $a, b$ are positive constants. A useful property of employing these kernel functions is the \textit{universal approximation property}, i.e., the RKHSs are dense in the space of all continuous functions over any compact set (see, e.g., \cite{universalkernel}). 
%Certainly, the choice of $\mathsf{k_i}$ affects the class of all functions 
%A useful choice for this kernel function may be the squared exponential (SE) kernel $\mathsf{k}_i (x, y) = e^{-\gamma \|x-y\| ^2}$, where $\gamma$ is an arbitrary positive constant. 
%In particular, it is known that the SE kernel satisfies the \textit{universal approximation property} \cite{universalkernel}, i.e., the RKHS corresponding to the SE kernel is dense in the space of all continuous functions over an arbitrary compact set. 
Hence, if the kernel function is selected as above, {any continuous function} can be estimated arbitrarily well by a function that lies in the RKHS. Note that there are indeed other kernels satisfying the universal approximation property, which may also be useful to be employed (see, e.g., \cite{universalkernel}). 
%Intuitively, this implies that {any} continuous function in an any compact set can be estimated arbitrary well by a function that lies in the RKHS (corresponding to the SE kernel). Note that there are indeed other kernels satisfying the universal approximation property, which can be also useful to be employed (see, e.g., \cite{universalkernel}). 
} \qedwhite
%When learning-based choice of the kernel function $\mathsf{k}_i$ 
%The choice of the kernel function $\mathsf{k}_i$ in \ras{rkhsd} is important since it affects the performance of the kernel-based learning. 
%space of 
%characterizes the class of the RKHS. 
\end{myrem}

\begin{myrem}
\normalfont 
\textcolor{black}{\textit{(On comparisons between \ras{rkhsd} and an alternative assumption)} An alternative assumption to \ras{rkhsd} made on $d_i$ is that $d_i$ is sampled from the Gaussian process $d_i \sim \mathcal{GP} (m(x), \mathsf{k}_i(x, x'))$, where $m(\cdot)$ is a given mean function (see, e.g., \cite{safelearning2}). %This assumption directly implies that the confidence interval is obtained in a probabilistic sense by utilizing the mean and the covariance of the posterior distribution (see, e.g., Eq.(13) in \cite{safelearning2}). 
Note that if $d_i \sim \mathcal{GP} (m(x), \mathsf{k}_i(x, x'))$, $\|d_i\| _{\mathsf{k}_i} = \infty$ holds almost surely \cite{wahba}. Hence, samples from the GP are rougher than the RKHS functions, in the sense that the GP assumption deals with the case $\|d_i\| _{\mathsf{k}_i} = \infty$. 
%Hence, samples from the GP are rougher than the RKHS functions, in the sense that $\|d_i\| _{\mathsf{k}_i} = \infty$. 
%and therefore, samples from the GP are rougher than the RKHS functions \cite{srinivas2}. 
%This assumption can deal with the case for $\|d_i\| _{\mathsf{k}_i} = \infty$, since the function that follows the GP satisfies $\|d_i\| _{\mathsf{k}_i} = \infty$ almost surely. %Hence, the former assumption may be more restrictive than the second one, in the sense that the induced norm should be bounded. 
On the other hand, the GP assumption $d_i \sim \mathcal{GP} (m(x), \mathsf{k}_i(x, x'))$ requires the knowledge that the unknown function follows the GP; not only the kernel function $\mathsf{k}_i$ but also how $d_i$ is sampled should be known apriori. 
In contrast, \ras{rkhsd} deals with all functions uniformly satisfying $\|d_i\| _{\mathsf{k}_i} < \infty$. Hence, the class of all functions satisfying the GP assumption does not include the class of all functions satisfying \ras{rkhsd}, and vice versa (i.e., neither the former nor the latter assumption is restrictive with respect to the other).} \qedwhite 
%In this paper, we make use of \ras{rkhsd} (rather than the GP assumption), since, as we will see later, it allows us to derive a deterministic error bound on $d_i$ (see Section~4.1).} \qedwhite 
%one might consider 
\end{myrem}

%\begin{myrem}
%\normalfont 
%For simplicity of presentation, we assume that the unknown function $d$ is only dependent on $x$. However, as considered in \cite{felix1}, the approach that will be presented in this paper can be easily extended to the case where $d$ depends on both $x$ and $u$. Indeed, this is achieved by taking the training input as $X_T= [[x^\mathsf{T} _1, u^{\mathsf{T}} _1]^\mathsf{T}, \ldots, [x^\mathsf{T} _T, u^\mathsf{T} _T]^\mathsf{T}]$ instead of ${{ X}}_T = \left [{x}_1, {x}_2, \ldots, {x}_{{T}} \right ]$. %\qedwhite
%\end{myrem}

\subsection{Controlled invariant set and safety controller}
A sequence $x_0, x_1, x_2, \ldots \in \mathbb{R}^{n_x}$ is called a \textit{trajectory} of the system \req{dynamics}, if there exist $u_0, u_1, u_2, \ldots \in \mathcal{U}$, $v_0, v_1, v_2, \ldots \in \mathcal{V}$ such that $x_0 = \bar{x}$, $x_{t+1} = f(x_t, u_t) + d(x_t) +v_t$, $\forall t\in\mathbb{N}_{\geq 0}$. 
Moreover, a \textit{controller} is defined as a set-valued mapping from each state onto the set of control inputs, i.e., $C : \mathbb{R}^{n_x} \rightarrow 2^\mathcal{U}$.
\textcolor{black}{Given $C$, a \textit{controlled trajectory} is defined as \textcolor{black}{any} trajectory of the system \req{dynamics}, $x_0, x_1, x_2, \ldots \in \mathbb{R}^{n_x}$ with $u_t \in C(x_t)$, $\forall t\in\mathbb{N}_{\geq 0}$.} 

Now, denote by $\mathcal{X} \subset \mathbb{R}^{n_x}$ a \textit{safe set}, in which the trajectory of the system \req{dynamics} must stay for all times. It is assumed that $\mathcal{X}$ is compact and can be either convex or non-convex, and that $\bar{x} \in \mathcal{X}$. Based on the above, we define the notion of a controlled invariant set (see, e.g., \cite{blanchini1999a}) and the safety controller as follows: 
%In particular, a trajectory $x_0, x_1, x_2, \ldots \in \mathbb{R}^{n_x}$ is called \textit{safe} if $x_t \in \mathcal{X}$ for all $t\in\mathbb{N}$. 
%Moreover, we define the notion of a controlled invariant set as follows: 
%, which characterizes the region guaranteeing the existence of a controller that makes the trajectories safe. 
 %and the safety controller as follows: 
%safety as follows: 
\begin{mydef}
\normalfont 
%A trajectory $x_0, x_1, x_2, \ldots \in \mathbb{R}^{n_x}$ is called \textit{safe} if $x_t \in \mathcal{X}$, $\forall t\in\mathbb{N}_{\geq 0}$. 
A set ${\mathcal{X}_S} \subseteq \mathcal{X}$ is called a \textit{controlled invariant set} in $\mathcal{X}$, if there exists a controller $C_S: \mathbb{R}^{n_x} \rightarrow 2^\mathcal{U}$ such that the following holds: for every $x \in \mathcal{X} _S$, there exists $u \in C_S(x)$ such that for every $v \in \mathcal{V}$, $f(x, u) + d(x) + v \in \mathcal{X} _S$. The controller $C_S$ is called a \textit{safety controller}. \qedwhite
%for \textit{every} controlled trajectory $x_0, x_1, x_2, \ldots \in \mathbb{R}^{n_x}$ induced by $C_S$, we have $x_t \in \mathcal{X}_S$, $\forall t\in\mathbb{N}_{\geq 0}$. 
%for every $x \in \mathcal{X} _S$, there exists $u \in C_S(x)$ such that for every $v \in \mathcal{V}$, $f(x, u) + d(x) + v \in \mathcal{X} _S$. The controller $C_S$ is called a \textit{safety controller}. \qedwhite
%for \textit{every} controlled trajectory $x_0, x_1, x_2, \ldots \in \mathbb{R}^{n_x}$ induced by $C$, we have $x_t \in \mathcal{X}$, $\forall t\in\mathbb{N}_{\geq 0}$
%Moreover, a controller $C$ is called a \textit{safety controller} if for \textit{every} controlled trajectory $x_0, x_1, x_2, \ldots \in \mathbb{R}^{n_x}$ induced by $C$, we have $x_t \in \mathcal{X}$, $\forall t\in\mathbb{N}_{\geq 0}$. \qedwhite 
\end{mydef}
That is, ${\mathcal{X}_S}$ is called a controlled invariant set if there exists a controller $C_S$ such that every controlled trajectory induced by $C_S$ (starting from anywhere in $\mathcal{X}_S$) stays in ${\mathcal{X}_S}$ for all times. 

%that makes every controlled trajectory stay in ${\mathcal{X}_S}$ for all times. 
%That is, the trajectory is called safe if it stays in the safety set for all times, and the {safety controller} makes every controlled trajectory of \req{dynamics} fulfill the safety. 

\subsection{The goal of this paper and overview of the approach}

The goal of this paper is to construct a \textit{symbolic model} of the control system \req{dynamics}, which indicates an abstract expression of \req{dynamics}. In particular, due to the existence of the unknown function $d$, we here propose a \textit{learning-based} approach, in which the symbolic model is constructed by learning the unknown function $d$ from training data. 
%develop a \textit{learning-based} approach towards symbolic abstractions for the control system \req{dynamics}. In particular, the symbolic model will be constructed by learning the unknown function $d$ from training data. 
%constructing a \textit{symbolic model}, which indicates an abstract expression of the control system \req{dynamics}. 
%In particular, due to the existence of the unknown function $d$, we develop a \textit{learning-based} approach, in which the symbolic model will be constructed by learning the unknown function $d$ from training data. 
Towards this end, we first provide an approach to construct a symbolic model for given training data (\rsec{symbolicsec}). 
%The main contributions of this paper are two fold. First, for a given set of training data, we provide an approach to construct a symbolic model, which represents an abstract expression of the control system. 
The symbolic model is constructed based on the GP regression and the concept of an $\varepsilon$-ASR; for details, see \rsec{symbolicsec}. Based on the symbolic model, we proceed by developing an overall algorithm that aims at collecting the training data from scratch and constructing the symbolic model (\rsec{mainsec}). In particular, we propose a \textit{safe exploration} algorithm, in which the trajectory of the system \req{dynamics} must stay in $\mathcal{X}$ for all times while collecting the training data and constructing the symbolic model. 
As we will see later, this is achieved by iteratively updating the symbolic model, controlled invariant set and the safety controller after each step of the state-space exploration; for details, see \rsec{mainsec}.  
%Second, we develop a \textit{safe exploration} algorithm, which aims at collecting the training data from scratch and constructing the symbolic model, while at the same time guaranteeing safety (i.e., the trajectory of the system \req{dynamics} stays in $\mathcal{X}$ for all times). 
%achieving a \textit{safe exploration}. 
%In other words, while collecting the training data and constructing the symblic model, the trajectory of the system \req{dynamics} must stay in $\mathcal{X}$ for all times. 
%This will be achieved by iteratively updating the symbolic model, controlled invariant set and the safety controller after each step of the state-space exploration; for details, see \rsec{mainsec}.  
%the controlled invariant set and the safety controller, and by applying the safety controller during the state-space exploration; for details, see \rsec{mainsec}.  
%a controlled invariant set and synthesizing a safety controller through 
%after each implementation of the state-space exploration (for details, see \rsec{mainsec}). 

\section{Constructing symbolic models {with Gaussian processes}}\label{symbolicsec}
In this section, we provide an approach to construct a {symbolic model} based on a given set of training data. 
In \rsec{asdsec}, we provide an approach to learn $d_i$ with the GP regression as well as a useful error bound on $d_i$ based on \ras{rkhsd}. In \rsec{constructsymbolicmodel}, we provide a way of how to construct a symbolic model from a given set of training data. 
In \rsec{synthesizesafetycontrol}, we provide a safety controller synthesis, which will be useful to achieve the safe exploration provided in the next section.
%Moreover, we provide a safety controller synthesis, which is useful to achieve the safe exploration. 
%that makes the trajectories of \req{dynamics} stay in $\mathcal{X}$ for all times. 
%The symbolic model will be constructed based on the training data for estimating the unknown function $d$, such that there exists an $\varepsilon$-ASR from the symbolic model to $S$.  
\subsection{\textcolor{black}{Learning $d$ with the GP regression}}\label{asdsec}
%Before constructing the symbolic model, we need to make a certain {assumption} on the unknown function $d$, since, otherwise, $d$ could be arbitrary given and there were no systematic ways to construct the symbolic model. 
In this paper, we estimate each element of $d$, i.e., $d_i$, $i\in\mathbb{N}_{1:n_x}$ ($d = [d_1, d_2, \ldots, d_{n_x}]^\mathsf{T}$) by the GP regression with the kernel function $\mathsf{k}_i$. 
To this end, for each $i\in\mathbb{N}_{1:n_x}$ let $\mathcal{D}_{T, i} = \{{{ X}}_T, { Y}_{T, i} \}$ be the set of input-output training data in order to estimate $d_i$, given by ${{ X}}_T = \left [{x}_1, {x}_2, \ldots, {x}_{{T}} \right ]$, ${ Y}_{T, i} = [y_{1,i}, y_{2,i}, \ldots, y_{T,i} ]^\mathsf{T}$, 
where $T\in\mathbb{N}_{>0}$ is the number of training data points and $y_{t, i} = x_{t+1, i} - f_i (x_t, u_t)$, $\forall t \in \mathbb{N}_{1:{T}}$ are the training outputs, with $x_{t, i}$ and $f_i (x_t, u_t)$ being the $i$-th element of $x_{t}$ and $f(x_t, u_t)$, respectively. 
\textcolor{black}{Note that we have $y_{t, i} = x_{t+1, i} - f_i (x_t, u_t) = d_i (x_t) + {v}_{t, i}$, where ${v}_{t, i}$ denotes the $i$-th ($i\in\mathbb{N}_{1:n_x}$) element of ${v}_{t}$ with $|{v}_{t, i}| \leq \sigma_v$. Hence, $y_{t, i}$ represents the noisy output of $d_i (x_t)$ with the additive noise bounded by $\sigma_v$.} 
\textcolor{black}{As above, the realization of the additive noise sequence is uniformly bounded by $\sigma_v$, i.e., $|v_{t, i}| \leq \sigma_v$, $t \in \mathbb{N}_{\geq 0}$. 
When learning the unknown function, on the other hand, it is \textit{approximated} that the additive noise is drawn independently from $\mathcal{N}(0, \sigma^2 _v)$, aiming at employing the GP regression. Moreover, it is assumed for simplicity that the mean function for the GP prior is zero (i.e., $m(x) \equiv 0$ for all $x$ in \rsec{gpsec}).} 
%follows the Gaussian distribution with zero mean and variance of $\sigma^2_v$, i.e., $\mathcal{N}(0, \sigma^2_v)$ 
%, when learning $d_i$ with the GP, 
%Note that we have $x_{t+1, i} - f_i (x_t, u_t) = d_i (x_t) + \tilde{v}_{t, i}$, $\forall i\in\mathbb{N}_{1:n_x}$, where $\tilde{v}_{t} = - v_t$ and $\tilde{v}_{t, i}$ denotes the $i$-th element of $\tilde{v}_t$ with $|\tilde{v}_{t, i}|\leq \sigma_v$. Hence, $y_{t, i}$ represents the noisy output of $d_i (x_t)$ with the additive noise bounded by $\sigma_v$. 
%Moreover, let $K_{T,i}$ denote the covariance matrix for the kernel function $\mathsf{k}_i$ to characterize $d_i$. 
%\textcolor{black}{Moreover, while true sequence of the additive noise $v_t =[v_{t, 1}, \ldots, v_{t, n_x}]^\mathsf{T}$ is bounded (i.e., $\|v_t\|_\infty \leq \sigma_v$ for all $t\in\mathbb{N}_{\geq 0}$), the GP estimate of $d_i$ \textit{assumes} that the additive noise follows the Gaussian distribution with zero mean and variance of $\sigma^2_v$, i.e., $\mathcal{N}(0, \sigma^2_v)$ (see \cite{srinivas2}).} 
Thus, the mean and the variance for the GP model of $d_i$ with an arbitrary input ${x} \in \mathbb{R}^{n_x}$, denoted as ${\mu}_{i} (x; \mathcal{D}_{T, i})$ and ${\sigma}^2_{i} ({{x}}; \mathcal{D}_{T, i})$, are computed by 
\begin{align}
{\mu}_{i} (x; \mathcal{D}_{T, i}) &= \mathsf{{k}}^\mathsf{*T} _{T, i} ({x}) (K_{T,i} + \sigma^2 _v {I})^{-1} {Y}_{T, i}, \label{mudi} \\ 
{\sigma}^2_{i} ({{x}}; \mathcal{D}_{T, i}) &= \mathsf{k}_i ({ x} , { x} ) \notag \\ 
                                                     &\ \ \ \ - \mathsf{k}^\mathsf{*T} _{T, i} ({ x}) (K_{T,i} + \sigma^2 _v {I})^{-1} \mathsf{k}^*_{T, i} ({ x} ), \label{sigmadi}
\end{align}
where $K_{T,i}$ denote the covariance matrix for the kernel function $\mathsf{k}_i$ and $\mathsf{k}^*_{T, i} ({ x}) = \left [\mathsf{k}_i ({ x} ,{ x}_1), \ldots, \mathsf{k}_i ({ x} , { x}_T)\right]^\mathsf{T}$.

 Now, recall that the unknown function $d_i$ lies in the RKHS corresponding to $\mathsf{k}_i$ (\ras{rkhsd}). 
 %is estimated by the GP regression, where the mean and the variance are computed by \req{mudi}, \req{sigmadi}. Based on these expressions, 
 Using this assumption, we can derive an \textit{error bound} on $d_i$, representing how the GP posterior mean $\mu_i$ differs from the ground truth $d_i$:  %the true value: 
\begin{mylem}\label{boundlemma}
\normalfont
Suppose that \ras{rkhsd} holds, and let $\mathcal{D}_{T, i}=\{{{ X}}_T, { Y}_{T, i} \}$ be the training data for $d_i$ with ${{ X}}_T = \left [{x}_1, {x}_2, \ldots, {x}_{{T}} \right ]$ and ${ Y}_{T, i} = [y_{1,i}, y_{2,i}, \ldots, y_{T,i} ]^\mathsf{T}$ for $T \in \mathbb{N}_{>0}$. Then, for all $x \in \mathbb{R}^{n_x}$ and $T \in \mathbb{N}_{>0}$, it follows that $d_i (x) \in \mathcal{Q}_{i} (x; \mathcal{D}_{T, i})$, where 
\begin{align}\label{qdi}
\mathcal{Q}_{i} (x; \mathcal{D}_{T, i}) = \left[{\mu}_{i} ({{ x}}; \mathcal{D}_{T, i}) \pm \beta_{T, i} {\sigma}_{i} ({{ x}}; \mathcal{D}_{T, i})\right]
\end{align}
with $\beta_{T, i} = \sqrt{B^2_i - {Y}^\mathsf{T} _{T, i} (K_{T,i}+{\sigma}^2 _v I)^{-1}{Y} _{T, i} + T}$. 
 \qedwhite 
\end{mylem}
%\rlem{boundlemma} has been already derived in the proof of Lemma~7.2 in \cite{srinivas2}, and thus we only provide the overview of this proof in the Appendix. 
For the proof, see the Appendix. 
\rlem{boundlemma} means that $d_i (x)$ is shown to be in the interval set $\mathcal{Q}_{i} (x; \mathcal{D}_{T, i})$, which can be computed based on the training data for $d_i$. 
\begin{myrem}\label{comppreviousrem}
\normalfont
Note that the previous methods of learning-based controller synthesis with the GP regression (e.g., \cite{felix1,felix2}) make use of the probabilistic error bound characterized by the notion of an \textit{information gain}; see Theorem~3 in \cite{srinivas2}. For example, \cite{felix1} employs the following error (or regret) bound: 
\begin{align}\label{dierror}
  d_i (x) \in & [{\mu}_{i} ({{ x}}; \mathcal{D}_{T, i}) \notag \\
  &\pm \sqrt{2\|d_i\|^2_{\mathsf{k}_i} +300 \gamma_T \log^3 (T/\delta) } {\sigma}_{i} ({{ x}}; \mathcal{D}_{T, i})]
\end{align}
which holds for all $T \in \mathbb{N}_{\geq 0}$ with probability at least $1-\delta$ ($0<\delta<1$), where $\gamma_T$, $T \in \mathbb{N}_{\geq 0}$ denote the information gain. 
In contrast to this bound, in this paper we provide the error bound in the \textit{deterministic} form as in \rlem{boundlemma}. 
Note that this bound is a direct consequence from computing the upper bound of the RKHS norm of the error ${\mu}_{i} (\cdot ; \mathcal{D}_{T, i}) - d_i(\cdot)$ with respect to the kernel $\mathsf{k}_{T,i}(x,x') = \mathsf{k}_i ({ x} , { x}' )- \mathsf{k}^\mathsf{T} _{T,i*} ({ x} ) (K_{T,i} + \sigma^2 _v {I})^{-1} \mathsf{k}_{T,i*} ({ x}')$, which has been derived in the proof of Lemma~7.2 in \cite{srinivas2} (\textcolor{black}{in particular, see the first equation in the left column of page 3261 in \cite{srinivas2}}), and see also Appendix~C in this paper. 
\textcolor{black}{The probabilistic error bound \req{dierror} is more conservative than the deterministic one of \rlem{boundlemma} in the following sense. 
Note that \req{dierror} achieves a \textit{deterministic} bound by setting $\delta \rightarrow 0$. However, setting $\delta \rightarrow 0$ in \req{dierror} implies $d_i (x) \in [-\infty, \infty]$ for all $T \in \mathbb{N}_{>0}$, which leads to an \textit{unbounded} interval $\mathbb{R}$ (and is thus not useful for constructing a symbolic model). 
On the other hand, the error bound obtained in Lemma~2 is deterministic and \textit{always bounded} (i.e., $d_i(x) \in \mathcal{Q}_{i} (x; \mathcal{D}_{T, i})  \subset \mathbb{R}$ holds for all $T \in \mathbb{N}_{>0}$).}
Such conservativeness might arise due to the fact that, in \cite{srinivas2} the probabilistic error bound \req{dierror} has been derived as a \textit{sufficient} condition to the deterministic one given in \rlem{boundlemma} (or Lemma~7.2 in \cite{srinivas2}). 
%As shown in \req{dierror}, setting $\delta \rightarrow 0$ corresponds to the deterministic error bound. 
%By employing the deterministic error bound
%\textcolor{black}{As shown in \cite{srinivas2}, the probabilistic error bound \req{dierror} has been obtained as a \textit{sufficient condition} to the deterministic one in \rlem{boundlemma}. Hence, a \textit{smaller} error bound can be obtained by applying Lemma~2 of our manuscript than the probabilistic one given in the previous works.}
%{As a simple intuition to see this, note that \req{dierror} achieves a deterministic bound by setting $\delta \rightarrow 0$. However, from \req{dierror} setting $\delta \rightarrow 0$ implies $d_i (x) \in [-\infty, \infty]$ for all $T \in \mathbb{N}_{>0}$, leading to an \textit{unbounded} interval. On the other hand, in our case, the \textit{bounded} interval is obtained for all $T \in \mathbb{N}_{>0}$ as shown in \rlem{boundlemma}. 
{%More technically, the conservativeness of the probabilistic bound \req{dierror} can be shown from several inequalities in \cite{srinivas2}. 
To see this, note that the error bound obtained in Lemma~7.2 of \cite{srinivas2} was {further upper bounded} by using the information gain $\gamma_T$ (see the second to the third inequality in the top of the right column of page~3261 in \cite{srinivas2}), as well as the concentration inequalities (see the second to the third inequality in the bottom of the right column of page~3261). 
Hence, from the above upper boundings, the probabilistic error bound has been obtained as the sufficient condition to the deterministic one given in \rlem{boundlemma}. }
\textcolor{black}{In this paper, we will make use of the deterministic error bound in \rlem{boundlemma} instead of the probabilistic one, since it allows us to derive an error bound that can get smaller as the number of the training data increases (see below for details). 
Such a property is useful to show that the controlled invariant set can enlarge as the number of training data increases, and, moreover, we can provide some computationally efficient algorithms for updating the symbolic models and the safety controller synthesis (see Section~5.1).} 
%in which redundant transitions can be iteratively removed while preserving alternating simulation relations. 
%will be 
%it allows us to follow the deterministic notion of an $\varepsilon$-ASR in order to construct the symbolic model. 
%In particular, we can obtain a error bound that never grows and potentially gets smaller as the number of the training data increases (see below), 
%tighter error bounds
\qedwhite 
\end{myrem}

%% The Appendices part is started with the command \appendix;
%% appendix sections are then done as normal sections
%% \appendix

\textcolor{black}{Now, for every $T \in \mathbb{N}_{> 0}$, it follows from Lemma~2 that $d_i (x) \in \mathcal{Q}_{i} (x; \mathcal{D}_{t, i})$ for all $t\in\mathbb{N}_{1:T}$. 
Thus, for every $T \in \mathbb{N}_{> 0}$, we have $d_i (x) \in \bigcap^T _{t=1} \mathcal{Q}_{i} (x; \mathcal{D}_{t, i})$. 
Therefore, for every $T \in \mathbb{N}_{> 0}$, we have $d_i (x) \in \mathcal{R}_{i} (x; \mathcal{D}_{T, i})$, where $\mathcal{R}_{i} (x; \mathcal{D}_{T, i}) = \bigcap^T _{t=1} \mathcal{Q}_{i} (x; \mathcal{D}_{t, i})$.} 
%Now, for any $T \in \mathbb{N}_{> 0}$, let $\mathcal{R}_{i} (x; \mathcal{D}_{T, i}) \subset \mathbb{R}$ be the interval set defined as the intersections of all $\mathcal{Q}_{i} (x; \mathcal{D}_{t, i})$, $t \in \mathbb{N}_{1:T}$, i.e., $\mathcal{R}_{i} (x; \mathcal{D}_{T, i}) = \bigcap^T _{t=1} \mathcal{Q}_{i} (x; \mathcal{D}_{t, i})$. Then, since $d_i (x) \in \mathcal{Q}_{i} (x; \mathcal{D}_{t, i})$ for all $t\in\mathbb{N}_{1:T}$, it follows that $d_i (x) \in \mathcal{R}_{i} (x; \mathcal{D}_{T, i})$. 
From the definition of $\mathcal{R}_{i} (x;  \mathcal{D}_{T, i})$, it follows that $\mathcal{R} _{i} (x; \mathcal{D}_{1, i}) \supseteq \mathcal{R} _{i} (x; \mathcal{D}_{2, i}) \supseteq \mathcal{R} _{i} (x; \mathcal{D}_{3, i}) \supseteq \cdots$. Let $\overline{r} _{i}(x; \mathcal{D}_{T, i}), \underline{r}_{i}(x; \mathcal{D}_{T, i}) \in \mathbb{R}$ be given by 
\begin{align}
\overline{r} _{i}(x; \mathcal{D}_{T, i}) &= \max\{ r\in\mathbb{R} \ |\ r \in \mathcal{R}_{i} (x; \mathcal{D}_{T, i})\}, \label{rmax}\\
\underline{r} _{i}(x ; \mathcal{D}_{T, i}) & = \min\{ r \in \mathbb{R}\ |\ r \in \mathcal{R}_{i} (x; \mathcal{D}_{T, i})\}. \label{rmin}
\end{align}
Since $\mathcal{R} _{i} (x; \mathcal{D}_{1, i}) \supseteq \mathcal{R} _{i} (x; \mathcal{D}_{2, i}) \supseteq \cdots$, it follows that $\overline{r} _{i}(x; \mathcal{D}_{T, i})$ and $\underline{r} _{i}(x; \mathcal{D}_{T, i})$ are \textit{non-increasing} and \textit{non-decreasing} with respect to $T$ (for fixed $x$), respectively. 
Thus, $d_i (x) \in \mathcal{R}_{i} (x; \mathcal{D}_{T, i})$ implies that the error bound on $d_i$ never grows or potentially gets smaller as the number of the training data increases. 
Note also that $d_i ({{ x}}) \in \mathcal{R}_{i} (x; \mathcal{D}_{T, i})$ implies $|d_i ({{ x}}) - \hat{d}_i (x; \mathcal{D}_{T, i})| \leq \Delta_{i} (x; \mathcal{D}_{T, i})$, 
where 
\begin{align}
\hat{d}_i (x; \mathcal{D}_{T, i}) &= 0.5 \left({\overline{r} _{i}(x; \mathcal{D}_{T, i}) + \underline{r} _{i}(x; \mathcal{D}_{T, i})} \right), \label{gihat} \\ 
\Delta_{i} (x; \mathcal{D}_{T, i}) &= 0.5 \left({\overline{r} _{i}(x; \mathcal{D}_{T, i}) - \underline{r} _{i}(x; \mathcal{D}_{T, i})}\right). \label{deltahat}
\end{align}

\subsection{Constructing symbolic models from training data}\label{constructsymbolicmodel}
Let us now construct a symbolic model of the system \req{dynamics} provided that the training data is obtained. We start by showing that the system \req{dynamics} can be described within the class of a transition system (\rdef{transystem}) as follows: 
%tFollowing \rdef{transystem}, we can describe the symtem \req{dynamics} within the class of a transition system as follows: 
\begin{mydef}\label{plantsystem}
\normalfont
A \textit{transition system} induced by the system \req{dynamics} is a quadruple $S = (\mathbb{R}^{n_x}, x_0, \mathcal{U}, G)$, where: 
\begin{itemize}
\item $\mathbb{R}^{n_x}$ is a set of states; 
\item $x_0 \in \mathbb{R}^{n_x}$ is an initial state;  
\item $\mathcal{U} \subset \mathbb{R}^{n_u}$ is a set of inputs; 
\item $G : \mathbb{R}^{n_x} \times \mathcal{U} \rightarrow 2^{\mathbb{R}^{n_x}}$ is a transition map, where $x^+ \in G (x, u)$ iff there exists $v\in \mathcal{V}$ such that $x^+ = f (x, u) + d(x) + v$.  \qedwhite 
\end{itemize} 
\end{mydef}

Based on the transition system $S$, a \textit{symbolic model} of $S$ is constructed by discretizing the state and the input spaces, whose transitions are defined based on the training data $\mathcal{D}_{T, i}$, $i\in\mathbb{N}_{1:n_x}$.  
More specifically, the symbolic model is constructed with a tuple $\mathsf{q} =  (\mathcal{D}_{T}, \eta_x, \eta_u, \varepsilon)$, where 
\begin{itemize}
\item $\mathcal{D}_{T} = \{ \mathcal{D}_{T, 1}, \ldots, \mathcal{D}_{T, n_x} \}$ is the set of training data; 
\item $\eta_x \in \mathbb{R}_{>0}$ is the discretization parameter for the state space $\mathbb{R}^{n_x}$;
\item \textcolor{black}{$\eta_u \in \mathbb{R}_{>0}$ is the discretization parameter for the input space $\mathcal{U}$;} 
\item $\varepsilon \in \mathbb{R}_{>0}$ is the parameter for the precision. 
\end{itemize}

The corresponding symbolic model is denoted as $S_{\mathsf{q}}$ and formally defined as follows: 
\begin{mydef}\label{symbolicmodel}
\normalfont 
Let $S = (\mathbb{R}^{n_x}, x_0, \mathcal{U}, G)$ be the transition system induced by the system \req{dynamics}. Given $\mathsf{q} =  (\mathcal{D}_{T}, \eta_x, \eta_u, \varepsilon)$, a symbolic model of $S$ is defined as a quadruple $S_{\mathsf{q}} =  (\mathcal{X}_{\mathsf{q}}, x_{\mathsf{q}0}, \mathcal{U}_{\mathsf{q}}, {G}_{\mathsf{q}})$, where 
\begin{itemize}
\item $\mathcal{X}_{\mathsf{q}} = [\mathbb{R}^{n_x}]_{\eta_x}$ is a set of states;
\item ${x}_{\mathsf{q}0} \in \mathcal{X}_{\mathsf{q}}$ is an initial state satisfying $x_{\mathsf{q}0} \in \mathsf{Nearest}_{\mathcal{X}_{\mathsf{q}}}(x_0)$; 
%$({x}_{\mathsf{q}0}, x_0) \in R(\varepsilon)$; 
\item $\mathcal{U}_{\mathsf{q}} = [\mathcal{U}]_{\eta_u}$ is a set of inputs; 
\item ${G}_{\mathsf{q}}: \mathcal{X}_{\mathsf{q}} \times \mathcal{U}_{\mathsf{q}} \rightarrow  2^{{\mathcal{X}_{\mathsf{q}}}}$ is a transition map, where ${x}^+ _{\mathsf{q}}  \in {G}_{\mathsf{q}} ({x}_\mathsf{q}, {u}_\mathsf{q})$ iff ${x}^+ _{\mathsf{q}, i} \in [\underline{h}_i ({x}_\mathsf{q}, {u}_\mathsf{q}; \mathcal{D}_{T, i}), \overline{h}_i ({x}_\mathsf{q}, {u}_\mathsf{q}; \mathcal{D}_{T, i})]$, $\forall i\in\mathbb{N}_{1:n_x}$,
%\begin{align}
%{x}^+ _{\mathsf{q}, i} \in [\underline{h}_i ({x}_\mathsf{q}, {u}_\mathsf{q}; \mathcal{D}_{T, i}), \overline{h}_i ({x}_\mathsf{q}, {u}_\mathsf{q}; \mathcal{D}_{T, i})], \ \forall i\in\mathbb{N}_{1:n_x}, \notag
%\end{align}
where ${x}^+ _{\mathsf{q}, i}$ is the $i$-th element of ${x}^+ _\mathsf{q} $, and 
\begin{align}
\overline{h}_i ({x}_\mathsf{q}, {u}_\mathsf{q}; \mathcal{D}_{T, i}) = & \overline{r} _{i} ({x}_\mathsf{q}; \mathcal{D}_{T, i}) + f_i ({x}_\mathsf{q}, {u}_\mathsf{q})+\sigma_v \notag \\ 
&+ \left( L_f \varepsilon + L_i \sqrt{\varepsilon} + \eta_x\right) \label{overh}\\ 
\underline{h}_i ({x}_\mathsf{q}, {u}_\mathsf{q}; \mathcal{D}_{T, i}) = & \underline{r} _{i} ({x}_\mathsf{q}; \mathcal{D}_{T, i}) + f_i ({x}_\mathsf{q}, {u}_\mathsf{q})-\sigma_v \notag \\ 
&-\left( L_f \varepsilon + L_i \sqrt{\varepsilon} + \eta_x\right). \label{underh}
\end{align} 
\end{itemize} 
\end{mydef}
Recall that $L_f$ is the Lipschitz constant for the function $f$, and $L_i$ is defined in \rlem{lipschiz}. Moreover, $\overline{r} _{i}$, $\underline{r} _{i}$ are defined in \req{rmax} and \req{rmin}, respectively. As shown in \rdef{symbolicmodel}, the symbolic model provides an abstract expression of $S$, in the sense that it considers the transitions only among the \textit{discretized} points in the state and the input spaces. 
The following result indeed shows that there exists an $\varepsilon$-ASR from $S_{\mathsf{q}}$ to $S$: 
\begin{myprop}\label{asrresult}
\normalfont
Suppose that Assumptions~1,2 hold, and let $S = (\mathbb{R}^{n_x}, x_0, \mathcal{U}, G)$. Moreover, given $\mathsf{q} =  (\mathcal{D}_{T}, \eta_x, \eta_u, \varepsilon)$ with $\varepsilon \geq \eta_x$, let $S_{\mathsf{q}} = (\mathcal{X}_{\mathsf{q}}, x_{\mathsf{q}0}, \mathcal{U}_{\mathsf{q}}, {G}_{\mathsf{q}})$ be the symbolic model of $S$ in \rdef{symbolicmodel}. 
%Also, let $\widetilde{\Sigma}^\varepsilon _{A} = (\widetilde{X}, \widetilde{X}_0, \widetilde{U}, \widetilde{M}, \widetilde{G}_{A}, \widetilde{O}_A)$ be the symbolic model of $\Sigma_{A}$ defined in \rdef{symbolicmodel}, where $\varepsilon$ is the precision parameter with $\varepsilon \geq \eta_x$. 
Then, 
\begin{align}
R ( \varepsilon ) = \left\{ ({x}_\mathsf{q}, x) \in \mathcal{X}_{\mathsf{q}} \times \mathbb{R}^{n_x} \ |\ \| {x}_\mathsf{q} - x\|_{\infty} \leq \varepsilon \right\}
\end{align}
%\begin{align}\label{relation}
%R ( \varepsilon ) = \left\{ ({x}_\mathsf{q}, x) \in \mathcal{X}_{\mathsf{q}} \times \mathbb{R}^{n_x} \ |\ \| {x}_\mathsf{q} - x\|_{\infty} \leq \varepsilon \right\} 
%\end{align}
is an $\varepsilon$-ASR from ${S}_{\mathsf{q}}$ to $S$.  \qedwhite 
\end{myprop}

The proof follows in the same way to \cite{zamani2012a} and is thus given in the Appendix. 
%% \section{}
%% \label{}
In addition to the above, we also have the following result: 
\begin{mylem}\label{zeroasrlemma}
\normalfont
Let $\mathcal{D}_{T, i}=\{{{ X}}_T, { Y}_{T, i} \}$, $i\in\mathbb{N}_{1:n_x}$ be the training data with ${{ X}}_T = \left [{x}_1, {x}_2, \ldots, {x}_{{T}} \right ]$ and ${ Y}_{T, i} = [y_{1,i}, y_{2,i}, \ldots, y_{T,i} ]^\mathsf{T}$ for all $T \in \mathbb{N}_{>0}$, and let $\mathcal{D}_{T} = \{\mathcal{D}_{T,1},\ldots, \mathcal{D}_{T,n_x} \}$, $T \in \mathbb{N}_{>0}$. Moreover, for any $T_1, T_2 \in \mathbb{N}_{>0}$ with $T_1 \leq T_2$, let $\mathsf{q}_1 =  (\mathcal{D}_{T_1}, \eta_x, \eta_u, \varepsilon)$ and $\mathsf{q}_2 =  (\mathcal{D}_{T_2}, \eta_x, \eta_u, \varepsilon)$ and let $S_{\mathsf{q_1}}$ and $S_{\mathsf{q_2}}$ be the corresponding symbolic models according to \rdef{symbolicmodel}. Then, the relation 
\begin{align}\label{zerorelation}
R = \{ ({x}_{\mathsf{q}}, {x}' _{\mathsf{q}}) \in \mathcal{X}_{\mathsf{q}} \times \mathcal{X}_{\mathsf{q}} \ |\ {x}_{\mathsf{q}} = {x}' _{\mathsf{q}} \}
\end{align}
is a $0$-ASR from $S_{\mathsf{q_1}}$ to $S_{\mathsf{q_2}}$. \qedwhite 
\end{mylem}

\begin{pf}
Let the two symbolic models be given by $S_{\mathsf{q}_1} = (\mathcal{X}_{\mathsf{q}_1}, x_{\mathsf{q}_1 0}, \mathcal{U}_{\mathsf{q}_1}, {G}_{\mathsf{q}_1})$, $S_{\mathsf{q}_2} = (\mathcal{X}_{\mathsf{q}_2}, x_{\mathsf{q}_2 0}, \mathcal{U}_{\mathsf{q}_2}, {G}_{\mathsf{q}_2})$. 
Note that $\mathcal{X}_{\mathsf{q}_1} = \mathcal{X}_{\mathsf{q}_2} = [\mathbb{R}^{n_x}]_{\eta_x}$, $x_{\mathsf{q}_1 0} = x_{\mathsf{q}_2 0}$ and $\mathcal{U}_{\mathsf{q}_1} = \mathcal{U}_{\mathsf{q}_2} = [\mathcal{U}]_{\eta_x}$, since we use the same discretization parameters $\eta_x$, $\eta_u$ for both $\mathsf{q}_1$ and $\mathsf{q}_2$. Hence, the condition (C.1) in \rdef{asr2} holds. The condition (C.2) holds from the definition of $R$ \req{zerorelation}. 
To show the condition (C.3), let us recall that for every ${x}_{\mathsf{q}} \in [\mathbb{R}^{n_x}]_{\eta_x}$, $\overline{r} _{i} ({x}_{\mathsf{q}}; \mathcal{D}_{T, i})$ (resp. $\underline{r} _{i} (x_{\mathsf{q}}; \mathcal{D}_{T, i})$) is non-increasing (resp. non-decreasing) with respect to $T$. Hence, for every ${x}_{\mathsf{q}} \in [\mathbb{R}^{n_x}]_{\eta_x}$ and ${u}_{\mathsf{q}} \in [\mathcal{U}]_{\eta_u}$, we have 
\begin{align}
[\underline{h}_i  ({x}_{\mathsf{q}}, {u}_{\mathsf{q}}; & \mathcal{D}_{T_2, i}), \overline{h}_i ({x}_{\mathsf{q}}, {u}_{\mathsf{q}}; \mathcal{D}_{T_2, i})] \notag \\ 
                    &\subseteq [\underline{h}_i ({x}_{\mathsf{q}}, {u}_{\mathsf{q}}; \mathcal{D}_{T_1, i}), \overline{h}_i ({x}_{\mathsf{q}}, {u}_{\mathsf{q}}; \mathcal{D}_{T_1, i})], \label{heq}
\end{align}
or in other words, $G_{\mathsf{q}_2}({x}_{\mathsf{q}}, {u}_{\mathsf{q}}) \subseteq G_{\mathsf{q}_1} ({x}_{\mathsf{q}}, {u}_{\mathsf{q}})$. This directly means that the condition (C.3) in \rdef{asr2} holds. Therefore, it is shown that the relation \req{zerorelation} is a $0$-ASR from $S_{\mathsf{q_1}}$ to $S_{\mathsf{q_2}}$.
\end{pf}
\rlem{zeroasrlemma} implies that, since $G_{\mathsf{q}_2}({x}_{\mathsf{q}}, {u}_{\mathsf{q}}) \subseteq G_{\mathsf{q}_1} ({x}_{\mathsf{q}}, {u}_{\mathsf{q}})$ for every ${x}_{\mathsf{q}} \in [\mathbb{R}^{n_x}]_{\eta_x}$ and ${u}_{\mathsf{q}} \in [\mathcal{U}]_{\eta_u}$, the redundant transitions that are present in the symbolic model can be removed by increasing the number of the training data. 
This is due to the fact that the uncertainty (or the error bound) on the unknown function $d$ can be smaller as the training data increases (see \rsec{asdsec}). 

\subsection{Synthesizing a safety controller}\label{synthesizesafetycontrol}
%Given that the training data $\mathcal{D}_{T}$ is obtained, we now derive an algorithm to find a controlled invariant set and the corresponding safety controller. 
Given $\mathsf{q} =  (\mathcal{D}_{T}, \eta_x, \eta_u, \varepsilon)$, suppose that the symbolic model $S_{\mathsf{q}}$ is obtained according to \rdef{symbolicmodel}. 
%such that $R(\varepsilon)$ in \req{relation} is $\varepsilon$-ASR from $S_{\mathsf{q}}$ to $S$. 
Based on the symbolic model, we can find a controlled invariant set $\mathcal{X}_S$ in $\mathcal{X}$ and the corresponding safety controller $C_S$ by employing a \textit{safety game}, see, e.g., \cite{tabuadabook2009}. The algorithm of the safety game is illustrated in \ralg{synsafecon}. 
In the algorithm, the operator ${\rm Pre}_{S_{\mathsf{q}}}: 2^{\mathcal{X}_{\mathsf{q}}} \rightarrow 2^{\mathcal{X}_{\mathsf{q}}}$ is called a \textit{predecessor operator} and is defined by 
\begin{align}
\!\!\!{{\rm Pre}}_{S_{\mathsf{q}}} (\mathcal{Q}) = \{{x}_\mathsf{q} \in &\mathcal{Q} \ |\ \exists {u}_\mathsf{q} \in \mathcal{U}_{\mathsf{q}} : {G}_{\mathsf{q}} ({x}_\mathsf{q}, {u}_\mathsf{q}) \subseteq \mathcal{Q} \}, \label{pre} 
%& \forall {x}^+ _\mathsf{q} \in {G}_{\mathsf{q}} ({x}_\mathsf{q}, {u}_\mathsf{q}),\ {x}^+ _\mathsf{q}  \in \mathcal{Q}\}, \label{pre} 
\end{align}
for a given $\mathcal{Q} \subseteq \mathcal{X}_{\mathsf{q}}$. That is, ${{\rm Pre}}_{S_{\mathsf{q}}} (\mathcal{Q})$ is the set of all states in $\mathcal{Q}$, for which there exists a control input in $\mathcal{U}_{\mathsf{q}}$ such that all the corresponding successors are inside $\mathcal{Q}$. 
\textcolor{black}{The controlled invariant set $\mathcal{X}_{S}$ is computed based on the fixed point set of $\mathcal{Q}_\ell$ (i.e., $\mathcal{X}_{S, \mathsf{q}}$). 
In particular, if $\mathcal{X}_{S, \mathsf{q}}$ is non-empty, the controlled invariant set is computed based on the $\varepsilon$-ASR $R(\varepsilon)$ (\rline{xs}). 
On the other hand, if $\mathcal{X}_{S, \mathsf{q}}$ is empty, it indicates that the controlled invariant set is not found (and so we set $\mathcal{X}_{S} \leftarrow \varnothing$ as shown in \rline{Csnotfound}).}
%The controller for the symbolic model, which we denote by ${C}_{\mathsf{q}, S}$ (\rline{obtaincsq}), is obtained as follows: 
%for all ${x}_\mathsf{q} \in \mathcal{X} _{S, \mathsf{q}}$, where $\mathcal{X} _{S, \mathsf{q}}$ is defined in \ralg{synsafecon}.
%The set $\mathcal{X} _{S} \subseteq \mathcal{X}$ is then given by 
%\begin{align}\label{xs}
%\mathcal{X} _{S} = \{x \in \mathcal{X}\ |\ \exists {x}_\mathsf{q} \in \mathcal{X}_{S,\mathsf{q}}, ({x}_\mathsf{q}, x) \in R(\varepsilon) \}, 
%\end{align}
%where $\mathcal{X} _{S, \mathsf{q}}$ is defined in \ralg{synsafecon} (\rline{statexq}). 
%Intuitively, the set $\mathcal{X} _{S}$ computed in \req{xs} indicates a set contained in $\mathcal{X}$, for which a safety controller exists\footnote{Formally, the set $\mathcal{X} _{S}$ is called a \textit{controlled invariant set.}}. 
%Finally, the safety controller $C_{S}$ is obtained as \rline{Cs}.
Roughly speaking, $C_{S,\mathsf{q}}$ (\rline{Csq}) serves as a safety controller for the symbolic model ${S}_{\mathsf{q}}$, and the safety controller $C_S$ for $S$ is \textit{refined} based on the $\varepsilon$-ASR $R(\varepsilon)$ (\rline{Cs}). 
%which is utilized to \textit{refine} the safety controller $C_S$ for the original transition system $S$. 
Note that \ralg{synsafecon} is guaranteed to terminate after a finite number of iteration, since $[\mathsf{Interior}_\varepsilon (\mathcal{X})]_{\eta_x}$ and $[\mathcal{U}]_{\eta_u}$ are both finite. %Moreover, \ralg{synsafecon} can be implemented using several off-the-shelf tools, such as PESSOA \cite{pessoa}. 
The following result is an immediate consequence from the fact that $R(\varepsilon)$ is the $\varepsilon$-ASR from $S_{\mathsf{q}}$ to $S$ and thus the proof is omitted (see, e.g., \cite{tabuadabook2009}). 
%result states that the set $\mathcal{X}_S$ is shown to be a controlled invariant set and $C_S$ is the corresponding safety controller. 

\renewcommand{\algorithmicrequire}{\textbf{Input:}}
\renewcommand{\algorithmicensure}{\textbf{Output:}}
\begin{algorithm}[t]
\caption{$\mathsf{SafeCon}(S_{\mathsf{q}}, \mathcal{X})$ (safety controller synthesis).}\label{synsafecon}
\begin{algorithmic}[1]
{\small
\REQUIRE{$S_{\mathsf{q}}$ (symbolic model of $S$), $\mathcal{X}$ (safe set);}
\ENSURE{$\mathcal{X}_S$ (if $\varnothing \neq \mathcal{X}_S$, it yields a controlled invariant set in $\mathcal{X}$), $C_S$ (if $\varnothing \neq \mathcal{X}_S$, it yields a safety controller);}
\STATE $\ell \leftarrow 0$; 
\STATE $\mathcal{Q}_\ell \leftarrow [\mathsf{Interior}_\varepsilon (\mathcal{X})]_{\eta_x}$; 
\REPEAT 
\STATE $\ell \leftarrow \ell +1$; 
\STATE $\mathcal{Q}_{\ell} \leftarrow {\rm Pre}_{S_{\mathsf{q}}} (\mathcal{Q}_{\ell-1})$; \label{intersectq} 
\UNTIL{$\mathcal{Q}_{\ell-1} = \mathcal{Q}_\ell$}
\STATE $\mathcal{X}_{S, \mathsf{q}} \leftarrow \mathcal{Q}_\ell$; \label{statexq}  
\textcolor{black}{\IF{$\varnothing \neq \mathcal{X}_{S, \mathsf{q}}$} 
\STATE $\mathcal{X} _{S} \leftarrow \{x \in \mathcal{X} | \exists {x}_\mathsf{q} \in \mathcal{X}_{S,\mathsf{q}}, ({x}_\mathsf{q}, x) \in R(\varepsilon) \}$; \label{xs}
\STATE ${C}_{S, \mathsf{q}} ({x}_\mathsf{q}) \leftarrow \{{u}_\mathsf{q} \in \mathcal{U}_{\mathsf{q}} |  {G}_{\mathsf{q}} ({x}_\mathsf{q}, {u}_\mathsf{q}) \subseteq \mathcal{X} _{S, \mathsf{q}}\}$, $\forall {x}_\mathsf{q} \in \mathcal{X}_{S, \mathsf{q}}$; \label{Csq}
\STATE $C_{S} (x) \leftarrow \left \{C_{S,\mathsf{q}} ({x}_\mathsf{q}) | ({x}_\mathsf{q}, x) \in R(\varepsilon) \right \}$, $\forall x \in \mathcal{X}_{S}$; \label{Cs}
\ELSE 
\STATE $\mathcal{X} _{S} \leftarrow \varnothing$, $C_{S} (x) \leftarrow \varnothing$, $\forall x \in \mathcal{X}$ (which indicates that the controlled invariant set and safety controller are not found); \label{Csnotfound}
\ENDIF}
}
\end{algorithmic}
\end{algorithm}

\begin{mylem}\label{controlled_invariantlem2}
\normalfont
%Let $\mathsf{q} =  (\mathcal{D}_{T}, \eta_x, \eta_u, \varepsilon)$ and 
%For given suppose that the symbolic model $S_{\mathsf{q}}$ is constructed according to \rdef{symbolicmodel}. Moreover, 
Suppose that for given $S_\mathsf{q}$ and $\mathcal{X}$, \ralg{synsafecon} is implemented and $\mathcal{X}_{S} \neq \varnothing$. 
Then, $\mathcal{X}_{S}$ is a controlled invariant set in $\mathcal{X}$, and $C_S$ is the corresponding safety controller. \qedwhite
\end{mylem}

In addition to the above, we also have the following result: 
\begin{mylem}\label{controlled_invariantlem}
\normalfont
Let $\mathcal{D}_{T, i}=\{{{ X}}_T, { Y}_{T, i} \}$, $i\in\mathbb{N}_{1:n_x}$ be the training data with ${{ X}}_T = \left [{x}_1, {x}_2, \ldots, {x}_{{T}} \right ]$ and ${ Y}_{T, i} = [y_{1,i}, y_{2,i}, \ldots, y_{T,i} ]^\mathsf{T}$ for all $T \in \mathbb{N}_{>0}$, and let $\mathcal{D}_{T} = \{\mathcal{D}_{T,1},\ldots, \mathcal{D}_{T,n_x} \}$, $T \in \mathbb{N}_{>0}$. Moreover, for any $T_1, T_2 \in \mathbb{N}_{>0}$ with $T_1 \leq T_2$, let $\mathsf{q}_1 =  (\mathcal{D}_{T_1}, \eta_x, \eta_u, \varepsilon)$ and $\mathsf{q}_2 =  (\mathcal{D}_{T_2}, \eta_x, \eta_u, \varepsilon)$ and let $S_{\mathsf{q_1}}$ and $S_{\mathsf{q_2}}$ be the corresponding symbolic models according to \rdef{symbolicmodel}. 
In addition, let $\mathcal{X}_{S_1}, \mathcal{X}_{S_2}$ be the resulting controlled invariant sets by executing $\mathsf{SafeCon}(S_{\mathsf{q_1}}, \mathcal{X})$ and $\mathsf{SafeCon}(S_{\mathsf{q_2}}, \mathcal{X})$, respectively. 
%Suppose that $\mathsf{SafeCon}(S_{\mathsf{q_1}}, \mathcal{X})$ and $\mathsf{SafeCon}(S_{\mathsf{q_2}}, \mathcal{X})$ are executed according to \ralg{synsafecon}, and let $\mathcal{X}_{S_1}, \mathcal{X}_{S_2}\subseteq \mathcal{X}$ be the resulting controlled invariant sets, respectively. 
%Moreover, let $X^* _{S_1}$ and $X^* _{S_2}$ be the controlled invariant sets in $X_S$ by applying \ralg{synsafecon} with the symbolic models $S_{\mathsf{q_1}}$ and $S_{\mathsf{q_2}}$, respectively. 
Then, $\mathcal{X}_{S_1} \subseteq \mathcal{X}_{S_2}$. \qedwhite 
\end{mylem}
%\rlem{controlled_invariantlem} immediately follows from the fact that the relation \req{zerorelation} is $0$-ASR from $S_{\mathsf{q_1}}$ to $S_{\mathsf{q_2}}$ (see \rlem{zeroasrlemma}). 
In essence, \rlem{controlled_invariantlem} means that the controlled invariant set does not shrink or can be enlarged by increasing the number of training data. As previously mentioned, this is due to that the symbolic model becomes more and more accurate  (i.e., the redundant transitions are removed) as the training data increases, since the error bound on $d$ can be smaller as the training data increases. 
While \rlem{controlled_invariantlem} might trivially follow from the existence of a $0$-ASR from $S_{\mathsf{q}_1}$ to $S_{\mathsf{q}_2}$ (see \rlem{zeroasrlemma}), we here provide a detailed proof below, since the proof procedure will be useful to provide an approach to reduce the computational load for the safety controller synthesis (for details, see \rsec{reducecomsec}). 
%As will be seen in the next section, this lemma is utilized to show that the algorithm to learn the symbolic model is guaranteed to terminate in a finite number of iterations. 
%\begin{comment}
\begin{pf}
Let the two symbolic models be given by $S_{\mathsf{q}_1} = (\mathcal{X}_{\mathsf{q}_1}, x_{\mathsf{q}_1 0}, \mathcal{U}_{\mathsf{q}_1}, {G}_{\mathsf{q}_1})$, $S_{\mathsf{q}_2} = (\mathcal{X}_{\mathsf{q}_2}, x_{\mathsf{q}_2 0}, \mathcal{U}_{\mathsf{q}_2}, {G}_{\mathsf{q}_2})$. Then, from the proof of \rlem{zeroasrlemma}, it follows that $G_{\mathsf{q}_2}({x}_{\mathsf{q}}, {u}_{\mathsf{q}}) \subseteq G_{\mathsf{q}_1} ({x}_{\mathsf{q}}, {u}_{\mathsf{q}})$ for every ${x}_{\mathsf{q}} \in [\mathbb{R}^{n_x}]_{\eta_x}$ and ${u}_{\mathsf{q}} \in [\mathcal{U}]_{\eta_u}$. 
%Note that $\mathcal{X}_{\mathsf{q}_1} = \mathcal{X}_{\mathsf{q}_2} = [\mathbb{R}^{n_x}]_{\eta_x}$, $x_{\mathsf{q}_1 0} = x_{\mathsf{q}_2 0}$ and $\mathcal{U}_{\mathsf{q}_1} = \mathcal{U}_{\mathsf{q}_2} = [\mathcal{U}]_{\eta_x}$, since we use the same discretization parameters $\eta_x$, $\eta_u$ for $\mathsf{q}_1$ and $\mathsf{q}_2$.
Now, let $\mathcal{Q}_{1,\ell}$, $\mathcal{Q}_{2,\ell}$, $\ell = 0, 1, \ldots$ be the sets of $\mathcal{Q}_\ell$ obtained by executing $\mathsf{SafeCon}(S_{\mathsf{q_1}}, \mathcal{X})$ and $\mathsf{SafeCon}(S_{\mathsf{q_2}}, \mathcal{X})$, respectively. Note that $\mathcal{Q}_{1,0} = \mathcal{Q}_{2,0}$. 
Hence, it follows that 
\begin{align}
G_{\mathsf{q}_1}(x_\mathsf{q}, u_{\mathsf{q}}) \subseteq \mathcal{Q}_{1,0}\implies G_{\mathsf{q}_2}(x_\mathsf{q}, u_{\mathsf{q}}) \subseteq \mathcal{Q}_{2,0}.
\end{align} 
%\begin{align}
%$\forall {x}^+ _{\mathsf{q}}  \in {G}_{\mathsf{q}_1} ({x}_{\mathsf{q}}, {u}_{\mathsf{q}}),\ {x}^+_{\mathsf{q}} \in \mathcal{Q} _{1,0}$, 
%\end{align}
%it follows that $\forall {x}^+_{\mathsf{q}}  \in {G}_{\mathsf{q}_2} ({x}_{\mathsf{q}}, {u}_{\mathsf{q}}),\ {x}^+  \in \mathcal{Q}_{2,0}$. 
Thus, we obtain ${{\rm Pre}}_{S_{\mathsf{q}_1}} (\mathcal{Q}_{1,0}) \subseteq {{\rm Pre}}_{S_{\mathsf{q}_2}} (\mathcal{Q}_{2,0})$. 
Hence, from the fact that $\mathcal{Q}_{1,0} = \mathcal{Q}_{2,0}$ and \rline{intersectq} in \ralg{synsafecon}, it follows that $\mathcal{Q}_{1,1} \subseteq \mathcal{Q}_{2,1}$. 
%Therefore, $Q^{(0)} _1 \subseteq Q^{(0)} _2$ implies $Q^{(1)} _1 \subseteq Q^{(1)} _2$. 
By recursively applying the same reasoning as above, it then follows that $\mathcal{Q}_{1,\ell} \subseteq \mathcal{Q}_{2,\ell}$, $\ell = 0, 1, \ldots$. 
In other words, we have $\mathcal{X} _{S_1, \mathsf{q}} \subseteq \mathcal{X} _{S_2, \mathsf{q}}$ and namely, $\mathcal{X} _{S_1} \subseteq \mathcal{X} _{S_2}$. 
\end{pf}

\begin{myrem}
\normalfont
As we will see in the next section, the safety controller is utilized to achieve \textit{safe exploration}, where training data can be collected while guaranteeing safety (i.e., staying in $\mathcal{X}$ for all times). Note that while we focus here on synthesizing a safety controller, we can also synthesize controllers under other specifications, including those expressed by temporal logic formulas or automata on infinite strings. For example, the simulation result given in this paper considers fulfilling a linear temporal logic (LTL) specification for adaptive cruise control (ACC) based on a symbolic model obtained by applying the proposed approach; for details, see \rsec{simsec}. \qedwhite 
\end{myrem}

\section{Learning-based safe symbolic abstractions}\label{mainsec}
%\subsection{Symbolic models based on simulation relations}
%to synthesize a controller based on approximate simulation relations. 
\renewcommand{\algorithmicrequire}{\textbf{Input:}}
\renewcommand{\algorithmicensure}{\textbf{Output:}}
\begin{algorithm}[t]
\caption{Learning-based symbolic abstractions with safe exploration (overall, main algorithm).}\label{overall_alg}
\begin{algorithmic}[1]
{\small
\REQUIRE{${x}_0$ (initial state), ${C} _{S, {\rm init}}$ (initial safety controller), $\eta_x, \eta_u, \varepsilon$ (some parameters for the symbolic model), $T_{{\rm exp}} \in\mathbb{N}_{>0}$ (number of training data collected for each iteration of safe exploration);}
\ENSURE{$S_{\mathsf{q}_N}$ (symbolic model);}
\STATE $\mathcal{D}_{0,i} \leftarrow \varnothing$, $\forall i\in \mathbb{N}_{1:n_x}$; \label{initializetraining} 
\STATE$\mathcal{D}_{0} \leftarrow \{\mathcal{D}_{0,1}, \ldots, \mathcal{D}_{0,n_x}\}$; 
\STATE $T_{1} \leftarrow T_{\rm exp}$; \label{initializealltraining}
\STATE$\{x_{T_{1}}, \mathcal{D}_{T_{1}}\} \leftarrow \mathsf{SafeExp} (x_{0}, T_{{\rm exp}}, C_{S,\mathrm{init}}, \mathcal{D}_{0})$;
\STATE$\mathsf{q}_{1} \leftarrow \{\mathcal{D}_{T_{1}}, \eta_x, \eta_u, \varepsilon\}$; 
\STATE $S_{\mathsf{q}_{1}} \leftarrow (\mathcal{X}_{\mathsf{q}_{1}}, x_{\mathsf{q}_{1}0}, \mathcal{U}_{\mathsf{q}_{1}}, {G}_{\mathsf{q}_{1}})$ (\rdef{symbolicmodel});
\STATE$\{\mathcal{X} _{S,1}, C_{S,1}\} \leftarrow \mathsf{SafeCon} (S_{\mathsf{q}_{1}}, \mathcal{X})$ (\ralg{synsafecon}); \label{initializeinvset}

\textcolor{black}{\IF{$\varnothing \neq \mathcal{X}_{S,1}$}
\REPEAT 
\STATE$T_{N+1} \leftarrow T_{N} + T_{\rm exp}$;\label{tupdate}
\STATE$\{x_{T_{N+1}}, \mathcal{D}_{T_{N+1}}\} \leftarrow \mathsf{SafeExp} (x_{T_{N}}, T_{{\rm exp}}, C_{S,N}, \mathcal{D}_{T_{N}})$\label{nonexp}\\ 
\STATE$N \leftarrow N+1$; 
\STATE$\mathsf{q}_{N} \leftarrow \{\mathcal{D}_{T_{N}}, \eta_x, \eta_u, \varepsilon\}$\label{updatesym0}; 
\STATE $S_{\mathsf{q}_{N}} \leftarrow (\mathcal{X}_{\mathsf{q}_{N}}, x_{\mathsf{q}_{N}0}, \mathcal{U}_{\mathsf{q}_{N}}, {G}_{\mathsf{q}_{N}})$ (\rdef{symbolicmodel}); \label{updatesym} 
\STATE$\{\mathcal{X} _{S, N}, C_{S, N}\} \leftarrow \mathsf{SafeCon} (S_{\mathsf{q}_{N}}, \mathcal{X})$ (\ralg{synsafecon}); \label{synsafecontroller} 
\UNTIL{$\mathcal{X} _{S, N-1} = \mathcal{X}_{S, N}$}
\ENDIF}
}
\end{algorithmic}
\end{algorithm}

In this section we present an overall algorithm that aims at collecting the training data from scratch and constructing the symbolic model while achieving the safe exploration. 
%learning framework that jointly learns the unknown function $d(\cdot)$ as well as the controlled invariant set and the corresponding safety controller. 
%present a concrete framework to solve \rpro{problem}. %During execution, the system explores the set for collecting the training data while guaranteeing safety (with high probability), and gradually enlarge the controlled invariant set and synthesize the safety and reachability controllers. 
Before providing the algorithm, we need to make the following assumption: 
\begin{myas}\label{initial_as}
\normalfont
There exists a \textit{known} %controlled invariant set $\mathcal{X} _{S, {\rm init}} \subset \mathcal{X}$ with $x_0 \in \mathcal{X} _{S, {\rm init}}$ and the corresponding 
safety controller $C_{S, {\rm init}} : \mathcal{X} \rightarrow 2^{\mathcal{U}}$ such that any trajectory induced by $C_{S, {\rm init}}$ stays in the safety set $\mathcal{X}$ for all times. \qedwhite 
\end{myas}
\ras{initial_as} implies the existence of an initial safety controller, so that the training data can be collected at the initial phase. 
%and it allows us to explore the region and collect the training data while guarantee safety at the initial phase. 
{The initial safety controller $C_{S, {\rm init}}$ may be obtained by employing an expert or heuristically based on the nominal model $f(x_k, u_k)$; for details, see \rrem{computeinitialinv} below.} 
%to estimate $d$ and enlarge the controlled invariant set accordingly. 

\renewcommand{\algorithmicrequire}{\textbf{Input:}}
\renewcommand{\algorithmicensure}{\textbf{Output:}}
\begin{algorithm}[t]
\caption{$\mathsf{SafeExp}(x_{T_{N}}, T_{{\rm exp}}, C_{S,N}, \mathcal{D}_{T_{N}})$ (safe exploration).}\label{exploration}
\begin{algorithmic}[1]
{\small
\REQUIRE{$x_{T_{N}}$(current state), $T_{\rm exp}$ (number of training data collected for each iteration of safe exploration), $C_{S,N}$ (safety controller), \\ $\mathcal{D}_{T_{N}}$ (current training data);}
\ENSURE{$x_{T_{N+1}}$,$\mathcal{D}_{T_{N+1}}$ (updated current state and training data after the exploration);}
\STATE${X} \leftarrow \varnothing$; 
\STATE${Y}_i \leftarrow \varnothing$, $i \in \mathbb{N}_{1:n_x}$(initialize the new training data);
\FOR{$t =T_{N}: T_{N} + T_{\rm exp}-1$}
\STATE Compute $u_t \in C_{S,N} (x_t)$ by \req{usel1},\req{usel2}; \label{computeuline} 
\STATE Apply $u_t$ and measure the next state: $x_{t+1}= [x_{t+1, 1}, \ldots, x_{t+1, n_x}]^\mathsf{T}$; 
%Mesure the level of safety $z_{k+1} = g(x_{k+1}) + w_k$; \\
\STATE For all $i \in \mathbb{N}_{1:n_x}$, set the training data as follows: 
\begin{align}
{X} \leftarrow [{X},\ x_t],\ {Y}_i \leftarrow [{Y}_i,\ x_{t+1, i} - f_i (x_t, u_t)];
\end{align} 
\ENDFOR
\STATE $T_{N+1}\leftarrow T_{N}+T_{\rm exp}$; 
\STATE $\mathcal{D}_{T_{N+1}, i} \leftarrow \mathcal{D}_{T_{N}, i} \cup \{{X}, {Y}_i\}$, $i \in \mathbb{N}_{1:n_x}$; 
\STATE $\mathcal{D}_{T_{N+1}} \leftarrow \{\mathcal{D}_{T_{N+1}, 1}, \ldots, \mathcal{D}_{T_{N+1}, n_x}\}$; 
}
\end{algorithmic}
\end{algorithm}

The overall learning algorithm is shown in \ralg{overall_alg} and the details are described as follows. The algorithm starts by initializing the training data by applying the initial safety controller and then updating the controlled invariant set and the safety controller (\rline{initializetraining}--\rline{initializeinvset}). 
%The training data is initially set as the emply set (\rline{initializetraining},\rline{initializealltraining}). 
\textcolor{black}{Then, if $\varnothing \neq \mathcal{X}_{S,1}$ (i.e., $\mathcal{X}_{S,1}$ is a controlled invariant set in $\mathcal{X}$), we move on to the iteration (lines~9--16). In the iteration, we first update $T_{N+1}$ (\rline{tupdate}).} 
Roughly speaking, $T_N$, $N\in\mathbb{N}_{\geq 0}$ indicate the number of training data that has been collected until the $N$-th iteration of \ralg{overall_alg}. The algorithm proceeds by executing a \textit{safe exploration} algorithm $\mathsf{SafeExp}$ (\rline{nonexp}), which aims at collecting the new training data while guaranteeing safety. %is executed, which aims at collecting the new training data while guaranteeing safety. 
%where $T_{\rm exp}$ indicates the number of 
 %which aims at collecting the new training data while guaranteeing safety. 
%symbolic model and the safety controller are updated. 
%For each iteration in \ralg{overall_alg}, it executes $\mathsf{SafeExp}$ (\rline{nonexp}), which indicates a \textit{safe exploration} algorithm that aims at collecting the training data while guaranteeing safety. 
In detail, the safe exploration algorithm is shown in \ralg{exploration}. 
In the algorithm, the control input $u_t$ (\rline{computeuline}) is computed as follows:
%for $N=0$ (the initial exploration), $u_t$ is randomly selected from $C_{S, N} (=C_{S, {\rm init}})$. For $N>0$, 
\begin{numcases}
{\!\!\!\!\!\!\!\!\!\!\! u_t = }
  {\rm select\ arbitrarily\ from\ }C_{S,N} (x_t), ({\rm if}\ N=0), \label{usel1}\\ 
	\underset{{u \in C_{S,N} (x_t)}}{\arg\max}\ \sum_{i=1} ^{n_x} \sigma^2 _i (\hat{x}^+ _i ; \mathcal{D}_{T_{N}, i}),\ \ ({\rm if}\ N>0),\label{usel2}
\end{numcases}
%\begin{numcases}
%{u_t =} 
%{\underset{{u \in C_{S,N-1} (x_t)}}{\arg\max}\ \sum_{i=1} ^{n_x} \sigma^2 _i (\hat{x}^+ _i ; \mathcal{D}_{T_{N-1}, i}). \\
%aaa}
%\end{numcases}
where $\hat{x}^+ = f(x_t, u) + \hat{d}(x_t; \mathcal{D}_{T_{N}})$ with $\hat{d}(x_t; \mathcal{D}_{T_{N}}) = [\hat{d}_1(x_t; \mathcal{D}_{T_{N},1}), ..., \hat{d}_{n_x} (x_t; \mathcal{D}_{T_{N},n_x})]^\mathsf{T}$. 
Recall that $\sigma^2 _i$ and $\hat{d}_i$ are defined in \req{sigmadi} and \req{gihat}, respectively. 
That is, we select the control input randomly from $C_{S,N}$ for the initial exploration, and, otherwise, select from $C_{S,N}$ such that the corresponding (predictive) next state has the largest variance on $d$. 
By doing so, the system actively explores the state-space so as to reduce the uncertainty on $d$ and enlarge the controlled invariant set while guaranteeing safety. The exploration is given until it collects the new $T_{\rm exp}$ training data, and it outputs the new training data $\mathcal{D}_{T_{N+1}}$ and the current state $x_{T_{N+1}}$ after the exploration. 
Afterwards, the symbolic model $S_{\mathsf{q}_N}$ is updated with the new training data according to \rdef{symbolicmodel} (\rline{updatesym0}, \rline{updatesym} in \ralg{overall_alg}), and the controlled invariant set $\mathcal{X} _{S,N}$ and the safety controller $C_{S, N}$ are updated by $\mathsf{SafeCon}$ (\rline{synsafecontroller} in \ralg{overall_alg}). 
\textcolor{black}{The above procedure is iterated until the controlled invariant set converges, i.e., $\mathcal{X} _{S, N-1} = \mathcal{X}_{S, N}$ (see line~12 in Algorithm~2). Note that $\mathcal{X} _{S, N}$ is computed by refining the set of discretized states, i.e., $\mathcal{X} _{S,N} = \{x \in \mathcal{X}\ |\ \exists {x}_\mathsf{q} \in \mathcal{X}_{S,\mathsf{q}_N}, ({x}_\mathsf{q}, x) \in R(\varepsilon) \}$ (see line~8 in Algorithm~1). Hence, we have $\mathcal{X} _{S, N-1} = \mathcal{X}_{S, N}$ if and only if $\mathcal{X}_{S,\mathsf{q}_N} = \mathcal{X}_{S,\mathsf{q}_{N-1}}$. 
Since $\mathcal{X}_{S,\mathsf{q}_{N-1}}$ and $\mathcal{X}_{S,\mathsf{q}_N}$ are both finite and $\mathcal{X}_{S,\mathsf{q}_{N-1}} \subseteq \mathcal{X}_{S,\mathsf{q}_N}$ (for details, see the proof of Theorem~1 below), the condition $\mathcal{X}_{S,\mathsf{q}_N} = \mathcal{X}_{S,\mathsf{q}_{N-1}}$ can be checked in a finite time (i.e., check if every $x_{\mathsf{q}} \in \mathcal{X}_{S,\mathsf{q}_N}$ is contained in $\mathcal{X}_{S,\mathsf{q}_{N-1}}$).}
%Together with the symbolic model $S_{\mathsf{q}}$ the algorithm also returns the controlled invariant set and the safety controller, since they can be useful for the controller synthesis. 

Regarding the overall algorithm, we can conclude the following result: 
\begin{mythm}
\normalfont 
Suppose that Assumptions~1--3 hold and \ralg{overall_alg} is implemented. Then, \ralg{overall_alg} terminates after a finite number of iteration. 
Moreover, the relation $R ( \varepsilon ) = \left\{ ({x}_\mathsf{q}, x) \in \mathcal{X}_{\mathsf{q}} \times \mathbb{R}^{n_x} \ |\ \| {x}_\mathsf{q} - x\|_{\infty} \leq \varepsilon \right\}$
is an $\varepsilon$-ASR from $S_{\mathsf{q}_N}$ to $S$ for all $N\in\mathbb{N}_{> 0}$ until \ralg{overall_alg} terminates. 
%Moreover, $S_{\mathsf{q}}$ is the symbolic model such that \req{relation} is $\varepsilon$-ASR from $S_{\mathsf{q}}$ to $S$.  %$\mathcal{X} _S$ is the controlled invariant set and $C_S$ is the corresponding safety controller. 
In addition, the \textit{safe exploration} is achieved, i.e., 
during the implementation of \ralg{exploration}, it is shown that the trajectory of the system \req{dynamics} stays in the safe set $\mathcal{X}$ for all times. \qedwhite 
\end{mythm}

\begin{pf}
Let us first show that \ralg{overall_alg} terminates after a finite number of iteration. 
Given $N\in\mathbb{N}_{>0}$, let $\mathcal{X}_{S, \mathsf{q}_N}$ be the set of $\mathcal{X}_{S, \mathsf{q}}$ in \ralg{synsafecon} (\rline{statexq}) computed by executing $\mathsf{SafeCon} (S_{\mathsf{q}_N}, \mathcal{X})$. 
Since $T_{N + 1} \geq T_N$, and from the proof of \rlem{controlled_invariantlem}, we obtain $\mathcal{X}_{S, \mathsf{q}_N} \subseteq\mathcal{X}_{S, \mathsf{q}_{N+1}}$. In general, it follows that $\mathcal{X}_{S, \mathsf{q}_0} \subseteq \mathcal{X}_{S, \mathsf{q}_1} \subseteq \mathcal{X}_{S, \mathsf{q}_2} \subseteq \cdots$. 
Note that $\mathcal{X}_{S, \mathsf{q}_N} \subseteq  [\mathsf{Interior}_{\varepsilon}(\mathcal{X})]_{\eta_x}$ for all $N\in\mathbb{N}_{>0}$ and that $[\mathsf{Interior}_{\varepsilon}(\mathcal{X})]_{\eta_x}$ is finite. 
%where $\widetilde{X}_{S} = \widetilde{X}_{S,\ell} (= \widetilde{X}_{S,\ell+1} \cdots)$. 
Hence, there exists an $N' \in \mathbb{N}_{>0}$ such that $\mathcal{X}_{S, \mathsf{q}_{N'}} = \mathcal{X}_{S, \mathsf{q}_{N'+1}}$. 
This in turn implies that $\mathcal{X}_{S, N'} = \mathcal{X}_{S, N'+1}$, and, therefore, \ralg{overall_alg} terminates after a finite number of iteration. 
The fact that $R(\varepsilon)$ is an $\varepsilon$-ASR from $S_{\mathsf{q}_N}$ to $S$ for all $N\in\mathbb{N}_{> 0}$ (until \ralg{overall_alg} terminates) trivially holds from \rprop{asrresult}. 
Moreover, we can achieve the safe exploration, since control inputs are always chosen from the safety controller $C_{S, N}$. 
\end{pf}
%% For citations use: 
%%       \citet{<label>} ==> Jones et al. [21]
%%       \cite{<label>} ==> [21]
%%

%% If you have bibdatabase file and want bibtex to generate the
%% bibitems, please use
%%
%%  \bibliographystyle{elsarticle-num-names} 
%%  \bibliography{<your bibdatabase>}

%% else use the following coding to input the bibitems directly in the
%% TeX file.

Note that, in order to make the implementation of \ralg{overall_alg} tractable, it is only necessary to construct the symbolic model within the state-space $[\mathcal{X}]_{\eta_x}$. Specifically, the update of the symbolic model (\rline{updatesym} in \ralg{overall_alg}) is replaced by defining a new symbolic model $S _{D, \mathsf{q}_N}$: 
%Specifically, we define the new symbolic model by replacing \rline{updatesym} in \ralg{overall_alg} with
\begin{align}\label{sdn}
S _{D, \mathsf{q}_N} \leftarrow (\mathcal{X}_{D,\mathsf{q}_N}, x_{D, \mathsf{q}_N0}, \mathcal{U}_{D, \mathsf{q}_N}, {G}_{D, \mathsf{q}_N}),
\end{align}
where $\mathcal{X}_{D,\mathsf{q}_N} = [\mathcal{X}]_{\eta_x}$, $x_{D, \mathsf{q}_N0} = x_{\mathsf{q}_N0}$, $\mathcal{U}_{D, \mathsf{q}_N} = \mathcal{U}_{\mathsf{q}_N}$, and
${G}_{D, \mathsf{q}_N}: \mathcal{X}_{D, \mathsf{q}_N} \times \mathcal{U}_{D,\mathsf{q}_N} \rightarrow 2^{\mathcal{X}_{D,\mathsf{q}_N}}$, with ${x}^+ _{\mathsf{q}}  \in {G}_{D, \mathsf{q}_N} ({x}_{\mathsf{q}}, {u}_{\mathsf{q}})$ if and only if ${x}^+ _{\mathsf{q}}  \in {G}_{\mathsf{q}_N} ({x}_{\mathsf{q}}, {u}_{\mathsf{q}})$ and ${G}_{\mathsf{q}_N} ({x}_{\mathsf{q}}, {u}_{\mathsf{q}}) \subseteq [\mathcal{X}]_{\eta_x}$ (i.e., if ${G}_{D, \mathsf{q}_N}(x_\mathsf{q},u_\mathsf{q}) \nsubseteq [\mathcal{X}]_{\eta_x}$ then $u_\mathsf{q} \notin \mathcal{U}_{D,\mathsf{q}_N} (x_\mathsf{q})$). 
%Let us mention that if ${G}_{D, \mathsf{q}_N}(x_\mathsf{q},u_\mathsf{q}) \nsubseteq [\mathcal{X}]_{\eta_x}$ then $u_\mathsf{q} \notin \mathcal{U}(x_\mathsf{q})$. 
%Recalling that $\mathcal{X}_{D,\mathsf{q}_N} = [\mathcal{X}]_{\eta_x}$ and $\mathcal{X}_{\mathsf{q}_N} = [\mathbb{R}^{n_x}]_{\eta_x}$, 
It can be easily shown that the relation $R = \{ ({x}_{\mathsf{q}}, {x}' _{\mathsf{q}}) \in [\mathcal{X}]_{\eta_x} \times [\mathbb{R}^{n_x}]_{\eta_x} \ |\ {x}_{\mathsf{q}} = {x}' _{\mathsf{q}} \}$ is a $0$-ASR from $S _{D, \mathsf{q}_N}$ to $S _{\mathsf{q}_N}$. From this and the fact that the relation $R ( \varepsilon ) = \left\{ ({x}_\mathsf{q}, x) \in \mathcal{X}_{\mathsf{q}} \times \mathbb{R}^{n_x} \ |\ \| {x}_\mathsf{q} - x\|_{\infty} \leq \varepsilon \right\}$ is the $\varepsilon$-ASR from $S_{\mathsf{q}_N}$ to $S$, it is shown that the relation $R_D ( \varepsilon ) = \left\{ ({x}_{\mathsf{q}}, x) \in [\mathcal{X}]_{\eta_x} \times \mathbb{R}^{n_x} \ |\ \| {x}_{\mathsf{q}} - x\|_{\infty} \leq \varepsilon \right\}$
is an $\varepsilon$-ASR from $S _{D, \mathsf{q}_N}$ to $S$ (see, e.g., \cite{zamani2012a} for a detailed discussion). 
Hence, any controller synthesized for the symbolic model ${S}_{D, \mathsf{q}_{N}}$ can be refined to a controller for the original system $S$ satisfying the same specification. 

\begin{myrem}[On obtaining $C_{S, {\rm init}}$]\label{computeinitialinv}
\normalfont
\textcolor{black}{The initial safety controller $C_{S, {\rm init}}$ could be obtained in the following ways.
First, it can be obtained by utilizing an \textit{expert} (or, human) only at the initial exploration phase. When controlling a drone, for example, we let an expert control the drone and stay within a given safety set so as to collect the training data at the initial phase (hence, the initial safety controller corresponds to the one given by the expert). After the exploration by the expert, we update the safety controller by solving the safety game, and this safety controller is applied at the next iteration. 
\textcolor{black}{Note that an expert to collect the (initial) training data has been employed in many works of literature in the context of reinforcement learning, such as \textit{{learning from demonstration}, {imitation learning}}, \textit{etc}. 
Besides, an expert has also been utilized to provide some formal guarantees on the considered control objective (e.g., safety) in learning-based control (see, e.g., \cite{robey2020ver2}).}
The initial safety controller may also be designed based on the nominal model $f(x_k, u_k)$. That is, we design a stabilizing controller such that the origin (or some target point) inside the safety set is asymptotically stable with respect to the nominal system. Since the modeling error $d_i$ is bounded, it potentially leads to that the actual system stays locally around the origin (or the target point), and so it is enough to stay within the safety set $\mathcal{X}$. The above approach is suggested in several papers of safe learning with the GP regression (see, e.g., \cite{felix2}).} \qedwhite 
\end{myrem}

%\footnote{As stated in \cite{zamani2012a}, it is shown that the composition of the two ASRs is still an ASR.}. 
%Hence, the controlled invariant set as well as the safety controller can be still found in the same way as \ralg{synsafecon}. 

%an $\varepsilon$-ASR from $S _{D, \mathsf{q}}$ to $S$, since there exists a $0$-ASR from $S _{D, \mathsf{q}}$ to $S_{\mathsf{q}}$ (for details, see \cite{zamani2012a}). 
%there exists an $\varepsilon$-ASR from $S _{D, \mathsf{q}}$ to $S$, since there exists a $0$-ASR from $S _{D, \mathsf{q}}$ to $S_{\mathsf{q}}$ (for details, see \cite{zamani2012a}). 
%all ${u}_\mathsf{q}$-successors from ${x}_\mathsf{q}$ are included in $[\mathcal{X}]_{\eta_x}$. 
%while the transition map of the symbolic model $G_{\mathsf{q}}$ is defined over $\mathcal{X}_{\mathsf{q}}$, during the implementation of \ralg{overall_alg} it is only necessary to define the transition map over $[\mathcal{X}]_{\eta_x}$, since we aim at 
 %originally 
%has a countably infinite set of states $\mathcal{X}_{\mathsf{q}}$, it is only necessary to define the symbolic model within 

\subsection{Some approaches to efficient computation}\label{reducecomsec}
Since the symbolic model needs to be updated for \textit{every} $N$, the {whole re-computation} of this abstraction (as well as the safety controller synthesis) for every iteration clearly leads to a heavy computational load. Therefore, in this section we provide some techniques to reduce the computational load so as to make our approach more practical. Specifically, we propose the following two approaches to speed up the abstraction and controller synthesis procedures: 
\begin{itemize}
\item \textit{(Lazy abstraction):} 
It should be expected that, the transitions are necessary to be updated only for the states where the uncertainty (or the variance) on $d$ is sufficiently reduced by collecting the new training data. Hence, we propose a \textit{lazy abstraction} scheme, in which, starting from the initial abstraction, transitions from states in $[\mathcal{X}]_{\eta_x}$ are then updated \textit{only when} the reduction of the variance on $d$ is large enough. 
The update of the transitions allows to reduce the redundant transitions and hence and the abstraction becomes less conservative. In the proposed procedure, we do not have to recompute the abstraction for the whole states in $[\mathcal{X}]_{\eta_x}$, but only for the states on which new training data is collected. 
\item \textit{(Speeding up the computation of predecessors):} It should be expected that the main source of the heavy computation for the safety controller synthesis is the predecessor operator ${\rm Pre}_{S_\mathsf{q}} (\mathcal{Q}_\ell)$ (see \req{pre}); clearly, checking for \textit{every} state in $\mathcal{Q}_\ell$ if there exists a control input such that all the corresponding successors are in $\mathcal{Q}_\ell$ requires a heavy computation, as this operation needs to be done for every $\ell$ and $N$. Therefore, we propose an approach to reduce the computational load of computing this predecessor operator, in order to speed up the safety controller synthesis. In particular, we eliminate redundant computations of the predecessor operator by making use of the earlier computed predecessors. 
\end{itemize}

%It should be expected that, the transitions are necessary to be updated only for the states where the uncertainty (or variance) on $d$ is sufficiently reduced by collecting the new training data. 
%the error bound on $d$ will become especially smaller around the region where new training data is collected (since the variance computed by the GP regression becomes small around the training data), and so the redundant transitions will be particularly eliminated around that region. 
%Hence, we propose a \textit{local update} algorithm, in which the transitions are updated only for the states where the reduction of the variance on $d$ is large enough, while keeping the same transitions as the previous iteration for the other states. 
Regarding the first approach in the above, the update of the symbolic model (\rline{updatesym} in \ralg{overall_alg}) is replaced by defining a new symbolic model $\widetilde{S} _{D, \mathsf{q}_N}$:
%Specifically, we define the new symbolic model by replacing \rline{updatesym} in \ralg{overall_alg} with 
\begin{align}
\widetilde{S} _{D, \mathsf{q}_N} \leftarrow (\mathcal{X}_{D, \mathsf{q}_N}, x_{D,\mathsf{q}_N0}, \mathcal{U}_{D,\mathsf{q}_N}, \widetilde{G}_{D, \mathsf{q}_N}), \label{stilde}
\end{align}
for all $N\in\mathbb{N}_{\geq 1}$, where 
$\widetilde{G}_{D, \mathsf{q}_N}$ is the transition map that is (newly) constructed by applying \ralg{consym}. 
\renewcommand{\algorithmicrequire}{\textbf{Input:}}
\renewcommand{\algorithmicensure}{\textbf{Output:}}
\begin{algorithm}[t]
\caption{Derivation of $\widetilde{G}_{D, \mathsf{q}_N}$ for all $N\in \mathbb{N}_{ \geq 1}$ (lazy abstraction).}\label{consym}
\begin{algorithmic}[1]
{\small
\REQUIRE{$\mathcal{D}_{T_{1:N}}$(training data), $ \rho \in\mathbb{R}_{>0}$ (threshold to update transitions in $\widetilde{G}_{D, \mathsf{q}_{N}}$), $\widetilde{G}_{D, \mathsf{q}_{N-1}}$(transition map of $\widetilde{S}_{D, \mathsf{q}_{N-1}}$ (if $N>1$));}
\ENSURE{$\widetilde{G}_{D, \mathsf{q}_{N}}$ (transition map of $\widetilde{S}_{D, \mathsf{q}_{N}}$);}

\IF{$N=1$ (initial execution of \ralg{consym})}
\STATE $\mathcal{X}^c _{\mathsf{q}} \leftarrow \varnothing$; \label{updatexcq}
\ENDIF

\FOR{each $x_{\mathsf{q}} \in \mathcal{X}^c _{\mathsf{q}}$}\label{keepstart}
\FOR{each $u_{\mathsf{q}} \in [\mathcal{U}]_{\eta_u}$}
\STATE $\widetilde{G}_{D, \mathsf{q}_N} (x_{\mathsf{q}}, u_{\mathsf{q}}) \leftarrow \widetilde{G}_{D, \mathsf{q}_{N-1}} (x_{\mathsf{q}}, u_{\mathsf{q}})$;
\ENDFOR
\ENDFOR\label{keepend}

\smallskip

\FOR{each $x_{\mathsf{q}} \in [\mathcal{X}]_{\eta_x} \backslash \mathcal{X}^c _{\mathsf{q}}$}\label{updatestart}
\FOR{each $u_{\mathsf{q}} \in [\mathcal{U}]_{\eta_u}$}
\STATE $\mathcal{X}^+ _{\mathsf{q}} \leftarrow \{x^+ _{\mathsf{q}} \in [\mathbb{R}^{n_x}]_{\eta_x}\ |\ {x}^+ _{\mathsf{q}, i} \in [\underline{h}_i ({x}_\mathsf{q}, {u}_\mathsf{q}; \mathcal{D}_{T_N, i}), \overline{h}_i ({x}_\mathsf{q}, {u}_\mathsf{q}; \mathcal{D}_{T_N, i})], \forall i\in\mathbb{N}_{1:n_x} \}$; \label{localupdatestart} 
\IF{$\mathcal{X}^+ _{\mathsf{q}}\subseteq [\mathcal{X}]_{\eta_x}$}
\STATE $\widetilde{G}_{D, \mathsf{q}_N} (x_{\mathsf{q}}, u_{\mathsf{q}}) \leftarrow \mathcal{X}^+ _{\mathsf{q}}$; 
%\STATE $\Delta \leftarrow \sum_{i=1}^{n_x} (\bar{r}_i({x}_\mathsf{q}; \mathcal{D}_{T_N, i}) - \underline{r}_i({x}_\mathsf{q}; \mathcal{D}_{T_N, i}))$; 
\IF{$\Delta_i({x}_\mathsf{q}; \mathcal{D}_{T_N, i}) < \rho$ for all $i\in\mathbb{N}_{1:n_x}$}\label{deltadefalg}
\STATE $\mathcal{X}^c _{\mathsf{q}} \leftarrow \mathcal{X}^c _{\mathsf{q}}\cup \{x_{\mathsf{q}}\}$; 
\ENDIF
%\STATE $\mathcal{L} (x_{\mathsf{q}}) \leftarrow N$; \label{localupdateend2}
\ENDIF
\ENDFOR
\ENDFOR \label{updateend}
}
\end{algorithmic}
\end{algorithm}
\textcolor{black}{The core element of \ralg{consym} is the set $\mathcal{X}^c _{\mathsf{q}}$. This set is defined as the empty set at the initial execution of \ralg{consym} ($N=1$), i.e., $\mathcal{X}^c _{\mathsf{q}} = \varnothing$ and then it is updated for $N>1$ (as detailed below). Note that since $\mathcal{X}^c _{\mathsf{q}} = \varnothing$ for the initial execution, the procedure of \rline{keepstart}--\rline{keepend} is not implemented for $N=1$. As shown in \rline{updatestart}--\rline{updateend}, for each state in $[\mathcal{X}]_{\eta_x} \backslash \mathcal{X}^c _{\mathsf{q}}$ and each input in $[\mathcal{U}]_{\eta_u}$, the corresponding transition map $\widetilde{G}_{D, \mathsf{q}_N} (x_{\mathsf{q}}, u_{\mathsf{q}})$ is updated (see lines~11 and 13). More importantly, as shown in lines~14 and 15, if $\Delta_i({x}_\mathsf{q}; \mathcal{D}_{T_N, i}) < \rho$ holds for all $i\in\mathbb{N}_{1:n_x}$, then ${x}_\mathsf{q}$ is \textit{added} to $\mathcal{X}^c _{\mathsf{q}}$ ($\rho$ denotes a user-defined threshold). Recall that $\Delta_i({x}_\mathsf{q}; \mathcal{D}_{T_N, i})$ represents the length of the confidence interval, or \textit{uncertainty} for $d_i ({x}_\mathsf{q})$ given the training data $\mathcal{D}_{T_N, i}$ (see \req{deltahat}). 
Hence, if the uncertainty of the state ${x}_\mathsf{q}$ becomes small enough, then ${x}_\mathsf{q}$ is added to $\mathcal{X}^c _{\mathsf{q}}$. As shown in line~4 to 8, if ${x}_\mathsf{q}$ is added to $\mathcal{X}^c _{\mathsf{q}}$, the transitions from ${x}_\mathsf{q}$ is kept the same as the previous iteration afterwards (see line~6). That is, the transitions from ${x}_\mathsf{q}$ are no more updated once the corresponding uncertainty becomes small enough. 
This is reasonable because the states having small uncertainties will not have redundant transitions and so it is no longer necessary to update the transition map. Moreover, this will indeed speed up the construction of the transition map, since the transitions from some of the states are not necessary to be updated once their uncertainties become small.} 

In summary, the symbolic model is given by \req{stilde}, where the corresponding transition map $\widetilde{G}_{D, \mathsf{q}_N}$ is computed by executing \ralg{consym} for all $N\in\mathbb{N}_{\geq 1}$ (until \ralg{overall_alg} terminates). 
The computational complexity of \ralg{consym} is provided as follows. 
\textcolor{black}{For the initial execution of \ralg{consym} ($N=1$), we have $\mathcal{X}^c _{\mathsf{q}} = \varnothing$ and thus the transition map ${G}_{D, \mathsf{q}_N}$ is computed for all states in  $[\mathcal{X}]_{\eta_x}$ and all inputs in $[\mathcal{U}]_{\eta_u}$. Hence, the computational complexity of constructing the transition map is $\mathcal{O}(|[\mathcal{X}]_{\eta_x}|\  |[\mathcal{U}]_{\eta_u}| c(T_N))$, where $c(T_N)$ denotes the computational complexity of the one-step reachable states (\rline{localupdatestart} in \ralg{consym}). Here, the computational complexity of the one-step reachable states depends on the data size $T_N$, since the computations of $\underline{h}_i ({x}_\mathsf{q}, {u}_\mathsf{q}; \mathcal{D}_{T_N, i})$ and $\overline{h}_i ({x}_\mathsf{q}, {u}_\mathsf{q}; \mathcal{D}_{T_N, i})$ involve the computations of the GP mean and variance. 
For example, standard computation of the GP mean/variance requires a cubic complexity $\mathcal{O}(T_N^3)$ due to the inversion of the $T_N \times T_N$ matrix. 
Note that the construction of the symbolic model for the \textit{known} dynamics requires $\mathcal{O}(|[\mathcal{X}]_{\eta_x}|\  |[\mathcal{U}]_{\eta_u}|)$, because we need to define the transition maps for every pair of the state and the control input $(x_{\mathsf{q}}, u_{\mathsf{q}} )\in [\mathcal{X}]_{\eta_x} \times [\mathcal{U}]_{\eta_u}$. Hence, the computational complexity of our approach additionally requires the multiplication of $c(T_N)$ (in contrast to the one of the abstraction scheme with the known dynamics). This is clear because the dynamics is here estimated by a {non-parametric} (or, GP) model based on training data.} 
\textcolor{black}{Now, consider $N>1$. As shown in \ralg{consym}, if $x_{\mathsf{q}} \in [\mathcal{X}]_{\eta_x} \backslash \mathcal{X}^c _{\mathsf{q}}$, transitions are re-computed for all $u_{\mathsf{q}} \in [\mathcal{U}]_{\eta_u}$ (\rline{updatestart}--\rline{updateend}), and otherwise, transitions from $x_{\mathsf{q}}$ are directly set as the previous ones of $N-1$ (\rline{keepstart}--\rline{keepend}). 
Hence, the computational complexity of \ralg{consym} is 
\begin{align*}
&\mathcal{O}\Bigl( \underbrace{ (|[\mathcal{X}]_{\eta_x}| - |\mathcal{X}^c _{\mathsf{q}}|)\cdot |[\mathcal{U}]_{\eta_u}| \cdot  c(T_N)}_{\rline{updatestart}-\rline{updateend}} + \underbrace{ |\mathcal{X}^c _{\mathsf{q}}|\cdot  |[\mathcal{U}]_{\eta_u}|}_{\rline{keepstart}-\rline{keepend}}\Bigr)\notag \\ 
&\!\!\!\! = \mathcal{O} \biggl (|[\mathcal{X}]_{\eta_x}|\cdot  |[\mathcal{U}]_{\eta_u}|\cdot c(T_N) -  |\mathcal{X}^c _{\mathsf{q}}|\cdot  |[\mathcal{U}]_{\eta_u}| \bigl(c(T_N) -1  \bigr) \biggr ).
\end{align*}
This implies that \ralg{consym} becomes faster as the cardinality of $\mathcal{X}^c _{\mathsf{q}}$ becomes larger, i.e., the number of states having small uncertainties is larger. 
Therefore, it is expected that the execution time of \ralg{consym} will be shorter as the state-space exploration progresses and the uncertainty on the unknown function $d$ becomes smaller.} 

\textcolor{black}{
The memory requirement of Algorithm~4 is $\mathcal{O}(|[\mathcal{X}]_{\eta_x}]|^2 \cdot |[\mathcal{U}]_{\eta_u}|)$, since for each $x_{\mathsf{q}}\in [\mathcal{X}]_{\eta_x}$ and $u_{\mathsf{q}} \in [\mathcal{U}]_{\eta_u}$ the set of the corresponding successors in $[\mathcal{X}]_{\eta_x}$ needs to be stored in the memory. 
Although we here provide the {worst case} memory analysis, %we cannot show from this result that the memory requirement decreases as the iteration progresses. 
%However, 
we have the potential to actually reduce the memory requirement as $N$ increases, since the redundant transitions are removed more and more as the iteration progresses. For this clarification, see the numerical experiment of Section~6, in which we have shown that the memory requirement to save all the transitions in the symbolic model (i.e., Algorithm~4) becomes smaller as $N$ increases.}

The following result shows that the existence of an $\varepsilon$-ASR is still guaranteed from $\widetilde{S}_{D, \mathsf{q}_N}$ to $S$. 
\begin{mythm}%\label{controlled_invariantlem2}
\normalfont
Suppose that Assumptions~1--3 hold and \ralg{overall_alg} is implemented, in which the symbolic model is given by \req{stilde} whose transition map $\widetilde{G}_{D, \mathsf{q}_N}$ is computed by executing \ralg{consym} for all $N\in\mathbb{N}_{\geq 1}$. Then, for every $N\in \mathbb{N}_{> 0}$, the relation $R_D ( \varepsilon ) = \{ ({x}_{\mathsf{q}}, x) \in [\mathcal{X}]_{\eta_x} \times \mathbb{R}^{n_x}\ |\ \| {x}_{\mathsf{q}} - x\|_{\infty} \leq \varepsilon \}$
 is an $\varepsilon$-ASR from $\widetilde{S}_{D, \mathsf{q}_N}$ to $S$. \qedwhite 
%Let $\mathsf{q} =  (\mathcal{D}_{T}, \eta_x, \eta_u, \varepsilon)$ and suppose that the symbolic model $S_{\mathsf{q}}$ is obtained such that the relation $R(\varepsilon)$ in \req{relation} is $\varepsilon$-ASR from $S_{\mathsf{q}}$ to $S$. Moreover, suppose that \ralg{synsafecon} is implemented and $\mathcal{X}_{S} \neq \varnothing$. Then, the controller $C_S$ as derived in \req{Cs} is a safety controller. \qedwhite
\end{mythm}

\begin{pf}
The result follows by induction. For $N=1$, $R_D$ is the $\varepsilon$-ASR from $\widetilde{S}_{D, \mathsf{q}_N}$ to $S$, since $\widetilde{S}_{D, \mathsf{q}_N} = {S}_{D, \mathsf{q}_N}$ and $R_D$ is the $\varepsilon$-ASR from ${S}_{D, \mathsf{q}_N}$ to $S$ (see the discussion after Theorem~1). 
For a given $N \in \mathbb{N}_{>1}$, assume that $R_D$ is the $\varepsilon$-ASR from $\widetilde{S}_{D, \mathsf{q}_N}$ to $S$, and suppose that, at the next iteration $N+1$, $\widetilde{G}_{D, \mathsf{q}_{N+1}}$ is given by \ralg{consym}. 
In what follows, it is shown that there exists a $0$-ASR
from $\widetilde{S}_{D, \mathsf{q}_{N}}$ to $\widetilde{S}_{D, \mathsf{q}_{N+1}}$. 
%From \ralg{consym}, for every $x_{\mathsf{q}}\in [\mathcal{X}]_{\eta_x}, u_{\mathsf{q}}\in[\mathcal{U}]_{\eta_u}$ with $\widetilde{G}_{D, \mathsf{q}_{N}}(x_{\mathsf{q}}, u_{\mathsf{q}}) \neq \varnothing$, there exists $N' \leq N$ such that $\widetilde{G}_{D, \mathsf{q}_N} (x_{\mathsf{q}}, u_{\mathsf{q}}) = \{x^+ _{\mathsf{q}} \in [\mathcal{X}]_{\eta_x}$ $| {x}^+ _{\mathsf{q}, i} \in [\underline{h}_i ({x}_\mathsf{q}, {u}_\mathsf{q}; \mathcal{D}_{T_{N'}, i}), \overline{h}_i ({x}_\mathsf{q}, {u}_\mathsf{q}; \mathcal{D}_{T_{N'}, i})], \forall i\in\mathbb{N}_{1:n_x} \}$. 
%Now, consider $N+1$ and suppose that $\widetilde{G}_{D, \mathsf{q}_{N+1}}$ is given by \ralg{consym}. 
From the derivation of $\widetilde{G}_{D, \mathsf{q}_{N+1}}$ in
\ralg{consym}, for every $x_{\mathsf{q}}\in [\mathcal{X}]_{\eta_x}, u_{\mathsf{q}}\in[\mathcal{U}]_{\eta_u}$ with $\widetilde{G}_{D, \mathsf{q}_{N}}(x_{\mathsf{q}}, u_{\mathsf{q}}) \neq \varnothing$, it follows either $\widetilde{G}_{D, \mathsf{q}_{N+1}}(x_{\mathsf{q}}, u_{\mathsf{q}}) = \widetilde{G}_{D, \mathsf{q}_{N}}(x_{\mathsf{q}}, u_{\mathsf{q}})$, or $\widetilde{G}_{D, \mathsf{q}_{N+1}} (x_{\mathsf{q}}, $$u_{\mathsf{q}})$$= \{x^+ _{\mathsf{q}} \in [\mathcal{X}]_{\eta_x}$ $| {x}^+ _{\mathsf{q}, i} \in [\underline{h}_i ({x}_\mathsf{q}, {u}_\mathsf{q}; \mathcal{D}_{T_{N+1}, i}),$ $\overline{h}_i ({x}_\mathsf{q}, {u}_\mathsf{q}; \mathcal{D}_{T_{N+1}, i})]$,$\forall i\in\mathbb{N}_{1:n_x}\}$. Note that for the latter case, we have 
\begin{align*}
[\underline{h}_i ({x}_\mathsf{q}, {u}_\mathsf{q}; & \mathcal{D}_{T_{N+1}, i}), \overline{h}_i ({x}_\mathsf{q}, {u}_\mathsf{q}; \mathcal{D}_{T_{N+1}, i})] \notag \\ 
&\subseteq [\underline{h}_i ({x}_\mathsf{q}, {u}_\mathsf{q}; \mathcal{D}_{T_{N'}, i}), \overline{h}_i ({x}_\mathsf{q}, {u}_\mathsf{q}; \mathcal{D}_{T_{N'}, i})], 
\end{align*}
for all $N' \leq N$, since $T_{N'} \leq T_N$ (see the proof of Lemma~3). %Thus, $\widetilde{G}_{D, \mathsf{q}_{N+1}} (x_{\mathsf{q}}, u_{\mathsf{q}})\subseteq \widetilde{G}_{D, \mathsf{q}_N} (x_{\mathsf{q}}, u_{\mathsf{q}})$. 
Hence, for every $x_{\mathsf{q}}\in [\mathcal{X}]_{\eta_x}, u_{\mathsf{q}}\in[\mathcal{U}]_{\eta_u}$ with $\widetilde{G}_{D, \mathsf{q}_{N}}(x_{\mathsf{q}}, u_{\mathsf{q}}) \neq \varnothing$, it follows that $\widetilde{G}_{D, \mathsf{q}_{N+1}} (x_{\mathsf{q}}, u_{\mathsf{q}})\subseteq \widetilde{G}_{D, \mathsf{q}_N} (x_{\mathsf{q}}, u_{\mathsf{q}})$. This implies that the relation $R = \left\{ ({x}_{\mathsf{q}}, {x}'_{\mathsf{q}}) \in [\mathcal{X}]_{\eta_x} \times [\mathcal{X}]_{\eta_x} \ |\ {x}_{\mathsf{q}} = {x}'_{\mathsf{q}} \right\}$ 
%This implies that the relation $R \subseteq [\mathcal{X}]_{\eta_x} \times [\mathcal{X}]_{\eta_x}$ defined by $({x}_\mathsf{q}, {x}'_\mathsf{q}) \in R$ if and only if $x_{\mathsf{q}} = x' _{\mathsf{q}}$ is 
is a $0$-ASR from $\widetilde{G}_{D, \mathsf{q}_{N}}$ to $\widetilde{G}_{D, \mathsf{q}_{N+1}}$. Thus, from the assumption that $R_D$ is the $\varepsilon$-ASR from $\widetilde{S}_{D, \mathsf{q}_N}$ to $S$, $R_D$ is the $\varepsilon$-ASR from $\widetilde{S}_{D, \mathsf{q}_{N+1}}$ to $S$. 
%Therefore, \req{relation2} 
%It then follows that $R(\varepsilon)$ defined in \req{relation} is $\varepsilon$-ASR from $G_{D, \mathsf{q}_{N+1}}$ to $S$, since it is assumed that $R(\varepsilon)$ is $\varepsilon$-ASR from $S_{D, \mathsf{q}_N}$ to $S$. 
Therefore, it is inductively shown that $R_D$ is the $\varepsilon$-ASR from $\widetilde{S}_{D, \mathsf{q}_{N}}$ to $S$ for all $N\in\mathbb{N}_{> 0}$.
\end{pf}

Hence, any controller synthesized for the symbolic model $\widetilde{S}_{D, \mathsf{q}_{N}}$ can be refined to a controller for the original system $S$ satisfying the same specification. 

%\begin{comment}
\begin{myrem}\label{redundantrem}
\normalfont
$\widetilde{S}_{D,\mathsf{q}_N}$ can have more (redundant) transitions than $S_{D, \mathsf{q}_N}$, since the transitions of $\widetilde{S}_{D,\mathsf{q}_N}$ are updated only for some states, while in $S_{D, \mathsf{q}_N}$ these are updated for all states in $[\mathcal{X}]_{\eta_x}$. From \rlem{controlled_invariantlem}, this implies that using $S_{D, \mathsf{q}_N}$ may result in a larger controlled invariant set than using $\widetilde{S}_{D, \mathsf{q}_N}$, which may be a drawback of using $\widetilde{S}_{D, \mathsf{q}_N}$. Nevertheless, as will be illustrated in the numerical example in the next section (\rsec{simsec}), constructing $\widetilde{S}_{D, \mathsf{q}_N}$ should be more practical and useful than constructing ${S}_{D, \mathsf{q}_N}$, since it achieves a significant reduction of the computational load. \qedwhite 
\end{myrem}
\begin{myrem}
\normalfont
Let us mention that the use of lazy approaches has been previously used in the symbolic control literature (see the approaches proposed in \cite{girard2016a,hussein,adnane2018d} and a review of the lazy techniques in \cite{adnanethesis}). In these approaches, the refinement of the abstraction is done for the regions that are not able to achieve the safety specification (either by using finer discretizations or lower inputs). In this paper, the criteria of the refinement are different, since we are refining on the regions where we are able to collect new data. Moreover, the method of refinement is also different since we conserve the same discretizations, the same input, but we benefit from the supplementary knowledge on the un-modeled dynamics to reduce the redundant transitions. \qedwhite 
\end{myrem}

\renewcommand{\algorithmicrequire}{\textbf{Input:}}
\renewcommand{\algorithmicensure}{\textbf{Output:}}
\begin{algorithm}[t]
\caption{Derivation of ${\rm Pre}_{\widetilde{S}_{D,\mathsf{q}_N}}(\mathcal{Q}_{N, \ell})$ for all $N \in \mathbb{N}_{\geq 1}$, $\ell \in \mathbb{N}_{\geq 0}$ (speeding up the computation of the predecessors).}\label{compreduce}
\begin{algorithmic}[1]
{\small
\REQUIRE{$\widetilde{S}_{D,\mathsf{q}_N}$, $\mathcal{Q}_{N, \ell}$, and $\mathcal{Q}_{N-1, \ell+1}$ (available if $N >1$);}
\ENSURE{${\rm Pre}_{\widetilde{S}_{D,\mathsf{q}_N}}(\mathcal{Q}_{N, \ell})$;}
%$\mathcal{Q} \leftarrow \mathcal{Q}_{N-1, \ell+1}$;}
\IF {$N = 1$} 
\STATE $\mathcal{Q} \leftarrow \varnothing$,\ $\mathcal{Q}_{N-1, \ell+1} \leftarrow \varnothing$; 
\ELSE 
\STATE $\mathcal{Q} \leftarrow \mathcal{Q}_{N-1, \ell+1}$; 
\ENDIF
\FOR {each $x_{\mathsf{q}} \in \mathcal{Q}_{N,\ell} \backslash \mathcal{Q}_{N-1, \ell+1}$}
\FOR {each $u_{\mathsf{q}}\in [\mathcal{U}]_{\eta_u}$}
\IF{$\widetilde{G}_{D, \mathsf{q}_N} ({x}_\mathsf{q}, {u}_\mathsf{q})\subseteq \mathcal{Q}_{N,\ell}$} 
\STATE $\mathcal{Q} \leftarrow \mathcal{Q} \cup \{x_{\mathsf{q}}\}$; 
\ENDIF
\ENDFOR
\ENDFOR 
\STATE ${\rm Pre}_{\widetilde{S}_{D,\mathsf{q}_N}}(\mathcal{Q}_{N, \ell}) \leftarrow \mathcal{Q}$; 
}
\end{algorithmic}
\end{algorithm}

\begin{comment}
\begin{algorithm}[t]\label{compreduce}
{
\SetKwInOut{Input}{Input}
\SetKwInOut{Output}{Output}
\Input{$\widetilde{S}_{D,\mathsf{q}_N}$, $\mathcal{Q}_{N, \ell}$, $\mathcal{Q}_{N-1, \ell+1}$; } 
\Output{${\rm Pre}_{\widetilde{S}_{D,\mathsf{q}_N}}(\mathcal{Q}_{N, \ell})$;}
$\mathcal{Q} \leftarrow \mathcal{Q}_{N-1, \ell+1}$; \\
\For {each $x_{\mathsf{q}} \in \mathcal{Q}_{N,\ell} \backslash \mathcal{Q}_{N-1, \ell+1}$}{
%\uIf {$x_{\mathsf{q}} \in \mathcal{Q}^{N-1} _{\ell+1}$}{
%$\mathcal{Q} \leftarrow \mathcal{Q} \cup \{x_{\mathsf{q}}\}$; \\
%}
%\Else 
\For {each $u_{\mathsf{q}}\in [\mathcal{U}]_{\eta_u}$}{
{\If{$\widetilde{G}_{D, \mathsf{q}_N} ({x}_\mathsf{q}, {u}_\mathsf{q})\subseteq \mathcal{Q}_{N,\ell}$}{
$\mathcal{Q} \leftarrow \mathcal{Q} \cup \{x_{\mathsf{q}}\}$; 
}
}
}
}
${\rm Pre}_{\widetilde{S}_{D,\mathsf{q}_N}}(\mathcal{Q}_{N, \ell}) \leftarrow \mathcal{Q}$; 
   \caption{{Derivation of ${\rm Pre}_{\widetilde{S}_{D,\mathsf{q}_N}}(\mathcal{Q}_{N, \ell})$ for all $N \in \mathbb{N}_{>1}$ and $\ell \in \mathbb{N}_{\geq 0}$ (speeding up the computation of predecessors).}}
    }
\end{algorithm}
\end{comment}

Let us now proceed by reducing the computational load of the predecessor operator ${\rm Pre}_{\widetilde{S}_{D,\mathsf{q}_N}}$ in order to speed up the safety controller synthesis. 
To this end, let $\mathcal{Q}_{N, \ell}$, $\ell= 0, 1, \ldots$ denote the sequence of sets $\mathcal{Q}_\ell$, $\ell= 0, 1, \ldots$ in \ralg{synsafecon} by executing $\mathsf{SafeCon}(\widetilde{S}_{\mathsf{q}_{N}}, \mathcal{X})$. Then, it follows from $T_{N} \geq T_{N-1}$ for all $N \in \mathbb{N}_{> 0}$ that $\mathcal{Q}_{N-1,\ell+1} \subseteq \mathcal{Q}_{N,\ell+1} = {\rm Pre}_{\widetilde{S}_{D,\mathsf{q}_N}}(\mathcal{Q}_{N, \ell})$, for all $N \in \mathbb{N}_{> 0}$ and $\ell \in \mathbb{N}_{> 0}$ (see the proof of \rlem{controlled_invariantlem}). 
Hence, when we aim at computing ${\rm Pre}_{\widetilde{S}_{D,\mathsf{q}_N}}(\mathcal{Q}_{N, \ell})$, it is \textit{known} that $\mathcal{Q}_{N-1,\ell+1}$ is the subset of ${\rm Pre}_{\widetilde{S}_{D,\mathsf{q}_N}}(\mathcal{Q}_{N, \ell})$. This implies that all states in $\mathcal{Q}_{N-1,\ell+1}$ can be directly added to the predecessors for $\mathcal{Q}_{N, \ell}$ \textit{without} checking the existence of a control input such that all successors are in $\mathcal{Q}_{N, \ell}$ according to \req{pre}. 
%set as the subset of ${\rm Pre}_{\widetilde{S}_{D,\mathsf{q}_N}}(\mathcal{Q}_{N, \ell})$ \textit{without} checking the existence of a control input such that all successors are in $\mathcal{Q}_{N, \ell}$ according to \req{pre}. 

\textcolor{black}{Based on the above observation, we propose \ralg{compreduce} so as to speed up the computation of ${\rm Pre}_{\widetilde{S}_{D,\mathsf{q}_N}}(\mathcal{Q} _{N, \ell})$. %As shown in line~6 to line~12, 
In the algorithm, $\mathcal{Q}$ represents the set of  predecessors for $\mathcal{Q}_{N,\ell}$ which are mainly updated according to line~6--line~12. 
For $N=1$, $\mathcal{Q}$ is initialized by the empty set (line~2). 
In other words, \textit{all} states in $\mathcal{Q}_{N, \ell}$ (since $\mathcal{Q} _{N, \ell}\backslash \mathcal{Q}_{N-1, \ell+1} = \mathcal{Q} _{N, \ell}$) are evaluated to check the existence of a control input such that all the successors are in $\mathcal{Q}_{N, \ell}$ according to line~6 to line~12. 
For $N>1$, on the other hand, $\mathcal{Q}$ is initialized by $\mathcal{Q}_{N-1, \ell+1}$ (line~4). This is due to the fact that it is \textit{already known} that $\mathcal{Q}_{N-1,\ell+1}$ is a subset of the predecessors for $\mathcal{Q}_{N, \ell}$ (i.e., $\mathcal{Q}_{N-1,\ell+1}\subseteq {\rm Pre}_{\widetilde{S}_{D,\mathsf{q}_N}}(\mathcal{Q}_{N, \ell})$). 
Hence, only the states in $\mathcal{Q}_{N,\ell} \backslash \mathcal{Q}_{N-1, \ell+1}$ (instead of $\mathcal{Q}_{N,\ell}$) are necessary to be evaluated to check the existence of a control input such that all the successors are in $\mathcal{Q}_{N, \ell}$ according to line~6--12.}

In summary, during execution of $\mathsf{SafeCon}(\widetilde{S}_{\mathsf{q}_{N}}, \mathcal{X})$ (\rline{synsafecontroller} in \ralg{overall_alg}) for all $N\in\mathbb{N}_{\geq 1}$, ${\rm Pre}_{\widetilde{S}_{D,\mathsf{q}_N}}(\mathcal{Q}_{N, \ell})$, $\ell = 0, 1, ...$ are computed by \ralg{compreduce}. 
\textcolor{black}{The computational complexity of computing predecessors according to \ralg{compreduce} for $N=1$ is $\mathcal{O} \bigl(|\mathcal{Q}_{N,\ell}|\  |[\mathcal{U}]_{\eta_u}|\bigr)$. 
For $N>1$, we have $\mathcal{O} \bigl(|\mathcal{Q}_{N,\ell} \backslash \mathcal{Q}_{N-1, \ell+1}|\  |[\mathcal{U}]_{\eta_u}|\bigr)$. Hence, the computation of the predecessors becomes faster as the cardinality of $\mathcal{Q}_{N,\ell} \backslash \mathcal{Q}_{N-1, \ell+1}$ becomes smaller, or in other words, $\mathcal{Q}_{N,\ell}$ is closer to $\mathcal{Q}_{N-1, \ell+1}$, i.e., $\mathcal{Q}_{N,\ell} \approx \mathcal{Q}_{N-1, \ell+1}$. Note that we have $\mathcal{Q}_{N,\ell} = \mathcal{Q}_{N-1, \ell+1}$ if the controlled invariant set converges $\mathcal{X}_{S, N-1} = \mathcal{X}_{S, N}$ (i.e., $\mathcal{Q}_{N, \ell} = \mathcal{Q}_{N-1, \ell}$, $\forall \ell \in \mathbb{N}_{\geq 0}$) and $\mathcal{Q}_{N,\ell} = \mathcal{Q}_{N, \ell+1}$ (i.e., $\mathcal{Q}_{N,\ell}$ converges to a fixed point). Hence, it is expected that \ralg{compreduce} becomes faster as both the controlled invariant set and $\mathcal{Q}_{N, \ell}$ get closer to their fixed points.} \textcolor{black}{The memory requirement is $\mathcal{O}(|[\mathsf{Interior}_\varepsilon (\mathcal{X})]_{\eta_x}|)$, since it needs to store the set $\mathcal{Q}_{N,\ell} \subseteq [\mathsf{Interior}_\varepsilon (\mathcal{X})]_{\eta_x}$.} 

\section{Simulation results} \label{simsec}
In this section we illustrate the effectiveness of the proposed approach through a simulation of an adaptive cruise control (ACC) \cite{nilsson,ames,saoud2019contract}. The simulation has been conducted on Windows 10, Intel(R) Core(TM) 2.40GHz, 8GB RAM. The state vector is given by $x = [x_1, x_2, x_3]^\mathsf{T} \in \mathbb{R}^3$, where $x_1$ is the velocity of the leading vehicle, $x_2$ is the velocity of the following vehicle, and $x_3$ is the distance between the lead vehicle and the following vehicle. 
\textcolor{black}{Moreover, the input vector indicates the acceleration of the following car $u \in \mathbb{R}$. 
The dynamics is given by 
\begin{align*} 
{{x}}_{t+1}  =  &\underbrace{x_t + \Delta \left [
\begin{array}{c}
0  \\
u_t  \\
x_{1,t} - x_{2,t} 
\end{array}
\right ]}_{f(x_t, u_t)} + \underbrace{\Delta\left [
\begin{array}{c}
a_{l,t} \\
0 \\
0 
\end{array}
\right ]}_{v_t} \notag \\  
&\ \ + \underbrace{\Delta\left [
\begin{array}{c}
0 \\
-(\nu_{0} + \nu_{1} x_{2,t} + \nu_{2} x_{2,t}^2 )/M \\
0 
\end{array}
\right ]}_{d(x_t)}, %\label{acc}
\end{align*}
where $M$ is the weight of the following vehicles, $\Delta$ represents the sampling time, $a_{l, t}$ is the acceleration of the lead vehicle that is assumed to be the additive noise, and $\nu_{0}, \nu_{1}, \nu_{2}$ are the constants for the aerodynamic drag force, whose function (i.e., $(\nu_{0} + \nu_{1} x_{2,t} + \nu_{2} x_{2,t}^2)/M$) is assumed to be {unknown} apriori.} 
It is assumed that $\Delta = 1, M=1000, \nu_{0} = 60, \nu_{1} = 1.2, \nu_{2} = 1.0$.  Moreover, we assume that the velocity of the lead vehicle fulfills $15 \leq x_{1,t} \leq 25$ for all $t \in\mathbb{N}_{\geq 0}$, and its acceleration is bounded as $|a_{l, t}| \leq 0.2$ for all $t \in\mathbb{N}_{\geq 0}$. The safe set is given by $\mathcal{X} = \mathcal{Z}\backslash \mathcal{O}$, where 
$\mathcal{Z} = \{ x \in \mathbb{R}^3\ |\ 15 \leq x_1 \leq 25,\ 15 \leq x_2 \leq 25, 30 \leq x_3  \leq 80 \}$, $\mathcal{O} = \{ x \in \mathbb{R}^3\ |\ 2 x_2 \leq x_3 \}$. 
The input constraint set is $\mathcal{U} = \{ u \in \mathbb{R}\ |\ |u| \leq 1.0\}$. The initial state is given by $\bar{x} = [20, 20, 60]^\mathsf{T}$, and $\eta_x =\varepsilon = 0.2$, $\eta_u = 0.2$, $T_{\rm exp} = 30$. Moreover, during the implementation of \ralg{overall_alg}, we incorporate \ralg{consym} and \ralg{compreduce} with $\rho = 0.01$ so as to reduce the computational load of abstractions and the safety controller synthesis. \textcolor{black}{We used a squared-exponential $ \mathsf{k}(x_t, x_{t'}) = \exp (-{\alpha |x_{t}-x_{t'}|})$ with $\alpha = 1$. 
%As mentioned in Section~3.1, computing the upper bound of the RKHS norm $\|d_2\|_{\mathsf{k}}$ may be hard in general, and it could be estimated based on evaluating the outputs from the function $d_i$. 
The computed upper bound of the RKHS norm was $\|d_2\|_{\mathsf{k}} \leq 2.0$ (for details on how to obtain this bound, see Appendix~A).}

\begin{figure}[t]
\centering
\includegraphics[width=8.5cm]{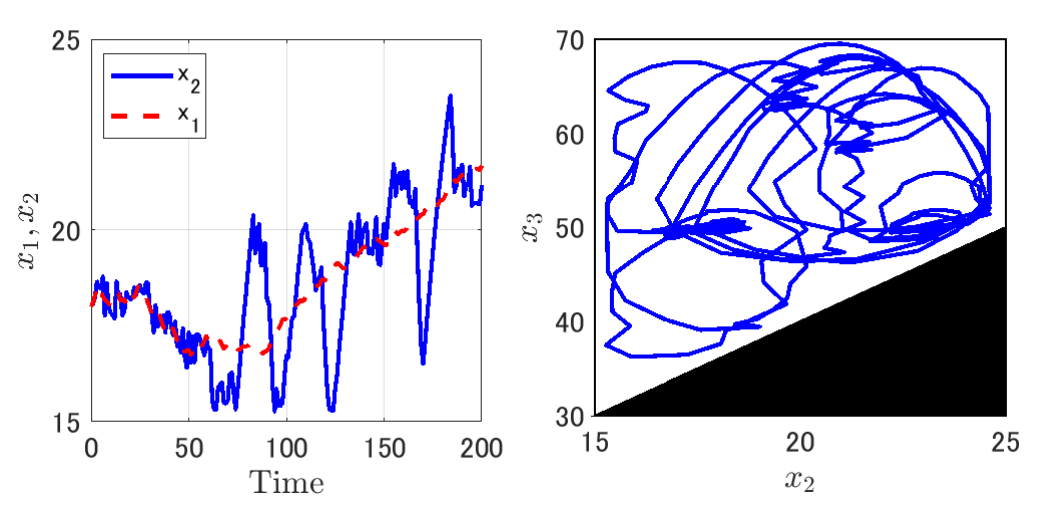}
\caption{The left figure illustrates the trajectories of $x_1$ and $x_2$ by \ralg{overall_alg}. The right figure illustrates the phase portrait in $x_2, x_3$ (the white region indicates the safe set $\mathcal{X}$).}
\label{simresult}
\end{figure}

\begin{figure}[t]
\centering
\includegraphics[width=8.5cm]{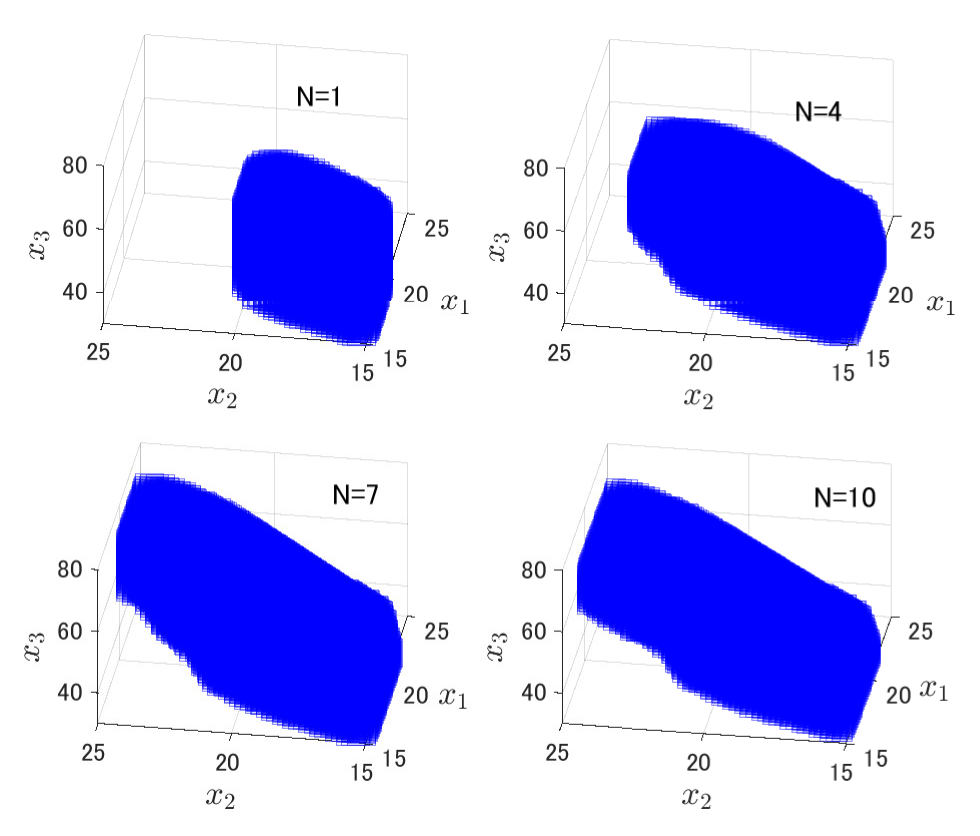}
\caption{\textcolor{black}{The computed controlled invariant set $\mathcal{X}_{S, N}$ for $N=1$ (upper left), $N=4$ (upper right), $N=7$ (lower left) and $N=10$ (lower right).}}
\label{invfig}
\end{figure}

\begin{figure}[b]
\centering
\includegraphics[width=8.5cm]{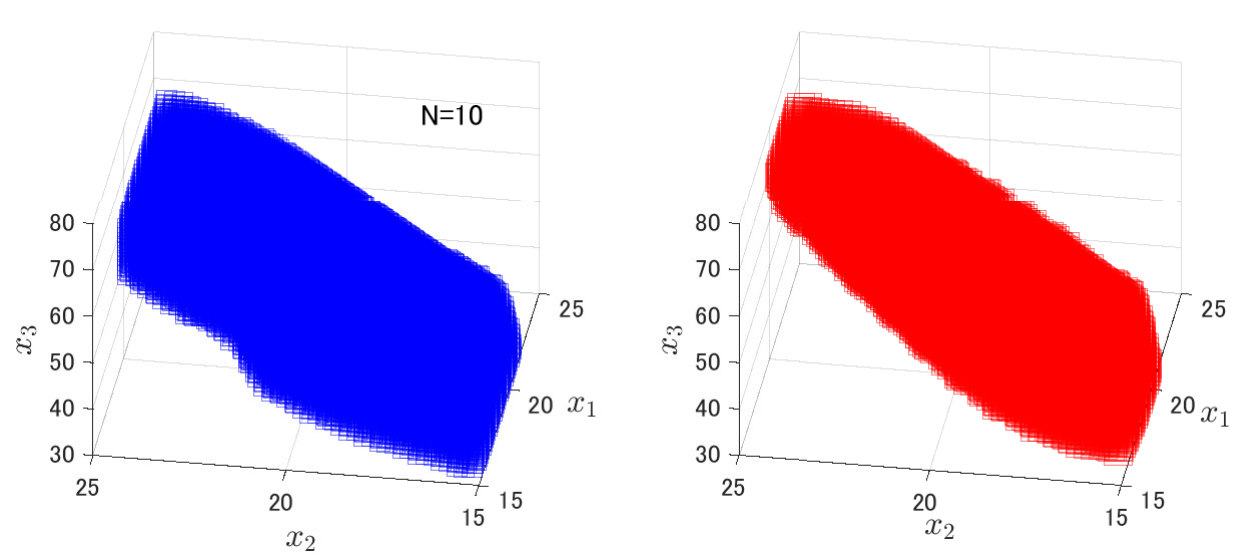}
\caption{The controlled invariant set finally obtained by applying the proposed approach (left) and the uniform disturbance-based approach (right).}
\label{proposeuniform}
\end{figure}

\textcolor{black}{For comparisons, we have also computed a 
symbolic model and a controlled invariant set by regarding $d_2(x_{2,t})=\nu_{0} + \nu_{1} x_{2,t} + \nu_{2} x_{2,t}^2/M$ as the \textit{uniform disturbance} (i.e., the aerodynamic drag force will not be learned from data). 
%and it is considered as the uniform disturbance. 
We assume that the uniform disturbance satisfies $0.30 \leq d_2 (x_{2, t}) \leq 0.71$, $\forall t \in \mathbb{N}$, since $\min_{x_2\in[15, 25]} d_2(x_2) = 0.30$ and $\max_{x_2\in[15, 25]} d_2(x_2) = 0.71$.
These lower and the upper bounds of $d_2$ have been utilized to construct the symbolic model and the controlled invariant set by following the abstraction procedure given in previous work, e.g., \cite{reissig}.}  

\rfig{simresult} shows the trajectories of $x_1$, $x_2$ by applying the proposed approach \ralg{overall_alg} and the phase portrait of $x_2$, $x_3$. The figure illustrates that the trajectories are always inside $\mathcal{X}$ (white region), showing the achievement of the safe exploration. 
The algorithm terminates at $N=10$. %and the resulting symbolic model $\widetilde{S} _{D, \mathsf{q}_N}$ has in total $35017$ states, $11$ inputs and $296727$ transitions. 
The computed controlled invariant sets for $N=1, 4, 7, 10$ are illustrated in \rfig{invfig}. 
%For comparisons, we also illustrate the controlled invariant set  with the assumption that $d$ is \textit{known} apriori. That is, the symbolic model is constructed based on the \textit{known} function $d$ (i.e., $\overline{r} _{i} ({x}_\mathsf{q}; \mathcal{D}_{T, i})$ and $\underline{r} _{i} ({x}_\mathsf{q}; \mathcal{D}_{T, i})$ in \req{overh} and \req{underh} are both replaced by $d_i ({x}_\mathsf{q})$), from which the controlled invariant set and the safety controller are computed. 
%the symbolic model, controlled invariant set and safety controller are computed based on the \textit{known} function $d$). 
%The result is shown in the lower right figure of \rfig{invfig}. 
The figure shows that the volume of the controlled invariant set is enlarged by collecting the training data according to \ralg{overall_alg}. 

\begin{figure}[t]
\centering
\includegraphics[width=8.7cm]{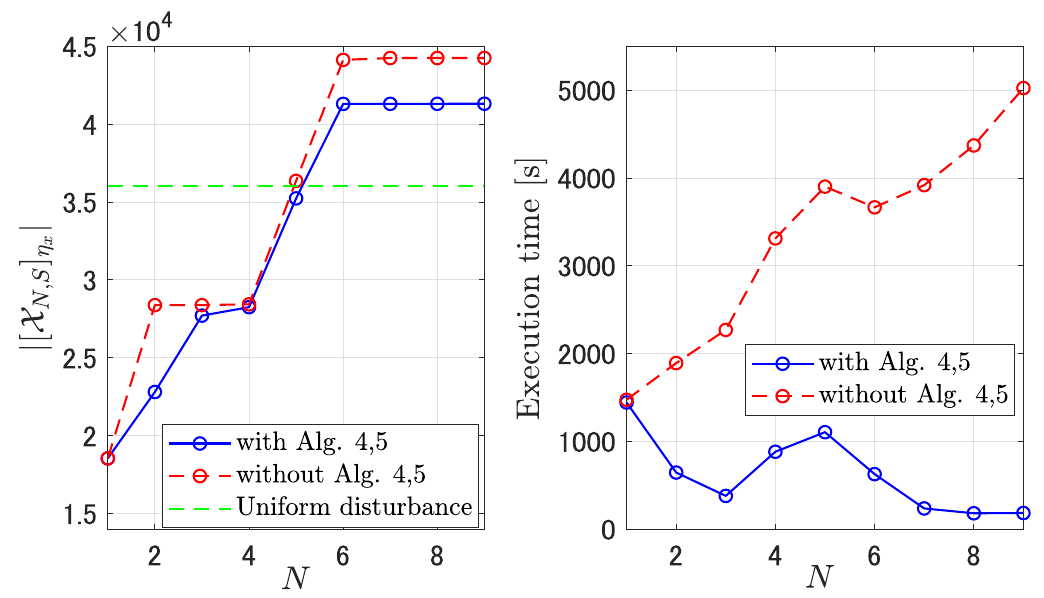}
\caption{The left figure indicates $|[\mathcal{X}_{S, N}]_{\eta_x}|$ with Algorithm~4 and 5 (blue solid), uniform disturbance-based approach (green dotted) and without Algorithm~4 and 5 (red dotted line). The right figure illustrates the total execution time to implement \rline{updatesym} and \rline{synsafecontroller} in \ralg{overall_alg}.}
\label{simresult2}
%\vspace{-0.4cm}
\end{figure}
\begin{figure}[t]
\centering
\includegraphics[width=7.5cm]{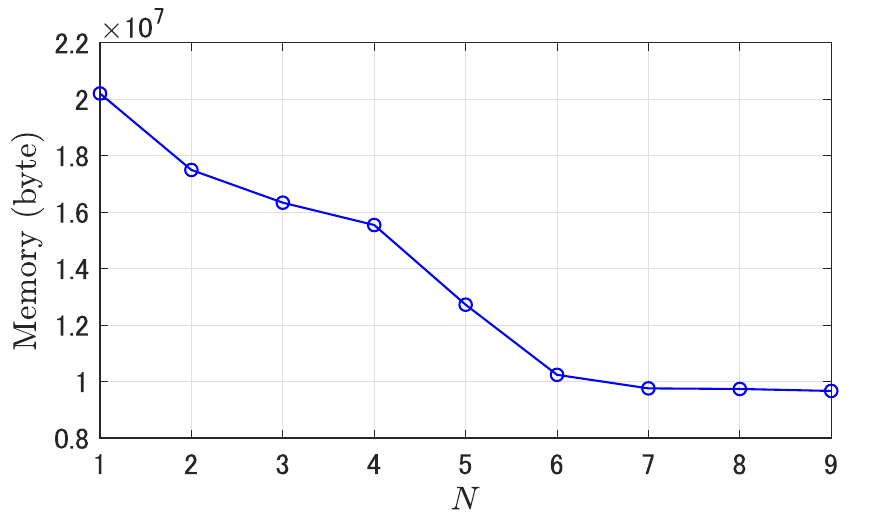}
\caption{\textcolor{black}{Required memory (in byte) to save all transitions for the symbolic model $S_{\mathsf{q},N}$ by the proposed approach.}}
\label{simresultmemory}
\end{figure}

\textcolor{black}{\rfig{proposeuniform} shows the controlled invariant set finally obtained by applying the proposed approach (which is equivalent to the lower right of \rfig{invfig}) and the uniform disturbance-based approach as described above. 
In addition, \rfig{simresult2} shows $|[\mathcal{X}_{S, N}]_{\eta_x}|$ (i.e., the cardinality or the number of states contained in $[\mathcal{X}_{S, N}]_{\eta_x}$) by applying the proposed approach against the number of iterations (blue dotted line) and the uniform disturbance-based approach (green dotted line). Note that the size of the controlled invariant set under the uniform disturbance-based approach is constant for all the iterations, since the unknown function is not learned. The figure shows that the controlled invariant set obtained by the proposed approach becomes larger than the uniform disturbance-based approach after $N=6$. This is because, by applying the proposed algorithm, the uncertainty of the unknown function becomes smaller as the iteration progresses, which results in reducing redundant transitions of the symbolic model (and thus enlarge the controlled invariant set); on the other hand, the uniform disturbance-based approach always considers the worst case effect of the disturbance, and thus the redundant transitions will not be removed. 
%while the uniform disturbance-based approach always considers the worst case effect that the disturbance has. 
Moreover, for further comparisons, we also implemented \ralg{overall_alg} \textit{neither} by employing \ralg{consym} (i.e., $S_{D, \mathsf{q}_N}$ in \req{sdn} is constructed for each $N$) \textit{nor} by employing \ralg{compreduce} for the safety controller synthesis, and the results are also plotted (red dotted lines). The right figure illustrates the total execution time to construct the symbolic model and the safety controller (i.e., the execution time to implement \rline{updatesym} and \rline{synsafecontroller} for each $N$ in \ralg{overall_alg}). The figure implies that the controlled invariant set by constructing the symbolic model $\widetilde{S}_{D, \mathsf{q}_N}$ is smaller than by constructing ${S}_{D, \mathsf{q}_N}$. As stated in \rrem{redundantrem}, this is due to the fact that in the former case the transitions are updated only for some states, while in the latter case these are updated for all states in $[\mathcal{X}]_{\eta_x}$. 
On the other hand, the total execution time (right figure in \rfig{simresult2}) by employing the former approach is shown to be significantly smaller than the latter approach, which illustrates the benefits of employing Algorithms~4 and~5.}
\textcolor{black}{Additionally, \rfig{simresultmemory} illustrates the required memory to save the symbolic model $S_{\mathsf{q},N}$ by applying the proposed approach (with Algorithms~4 and5). The figure shows that the required memory becomes smaller as $N$ increases, which is due to the fact that uncertainties of the unknown function are reduced and thus redundant transitions of the symbolic model are removed as the exploration progresses.}

\textcolor{black}{In the above, we have set the upper bound of the RKHS norm as $B_2 = 2$. 
In order to further analyze how the selection of $B_2$ affects the performance, we have implemented the proposed approach with different selections of $B_2$. 
\rtab{table:param} illustrates $|[\mathcal{X}_{S, N}]_{\eta_x}|$ obtained by applying the proposed approach with Algorithms 4 and 5 at the final iteration (i.e., when the controlled invariant set converges and Algorithm~2 terminates),
with $B_2$ being selected differently as $B_2 = 2, 50, 100, 200$. 
The table shows that, as $B_2$ is more conservatively chosen (i.e., it is selected larger), it results in obtaining smaller controlled invariant sets. Hence, it is of importance to select appropriate upper bound of the RKHS norm, i.e., it should be selected large enough so as to be the upper bound of $\|d_i\|_{\mathsf{k}_i}$, while it should be selected not too large so as not to be conservative, and investigating this trade-off should be given in our future work of research.} 

\begin{table}[tb]
  \caption{$|[\mathcal{X}_{S, N}]_{\eta_x}|$ by applying Algorithm~2 at the final iteration (i.e., when the controlled invariant set converges and Algorithm~2 terminated) with $B_2 = 2, 50, 100, 200$.}
  \label{table:param}
  \centering
  \begin{tabular}{lc}
    \hline
     $B_2$ & $|[\mathcal{X}_{S, N}]_{\eta_x}|$\\
     \hline
     $2$ & 41033 \\
     $50$ & 40210 \\ 
     $100$ & 39117  \\
     $200$ & 32617 \\
    \hline
  \end{tabular}
\end{table}

%Finally, \rtab{hoge} illustrates 
%implement \rline{updatesym} and \rline{synsafecontroller} for each $N$ in \ralg{overall_alg}. 

%and that it is getting closer to the case when $d$ is completely known apriori. 
%Moreover, \rtab{hoge} shows $|[\mathcal{X}_{S, N}]_{\eta_x}|$ (i.e., the cardinality or the number of states contained in $[\mathcal{X}_{S, N}]_{\eta_x}$) and the one for the case when $d$ is {completely known} apriori. 
%The result implies that the resulting controlled invariant set by applying \ralg{overall_alg} is smaller than the one for the case when $d$ is {known} apriori. %This may be due to the lack of 
%\ralg{overall_alg}, in which the symbolic model is constructed \textit{without} employing \ralg{consym} (i.e., $S_{D, \mathsf{q}_N}$ in \req{sdn} is constructed for each $N$) and the controlled invariant set is computed \textit{without} employing \ralg{compreduce}, and the results are also plotted in \rfig{simresult2}. 
%to compute $\widetilde{S}_{D, \mathsf{q}_N}$ is significantly reduced in contrast to the case ${S}_{D, \mathsf{q}_N}$, showing the usefulness of Algorithm~4 and 5. 
\begin{comment}
\begin{table}[b]
\begin{center}
\caption{\small Parameter settings} \label{params}
{ 
\begin{tabular}{l||l||l}
$\Delta: 2\ $s & $\nu_0$: 50\ N & $\nu_2$: 0.1\ Ns$^2$/m$^2$ \\
$M$: 1000\ kg & $\nu_1$: 2\ Ns/m& 
\end{tabular}}
\end{center}
\end{table}
\end{comment}
%\end{comment}

Now, using the learned symbolic model $\widetilde{S} _{D, \mathsf{q}_N}$, we can synthesize a controller satisfying complex control specifications, such as \textit{temporal logic formulas}. Following a correct-by-construction approach \cite{nilsson}, we encode the requirements for the ACC by the linear temporal logic (LTL). %First, let $\mathcal{X}_1, \mathcal{X}_2 \subset \mathcal{X}$ be given by $\mathcal{X}_1 = \{x \in \mathcal{X} \ |\ x^{*} _2 \leq x_3 / \omega^{*}\}$, $\mathcal{X}_2 = \mathcal{X}_1 \backslash \mathcal{X}_1$
First, consider two modes, called \textit{set-speed mode} and \textit{time-gap mode}. 
If the mode is in {set-speed mode}, the following vehicle must keep a given desired speed $x^* _2$ with some accuracy, i.e., $|x_2 - x^{*} _2 | \leq \epsilon_1$. If the mode is in time-gap mode, the following vehicle must achieve a desired {time headway} $\omega^{*}$ with some accuracy, i.e., $|x_3/x_2 - \omega^{*}| \leq \epsilon_2$. 
Let $\mathsf{mode}_1$, $\mathsf{mode}_2$ be atomic propositions, such that $\mathsf{mode}_1$ (resp. $\mathsf{mode}_2$) is satisfied if the mode is in set-speed mode (resp. the time-gap mode). It is assumed that $\mathsf{mode}_1$ (resp. $\mathsf{mode}_2$) is satisfied if the state is included in the set $\mathcal{X}_1 = \{x \in \mathcal{X} \ |\ x_3 \leq 60\}$ (resp. $\mathcal{X}_2 = \mathcal{X}_1 \backslash \mathcal{X}_1$). 
Let $\mathsf{spec}_1$, $\mathsf{spec}_2$ be the atomic propositions, such that $\mathsf{spec}_1$ (resp. $\mathsf{spec}_2$) is satisfied if $|x_2 - x^{*} _2 | \leq \epsilon_1$ (resp. $|x_3/x_2 - \omega^{*}| \leq \epsilon_2$). 
Moreover, let $\mathsf{safe}$ be the atomic proposition, such that it is satisfied if the state is included in $\mathcal{X}$.
\begin{figure}[t]
\centering
\includegraphics[width=8.6cm]{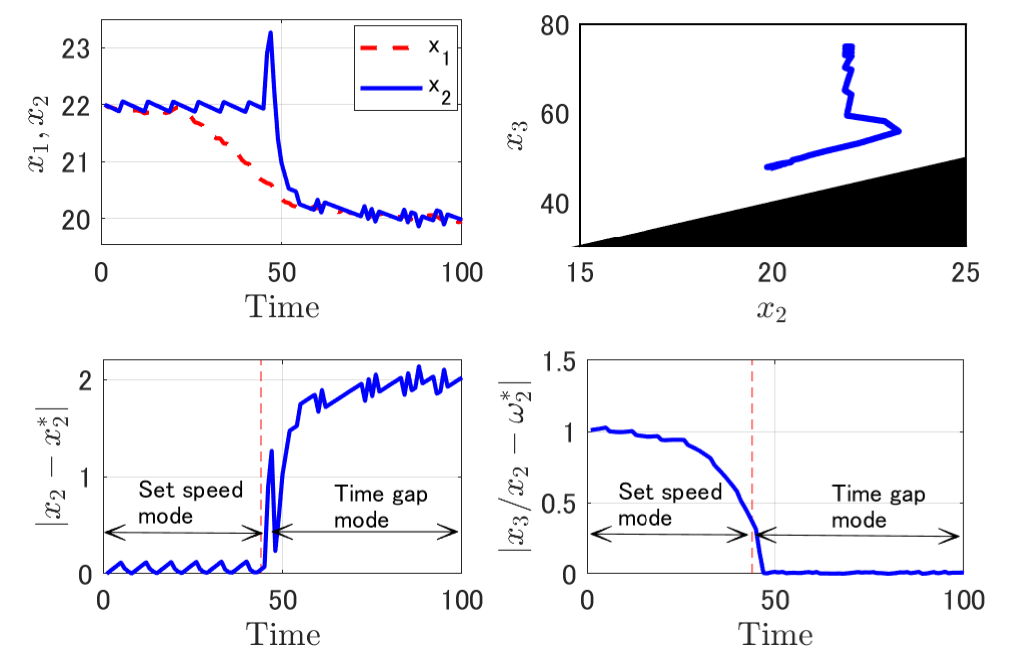}
\caption{The upper figures illustrate the trajectories of $x_1$ and $x_2$ (upper left) and the corresponding phase portrait in $x_2, x_3$ (upper right), by applying the synthesized controller satisfying $\psi$. The lower figures illustrate the absolute errors $|x_{2,t} - x^{*} _2 |$ (lower left) and $|x_{3,t}/x_{2,t} - \omega^{*}|$ (lower right) with $x^* _2 = 22$, $\omega^* = 2.4$.}
\label{simresult3}
%\vspace{-0.4cm}
\end{figure}
%Let $\mathcal{X}_1, \mathcal{X}_2 \subset \mathcal{X}$ be given by $\mathcal{X}_1 = \{x \in \mathcal{X} \ |\ x^{*} _2 \leq x_3 / \omega^{*}\}$, $\mathcal{X}_2 = \mathcal{X}_1 \backslash \mathcal{X}_1$, and suppose that the atomic proposition $M_i$ ($i\in\{1,2\}$) is associated to $\mathcal{X}_i$. Here, the atomic propositions $M_1$, $M_2$ are satisfied if and only if the mode is in set-speed mode and the time-gap mode, respectively. Let $B$ be the atomic proposition associated to $\mathcal{X}$. Moreover, let $\mathcal{Y}_1 =\{x\in \mathcal{X}\ |\ |x_2 - x^{*} _2 | \leq \epsilon_1\}$, $\mathcal{Y}_2 =\{x\in \mathcal{X}\ |\ |x_3/x_2 - \omega^{*}| \leq \epsilon_2\}$, and suppose that the atomic proposition $S_i$ ($i\in\{1,2\}$) is associated to $\mathcal{Y}_i$. 
%begin{comment}
%Our goal is then to synthesize a controller, such that the following specification $\psi$ is satisfied: 
Then, we encode the control specification by the LTL formula as follows: 
\begin{align}
\psi =\Box \mathsf{safe} \wedge \Box \wedge^2 _{i=1} (\mathsf{mode}_i \implies \bigcirc \bigcirc \mathsf{spec}_i), 
\end{align}
where $\Box$ and $\bigcirc$ are so-called the \textit{``always''} and \textit{``next''} temporal operators, respectively (see, e.g., \cite{baier}). In words, the state $x$ must always stay in the safe set $\mathcal{X}$, and if the mode is in set-speed mode (resp. time-gap mode), the following vehicle must achieve the desired speed in two time steps (resp. the desired time headway in two time steps). Note that the controller for the safety specification $\Box \mathsf{safe}$ has been already obtained after the implementation of \ralg{overall_alg}. The controller for the remaining part $\Box \wedge^2 _{i=1} (\mathsf{mode}_i \implies \bigcirc \bigcirc \mathsf{spec}_i)$ can be synthesized by a fixed point algorithm (see, e.g., \cite{nilsson}). The upper figures of \rfig{simresult2} indicate the state trajectories by employing the synthesized controller with $x^{*} _2 = 22, \omega^* = 2.4$ and $\epsilon_1 = 0.2, \epsilon_2 = 0.2$. 
Moreover, the lower figures indicate the sequences of the error $|x_2 - x^{*} _2 |$ and $|x_3/x_2 - \omega^{*}|$. %In the lower figures of \rfig{simresult3}, the gray areas indicate the time span when the mode is in set-speed mode (for the other regions, the mode is in time-gap mode). 
It can be verified that the formula $\psi$ is satisfied by applying the synthesized controller, showing the effectiveness of the proposed approach. 

\textcolor{black}{
\subsection{Some discussions on implementation issues}
In the proposed approach, the updates of the symbolic model, controlled invariant set and the safety controller are iteratively given as shown in Algorithm~2. 
While we proposed Algorithms~4 and 5 so as to reduce their computational costs, it might still take a long time to update them in practice. For instance, as shown in \rfig{simresult2}, Algorithms~4 and 5 took hundreds of seconds even with $N\geq 6$ (and thus, the updates could still be "slow" in terms of the online implementation). 
Nevertheless, it is argued that the proposed approach is still applicable and useful by employing the following technique; 
for each iteration $N \in \mathbb{N}_{\geq 0}$, during the updates of the symbolic model and the safety controller (i.e., during the execution of Algorithms~4 and 5), we can apply the safety controller obtained at the latest iteration (i.e., $C_{S,N-1}$), so that safety can be guaranteed even while updating the symbolic model and the safety controller. 
In other words, the construction of the symbolic model and the safety controller (Algorithms~4 and 5) could be done in an \textit{offline} fashion, while we keep collecting the training data online. 
%Hence, Algorithms~4 and 5 are still useful and applicable by employing the above technique, even though they could take a long time with respect to the online implementation. 
Note that, even though the updates of the symbolic model and the safety controller can be given offline as above, the proposed algorithms (Algorithms~4 and 5) are still necessary and useful, since they could lead to a significant reduction of the computational time in contrast to the case \textit{without} Algorithms~4 and 5 (see \rfig{simresult2}). 
In many practical scenarios, it is argued that the total run time to learn the safety controller is preferable to be as small as possible so as to reduce the operational cost (e.g., the energy consumption of the plant), and hence Algorithms~4 and 5 proposed in this paper are still beneficial. } 

\section{Conclusions and future works} 
In this paper, we propose a learning-based approach towards symbolic abstractions for nonlinear control systems. The symbolic model is constructed by learning the un-modeled dynamics from training data, and the concept of an $\varepsilon$-approximate alternating simulation relation. Moreover, the safe exploration has been achieved by iteratively updating the controlled invariant and the safety controller, employing the safety game. In addition, we provide several techniques to alleviate the computational load to construct the symbolic models and the controlled invariant set. Finally, we illustrate the effectiveness of the proposed approach through a simulation example of an adaptive cruise control. 

\textcolor{black}{In our problem setup, it is of great importance to compute the upper bound of the RKHS norm $\|d_i\|_{\mathsf{k}_i} \leq B_i$ since it has been utilized to construct the symbolic models. 
%As described in Appendix~A, $\|d_i\|_{\mathsf{k}_i}$ could be estimated at the initial exploration phase. However, if we cannot obtain enough training data, it could lead to poor estimation accuracy for $\|d_i\|_{\mathsf{k}_i}$. 
Hence, as described in Section~6, obtaining a large enough, yet not too conservative bound for $\|d_i\|_{\mathsf{k}_i}$ should be further investigated in the future. In addition, since there exist no outliers in our problem setup, investigating how these can affect (if they exist) the estimation accuracy of the unknown function as well as how to detect them should be further pursued in future work. }

%\bibliographystyle{unsrt}
% \bibliography{myrefs}

\appendix 
\section{\textcolor{black}{On computing an upper bound of $\|d_i\|_{\mathsf{k}_i}$}}
\textcolor{black}{Let $\mathcal{D}_{T, i} = \{{{ X}}_T, { Y}_{T, i} \}$ denote the set of the data with ${{ X}}_T = \left [{x}_1, {x}_2, \ldots, {x}_{{T}} \right ]$, ${ Y}_{T, i} = [y_{1,i}, y_{2,i}, \ldots, y_{T,i} ]^\mathsf{T}$ with $y_{t, i} = x_{t+1, i} - f_i (x_t, u_t)$, $\forall t \in \mathbb{N}_{1:{T}}$. 
Then, letting $d^{*}_{T,i} = [d_i (x_1), d_i (x_2), \ldots, d_i (x_T)]^\mathsf{T}$, it follows that $\sqrt{d^{*\mathsf{T}}_{T,i} K^{-1}_{T,i} d^{*}_{T,i}} \leq \|d_i\|_{\mathsf{k}_i}$ for all $T\in \mathbb{N}_{>0}$, and, moreover, $d^{*\mathsf{T}}_{T,i} K^{-1}_{T,i} d^{*}_{T,i}$ has the monotonicity property (that is, as $T$ increases, $d^{*\mathsf{T}}_{T,i} K^{-1}_{T,i} d^{*}_{T,i}$ increases); for details, see Appendix~A in \cite{emilio2021}. 
In other words, the more training data we get, the closer $\sqrt{d^{*\mathsf{T}}_{T,i} K^{-1}_{T,i} d^{*}_{T,i}}$ becomes to the true norm $\|d_i\|_{\mathsf{k}_i}$. 
The example provided in \cite{emilio2021} indeed illustrates that $\sqrt{d^{*\mathsf{T}}_{T,i} K^{-1}_{T,i} d^{*}_{T,i}}$ converges quickly to $\|d_i\|_{\mathsf{k}_i}$ as the number of the training data increases. The upper bound of the RKHS norm can thus be obtained by evaluating the convergence of $\sqrt{d^{*\mathsf{T}}_{T,i} K^{-1}_{T,i} d^{*}_{T,i}}$.}
%Thus, we can regard $\sqrt{d^{*\mathsf{T}}_{T,i} K^{-1}_{T,i} d^{*}_{T,i}}$ as a suitable estimate for $\|d_i\|_{\mathsf{k}_i}$, based on which the upper bound for $\|d_i\|_{\mathsf{k}_i}$can also be computed.} 

\begin{figure}[t]
\centering
\includegraphics[width=8.cm]{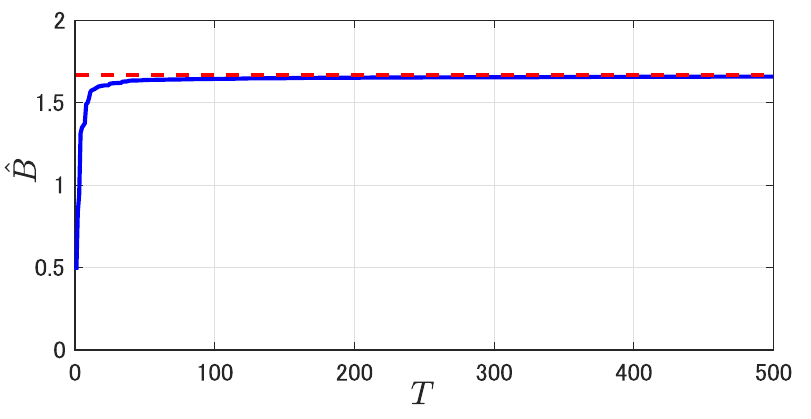}
\caption{\textcolor{black}{The quantities of $\sqrt{d^{*\mathsf{T}}_{T,2} K^{-1} _T d^{*}_{T,2}}$, $T=1,\ldots, 500$ obtained for the function $d_2$ used in the numerical experiment of Section~6, where $X_{T,2} = [x_{2,1}, \ldots, x_{2,T}]$ were randomly sampled from the interval $[15, 25]$. At $T=500$, we have $\hat{B} = 1.67$ (red dotted line).}}
\label{rkhsnorm}
\end{figure}

\textcolor{black}{For example, in the numerical experiment of Section~6, we obtained the upper bound of the RKHS norm via Monte-Carlo evaluations for $d_2 (x_{2,t}) = (\nu_{0} + \nu_{1} x_{2,t} + \nu_{2} x_{2,t}^2)/M$. \rfig{rkhsnorm} shows the quantity of $\hat{B}=\sqrt{d^{*\mathsf{T}}_{T,2} K^{-1} _T d^{*}_{T,2}}$, where $X_{T,2} = [x_{2,1}, \ldots, x_{2,T}]$ were randomly sampled from the interval $[15, 25]$. Based on this result, we took the upper bound of the RKHS norm conservatively as $\|d_2\|_{\mathsf{k}} \leq B_2 = 2$.} 

\section{Proof of \rlem{lipschiz}}
\noindent
From \ras{rkhsd}, we have $|d_i (x_1) - d_i (x_2) |^2 \leq \|d_i\|^2_{\mathsf{k}_i} \{\mathsf{k}_i (x_1, x_1)- 2 \mathsf{k}_i (x_1, x_2) + \mathsf{k}_i (x_2, x_2) \}$; see Lemma~4.28 in \cite{christmann}. 
Moreover, $|\mathsf{k}_i (x_1, x_1) - 2 \mathsf{k}_i (x_1, x_2) + \mathsf{k}_i (x_2, x_2)| \leq |\mathsf{k}_i (x_1, x_1) - \mathsf{k}_i (x_1, x_2)| + |\mathsf{k}_i (x_2, x_2) - \mathsf{k}_i (x_1, x_2)|$, and 
\begin{align}
|\mathsf{k}_i (x_1, x_1) &- \mathsf{k}_i (x_1, x_2)| \notag \\ 
&\leq \sup_{y \in \mathcal{X}} |\mathsf{k}_i (x_1, y) - \mathsf{k}_i (x_2, y)| \notag \\
&\leq \sup_{y \in \mathcal{X}} \| \partial \mathsf{k}_i (x, y)/\partial x\|_\infty \cdot \|x_1 - x_2 \|_{\infty} \notag \\ 
&\ \ \ \ = \| \partial \mathsf{k}_i/\partial x\|_\infty \|x_1 - x_2 \|_{\infty}. \notag
\end{align}
Similarly, we have $|\mathsf{k}_i (x_2, x_2) - \mathsf{k}_i (x_1, x_2)| \leq \| \partial \mathsf{k}_i/\partial x\|_\infty$ $\|x_1 - x_2 \|_{\infty}$. 
Hence, we have $|d_i (x_1) - d_i (x_2) |^2 \leq 2 \|d_i\|^2 _{\mathsf{k}_i} \| \partial \mathsf{k}_i/\partial x\|_\infty \|x_1 - x_2 \|_{\infty}$, completing the proof. 

\section{Proof of \rlem{boundlemma}}
%For simplicity, we only show the proof for the function $d_i$. 
%Given the training data $\mathcal{D}_{T, i} = \{X_n, Y_{n, i}\}$, 
%In the following derivations, we omit the subscript $i$ for simplicity. 
The proof follows the one of Lemma~7.2 in \cite{srinivas2}, while here we provide a more detailed derivation. 
First, letting $\alpha^*_{t,i} = (K_{t,i} + \sigma^2_v I)^{-1} Y_{t,i}$, we have $\mu_i (x; \mathcal{D}_{t,i}) = \alpha^{*\mathsf{T}}_{t,i} \mathsf{k}^*_{t,i} (x)$. 
Since $d_i$ lies in the RKHS, it is characterized by $d_i(x) = \sum_n \beta_{n,i} \mathsf{k}_i(x, x^{r}_{n,i})$ ($\beta_{n,i}$ are coefficients and $x^r_{n,i}$ are representer points). From the definition of the inner product (which we denote by $\langle \cdot \rangle_{\mathsf{k}_i}$), we have 
\begin{align*}
    \langle\mu_i (\cdot; \mathcal{D}_{T,i}), d_i \rangle_{\mathsf{k}_i} &= \sum_{t=1}^T \sum_{n} \beta_{n,i} \alpha_{t,i} \mathsf{k}_i(x_t, x^r_{n,i})  \\ 
    &= \sum_{t=1}^T \alpha_{t,i} \sum_{n} \beta_{n,i} \mathsf{k}(x_t, x^r_{n,i}) = d^{*\mathsf{T}}_{T,i} \alpha^* _{T,i} 
\end{align*}
 where we denote $\alpha^* _{T,i} = [\alpha_{1,i}, \alpha_{2,i}, \ldots, \alpha_{T,i}]^\mathsf{T}$ (i.e., $\alpha_{t,i}$ is the $t$-th element of $\alpha^* _{T,i}$) and $d^{*}_{T,i} = [d_i(x_1), \ldots, d_i(x_T)]^\mathsf{T}$. In addition, we have 
 \begin{align*}
 \|{\mu}_i (\cdot ; \mathcal{D}_{T,i})\|^2 _{\mathsf{k}_i} &= \langle {\mu}_i (\cdot ; \mathcal{D}_{T,i}), {\mu}_i (\cdot ; \mathcal{D}_{T,i})\rangle \\ 
 &= \sum_{t=1}^T \sum_{t'=1} ^T \alpha_{t,i} \alpha_{t',i} \mathsf{k}_i (x_t, x_{t'}) =  \alpha^{*\mathsf{T}} _{T,i} K_{T,i} \alpha^*_{T,i} \\
& = Y^\mathsf{T} _{T,i} (K_{T,i} + \sigma^2_v I)^{-1} K_{T,i} \alpha^* _{T,i} \\ 
&= Y^\mathsf{T} _{T,i} \left\{ I -\sigma^2_v (K_{T,i} + \sigma^2_vI)^{-1}\right\} \alpha^*_ {T,i} \\ 
&= Y^\mathsf{T} _{T,i} \alpha^* _{T,i} - \sigma^2_v Y^\mathsf{T} _{T,i} (K_{T,i} + \sigma^2_vI)^{-1} \alpha^* _{T,i}\\ 
&= Y^\mathsf{T} _{T,i} \alpha^* _{T,i} - \sigma^2_v\|\alpha^* _{T,i}\|^2, 
 \end{align*}
 where we used $(K_{T,i} + \sigma^2_v I)^{-1}K_{T,i} = I -\sigma^2_v (K_{T,i} + \sigma^2_vI)^{-1}$ (to see this, we have $I-(K_{T,i} + \sigma^2_v I)^{-1}K_{T,i} = (K_{T,i} +\sigma^2_vI )^{-1}(K_{T,i} +\sigma^2_vI ) -(K_{T,i} + \sigma^2_v I)^{-1}K_{T,i} = (K_{T,i} +\sigma^2_vI )^{-1} (K_{T,i} + \sigma^2_vI-K_{T,i}) = \sigma^2_v (K_{T,i} + \sigma^2_v I)^{-1}$). In addition, $\mathsf{k}^* _{T,i} (x_t) = [\mathsf{k}_i(x_t, x_1), \ldots,\mathsf{k}_i(x_t, x_T)]^\mathsf{T} = K_{T,i} \delta_t$, where $\delta_t$ is the $T$-dimentional vector whose element is 1 for the $t$-the element and 0 otherwise. Hence, 
 \begin{align*}
     {\mu}_i (x_t ; \mathcal{D}_{{T,i}}) &= \mathsf{k}^{*\mathsf{T}} _{T,i} (x_t)(K_{T,i} + \sigma^2_v I)^{-1} Y_{T,i} \\ 
     &= \delta_t^\mathsf{T} K_{T,i} (K_{T,i} + \sigma^2 I)^{-1} Y_{T,i}\\ 
     &= \delta_t^\mathsf{T} \left\{ I -\sigma^2_v (K_{T,i} + \sigma^2_vI)^{-1}\right\}Y_{T,i} \\
     &= \delta_t^\mathsf{T}\left\{Y_{T,i} -\sigma^2_v (K_{T,i} + \sigma^2_vI)^{-1}Y_{T,i}\right\}\\ 
     &=\delta_t^\mathsf{T}(Y_{T,i} - \sigma^2_v \alpha^*_{T,i})=y_{t,i} - \sigma^2_v \alpha_{t,i} 
 \end{align*}
where $y_{t,i}$ is the $t$-th element of $Y_{T,i}$, and again, we used $K_{T,i}(K_{T,i} + \sigma^2 _v I)^{-1} =\{ (K_{T,i} + \sigma^2 _v I)^{-1}K_{T,i} \}^\mathsf{T} =\{ I -\sigma^2 _v (K_{T,i} + \sigma^2_vI)^{-1}\}^\mathsf{T} =I -\sigma^2_v (K_{T,i} + \sigma^2_vI)^{-1}$. 
Let $\mathsf{k}_{{T,i}} : \mathbb{R}^{n_x} \times \mathbb{R}^{n_x} \rightarrow \mathbb{R}_{\geq 0}$ be given by $\mathsf{k}_{{T,i}}(x,x') = \mathsf{k}_i ({ x} , { x}' )- \mathsf{k}^{*\mathsf{T}}_{{T,i}} ({ x} ) (K_{T,i} + \sigma^2 _v {I})^{-1} \mathsf{k}^{*\mathsf{T}}_{{T,i}} ({ x}')$, 
and let $\|\cdot\|_{\mathsf{k}_{{T,i}}}$ be the RKHS norm corresponding to $\mathsf{k}_{{T,i}}$. 
Since we have $\|{\mu}_i (\cdot ; \mathcal{D}_{{T,i}}) - d_i(\cdot) \|^2 _{\mathsf{k}_{{T,i}}} = \|{\mu} (\cdot ; \mathcal{D}_{{T,i}}) - d_i(\cdot) \|^2 _{\mathsf{k}_i} + \sigma^{-2}_v \sum_{t=1}^T \{\mu_i(x_t; \mathcal{D}_{{T,i}}) -d_i(x_t)\}^2$ (see \cite{srinivas2}), we obtain 
\begin{align*}
& \| {\mu}_i (\cdot ; \mathcal{D}_{{T,i}}) - d_i(\cdot) \|^2 _{\mathsf{k}_{{T,i}}} \\
    & = \|{\mu}_i (\cdot ; \mathcal{D}_{{T,i}}) - d_i(\cdot) \|^2 _{\mathsf{k}_i} + \sigma^{-2}_v \sum_{t=1}^T \{\mu_i(x_t; \mathcal{D}_{{T,i}}) -  d_i(x_t)\}^2 \\ 
    &= \|{\mu}_i (\cdot ; \mathcal{D}_{T,i})\|^2_{\mathsf{k}_i} - 2 \langle \mu(\cdot; \mathcal{D}_{T,i}), d_i \rangle_{\mathsf{k}_i} + \|d_i\|^2_{\mathsf{k}_i} \\ 
    &\ \ \  + \sigma^{-2}_v \sum_{t=1}^T \{y_{t,i} - \sigma^2_v\alpha_{t,i} - d_i(x_t)\}^2 \\ 
    %&=\|d_i\|^2_{\mathsf{k}} +Y^\mathsf{T} _{T,i} \alpha^* _{T,i} - \sigma^2_v\|\alpha^* _{T,i}\|^2 - 2 d^{*\mathsf{T}}_{T,i} \alpha^*_{T,i} \\ 
    &=  \|d_i\|^2_{\mathsf{k}_i} +Y^\mathsf{T} _{T,i} \alpha^* _{T,i} - \sigma^2_v \|\alpha^* _{T,i}\|^2 - 2 d^{*\mathsf{T}}_{T,i} \alpha^*_{T,i} \\ 
    &\ \ \ + \sigma^{-2}_v (\|v^*_{T,i}\|^2 - 2\sigma^2_v v^{*\mathsf{T}}_{T,i}\alpha^{*}_{T,i} +\sigma^4_v \|\alpha^{*}_{T,i}\|^2) \\
    &=\|d_i\|^2_{\mathsf{k}} - 2 (d^{*}_{T,i} + v^*_{T,i})^\mathsf{T} \alpha^*_{T,i} + Y^\mathsf{T} _{T,i} \alpha^* _{T,i} + \sigma^{-2}_v \|v^*_{T,i}\|^2 \\ 
    &= \|d_i\|^2_{\mathsf{k}} - 2 Y^\mathsf{T} _T \alpha^* _T + Y^\mathsf{T} _T \alpha^* _T + \sigma^{-2}_v \|v^*_T\|^2 \\ 
    &= \|d_i\|^2_{\mathsf{k}_i} - {Y}^\mathsf{T} _{{T,i}} (K_{{T,i}}+{\sigma}^2 _v I)^{-1}{Y} _{{T,i}} + \sigma^{-2}_v \|v^*_{T,i}\|^2, 
\end{align*}
where we denote $v^*_{T,i} = [v_{1,i}, v_{2,i}, \ldots, v_{T,i}]^\mathsf{T}$ and $v_{t,i}$ is the $i$-th element of $v_t$. 
Thus, we have $\|{\mu}_{i} (\cdot ; \mathcal{D}_{T, i}) - d_i(\cdot) \|^2 _{\mathsf{k}_{T,i}}\leq B^2_i - {Y}^\mathsf{T} _{T, i} (K_{T,i}+{\sigma}^2 _v I)^{-1}{Y} _{T, i}+ \sigma^{-2} _v \sum^T _{t=1} v^2 _{t,i}$. 
%where $v_{t,i}$ is the $i$-th element of $v_t$ (for the above derivation, please see the proof of Lemma~7.2 in \cite{srinivas2}). 
Hence, using the fact that $|v_{t,i}| \leq \sigma_v$, $\forall t\in\mathbb{N}_{1:T}$, we have 
$\|{\mu}_{i} (\cdot ; \mathcal{D}_{T, i}) - d_i(\cdot) \|^2 _{\mathsf{k}_{T,i}} \leq B^2_i - {Y}^\mathsf{T} _{T, i} (K_{T,i}+{\sigma}^2 _v I)^{-1}{Y} _{T, i} + T$. 
Moreover, we have $|{\mu}_{i} (x ; \mathcal{D}_{T, i}) - d_i (x)| \leq \mathsf{k}_{T,i}(x, x)^{-1/2} \|{\mu}_{i} (\cdot ; \mathcal{D}_{T, i}) - d_i(\cdot)\|_{\mathsf{k}_{T,i}}= {\sigma}_{i} (x ; \mathcal{D}_{T, i}) \|{\mu}_{i} (\cdot ; \mathcal{D}_{T, i}) - d_i(\cdot)\|_{\mathsf{k}_{T,i}}$,
where the first inequality follows from the Cauchy-Schwarz inequality. Then, we obtain $|{\mu}_{i} (x ; \mathcal{D}_{T, i}) - d_i (x)| \leq \beta_{T,i} {\sigma}_{i} (x ; \mathcal{D}_{T, i})$ for all $T \in \mathbb{N}_{>0}$, completing the proof. 
%Hence, it follows that $d_i (x) \in \mathcal{Q}_{i} (x; \mathcal{D}_{T, i})$, for all $T \in \mathbb{N}_{> 0}$. %where $\mathcal{Q}_{i} (x; \mathcal{D}_{T, i})$ is the interval set given by \req{qdi}. %This implies that $d_i (x) \in \mathcal{R}_{i} (x; \mathcal{D}_{T, i})$, where $\mathcal{R}_{i} (x; \mathcal{D}_{T, i}) = \bigcap^T _{t=1} \mathcal{Q}_{i} (x; \mathcal{D}_{t, i})$. %$\mathcal{R}_{i}$ is the intersections of all $\mathcal{Q}_{i} (x; \mathcal{D}_{t, i})$, $t \in \mathbb{N}_{1:T}$.  

\section{Proof of \rprop{asrresult}}

%\begin{pf}
%Let $R(\varepsilon)$ be given by \req{relation} with $\varepsilon \geq \eta_x$. 
%Since ${X}_{{\mathsf{q}}0} \subseteq X_0$, for every $\tilde{x}_{0} \in {X}_{{\mathsf{q}}0}$ there exists ${x}_0 = \tilde{x}_{0} \in X_0$ such that $\|\tilde{x}_{0}  - x_0 \|_{\infty} = 0 \leq \varepsilon$. 
The condition (C.1) in \rdef{asr2} trivially holds from $(x_{\mathsf{q}0}, x_0) \in R(\varepsilon)$. 
Moreover, the condition (C.2) is satisfied from the definition of $R(\varepsilon)$. To check (C.3), consider any $({x}_{{\mathsf{q}}}, x) \in R(\varepsilon)$ and ${u}_\mathsf{q} \in \mathcal{U}_{\mathsf{q}}$. Let $u = {u}_\mathsf{q} \in \mathcal{U}$ and consider $x^+ \in G (x, u)$, implying that there exists $v \in \mathcal{V}$ with $v=[v_1,\ldots,v_{n_x}]^\mathsf{T}$ such that $x^+ _i = f_i({x}, {u}) + d_i (x)+v_i$ for all $i\in\mathbb{N}_{1:n_x}$. 
Pick ${x}^+ _\mathsf{q} \in \mathsf{Nearest}_{\mathcal{X}_{\mathsf{q}}} (x^+)$. 
It follows that $\| {x}^+ _\mathsf{q}  - x^+\|_\infty \leq \eta_x$, i.e., $({x}^+ _\mathsf{q} , x^+) \in R (\varepsilon)$. 
Now, let us show that ${x}^+ _{\mathsf{q}} \in G_{\mathsf{q}}({x}_{\mathsf{q}},{u}_{\mathsf{q}})$: 
\begin{align}
|&{x}^+ _{\mathsf{q},i} - f_i({x}_\mathsf{q}, {u}_\mathsf{q}) - \hat{d}_i ({x}_{\mathsf{q}}; \mathcal{D}_{T, i}) | \notag \\ 
&\leq |x^+ _i - f_i({x}_\mathsf{q}, {u}_\mathsf{q}) - \hat{d}_i ({x}_\mathsf{q}; \mathcal{D}_{T, i})| + \eta_x \notag \\  
& \leq |f_i({x}, {u}) + d_i (x)+v_i - f_i({x}_\mathsf{q}, {u}_\mathsf{q}) - \hat{d}_i ({x}_\mathsf{q}; \mathcal{D}_{T, i}) | + \eta_x \notag \\ 
&\leq L_f  \varepsilon + L_i \sqrt{\varepsilon} + \eta_x +\sigma_v + |d_i ({x}_\mathsf{q}) - \hat{d}_i ({x}_\mathsf{q}; \mathcal{D}_{T, i})|, \notag 
%& \leq  (L_f + L_g) \varepsilon + \eta_x + \beta_{T, i} {\sigma}_{N, i} (\tilde{x}, \tilde{u}).  \notag 
\end{align}
where $\hat{d}_i (x; \mathcal{D}_{T, i})$ is defined in \req{gihat}. 
 %we simply denote $\mu_{i}(\tilde{x}, \tilde{u}) = \mu_{i}(\tilde{x}, \tilde{u}; \mathcal{D}_{T, i})$. 
Hence, $|{x}^+ _{\mathsf{q}, i} - f_i({x}_\mathsf{q}, {u}_\mathsf{q}) - \hat{d}_i ({x}_\mathsf{q}; \mathcal{D}_{T, i}) | \leq L_f  \varepsilon + L_i \sqrt{\varepsilon} + \eta_x + \sigma_v+{\Delta}_{i} ({x}_\mathsf{q}; \mathcal{D}_{T, i})$,
%\begin{align}
%|{x}^+ _{\mathsf{q}, i} &- f_i({x}_\mathsf{q}, {u}_\mathsf{q}) - \hat{d}_i ({x}_\mathsf{q}; \mathcal{D}_{T, i}) | \notag \\ 
%&\leq (L_f + L_i  ) \varepsilon + \eta_x + \sigma_v+{\Delta}_{d_i} ({x}_\mathsf{q}; \mathcal{D}_{T, i}), 
%\end{align}
where ${\Delta}_{i} ({x}_\mathsf{q}; \mathcal{D}_{T, i})$ is defined in \req{deltahat}. 
From the above, ${x}^+ _{\mathsf{q},i} \in [\underline{h}_i ({x}_\mathsf{q}, {u}_\mathsf{q}; \mathcal{D}_{T, i}), \overline{h}_i ({x}_\mathsf{q}, {u}_\mathsf{q}; \mathcal{D}_{T, i})]$, $\forall i\in\mathbb{N}_{1:n_x}$, which implies ${x}^+ _{\mathsf{q}} \in G_{\mathsf{q}}({x}_{\mathsf{q}}, {u}_{\mathsf{q}})$ with $({x}^+ _{\mathsf{q}},x^+)\in R(\varepsilon)$. 
%This implies that there exists ${x}^+ _\mathsf{q} \in  {G}_{\mathsf{q}} ({x}_\mathsf{q}, {u}_\mathsf{q})$ such that $({x}^+ _\mathsf{q} , x^+) \in R(\varepsilon)$. 
Hence, $R(\varepsilon)$ is an $\varepsilon$-ASR from $S_{\mathsf{q}}$ to $S$. 
%\end{pf}

%\bibliographystyle{unsrt}
%\bibliographystyle{elsarticle-num}
%\bibliographystyle{model5-names}
%\bibliography{myrefs}
\end{document}